\documentclass{aa}  
\usepackage{graphicx}
\usepackage[colorlinks=true, linkcolor=blue, citecolor=blue, urlcolor=blue]{hyperref}
\usepackage{amsmath}	% Advanced maths commands
\usepackage{physics,amsmath}
\usepackage[dvipsnames, table]{xcolor}
\usepackage{newtxtext,ulem}
\usepackage[varvw]{newtxmath} % change v which looks like \nu to \varv
\everymath{\thickmuskip=2mu minus 2mu} % reduce spaces in all inline math !!!
\graphicspath{{figs/}}
\usepackage{defs}
\begin{document}
%%%%%%%%%%%%%%%%%%%%%%%%%%%%%%%%%%%%%%%%
\title{Probing the Cosmic Web with Fast Radio Bursts}
\subtitle{I. Scattering}
\titlerunning{Probing the CW with FRBs -- I. Scattering}
%%%%%%%%%%%%%%%%%%%%%%%%%%%%%%%%%%%%%%%%
\author{Sharon Lapiner\inst{1}
    \fnmsep\thanks{Corresponding author: sharon.lapiner@mail.huji.ac.il}
    \and Nir Mandelker\inst{1}
    \and Paz Beniamini\inst{2,3,4}
    \and S. Peng Oh\inst{5}
    }
\authorrunning{S. Lapiner et al.}
\institute{
    Racah Institute of Physics, The Hebrew University, Jerusalem 91904 Israel
    \and 
    Astrophysics Research Center of the Open University (ARCO), The Open University of Israel, P.O. Box 808, Ra’anana 43537, Israel
    \and 
    Department of Natural Sciences, The Open University of Israel, P.O Box 808, Ra'anana 4353701, Israel
    \and
    Department of Physics, The George Washington University, 725 21st Street NW, Washington, DC 20052, USA
    \and
    Department of Physics, University of California, Santa Barbara, CA 93106, USA
}
%%%%%%%%%%%%%%%%%%%%%%%%%%%%%%%%%%%%%%%%
% \date{Received xxxx xx, 2026}
% @@@@@@@@@@@@@@@@@@@@@@@@@@@@@@@@@@@@@@@@@
% @@@@@@@@@@@@@@@@@@@@@@@@@@@@@@@@@@@@@@@@@
\abstract
{
We study the formation of multiphase gas in the post-accretion-shock regions of cosmic sheets, filaments, and the circumgalactic medium (CGM) of haloes, i.e., cosmic web objects (CWOs). Local instabilities in the hot medium result in fragmentation and cooling, eventually forming small-scale overdensities with temperatures of $\sim10^{4}\Kel$ in pressure equilibrium with the hot environment.
Such dense, ionised inhomogeneities can affect the propagation of radio waves from fast radio bursts (FRBs), thereby offering us a way to probe their presence and properties in CWOs through scattering signatures in the observed FRB flux.
We find that high-$z$ filaments \& sheets have a negligible contribution to the total observed scattering. 
The high rates of FRBs expected even at high redshifts may still allow detection from high-temperature filaments along rare sightlines, and we suggest other methods for such systems in a companion paper.
Our model further predicts that if turbulent cloudlets exist in the CGM of intervening massive haloes with a volume-filling fraction of $\fv\gsim10^{-3}$, they are expected to cause considerable cumulative scattering along an average sightline, resulting in a significant correlation between the total scattering time and source redshifts.
The lack of such a correlation in current observations may imply that the cool gas in the CGM has substantial non-thermal pressure, reducing its density, or significant damping of small-scale density fluctuations. Forthcoming localised FRB samples can map these constraints into bounds on volume-filling fractions, densities, cloud sizes, and the strength of turbulence. 
}
% @@@@@@@@@@@@@@@@@@@@@@@@@@@@@@@@@@@@@@@@@
% @@@@@@@@@@@@@@@@@@@@@@@@@@@@@@@@@@@@@@@@@
% @@@@@@@@@@@@@@@@@@@@@@@@@@@@@@@@@@@@@@@@@
% @@@@@@@@@@@@@@@@@@@@@@@@@@@@@@@@@@@@@@@@@
\maketitle
\nolinenumbers
% @@@@@@@@@@@@@@@@@@@@@@@@@@@@@@@@@@@@@@@@@
\section{Introduction}
% @@@@@@@@@@@@@@@@@@@@@@@@@@@@@@@@@@@@@@@@@

% CGM and IGM: Definitions, importance, observations with absorption and emission
Galaxies contain only a small fraction of the Universe's baryons, accounting for both stars and the interstellar medium (ISM) \citep[e.g.][]{Peeples14, Tumlinson2017, Wechsler_Tinker18}. The majority of baryons reside in either the \emph{circumgalactic medium} (CGM, gas outside galaxies but within dark matter halos) or the \emph{intergalactic medium} (IGM, gas outside dark matter halos). Besides their importance for the cosmic baryon budget, the physical properties of the C/IGM offer valuable insight into galaxy evolution, supplying galaxies with fresh gas and acting as a reservoir for their ejected, enriched gas \citep[the cosmic baryon cycle, e.g.][]{Putman2012, McQuinn16, Tumlinson2017}, and can provide important cosmological constraints on the growth of large scale structure and the nature of dark matter \citep[e.g.][]{Rauch98, Viel13, Lidz_Malloy14, McQuinn16, Eilers18}.

% Cosmic Web - definition, evidence, importance
\smallskip
In the IGM, on large $>\Mpc$ scales, both dark matter and gas are arranged in an intricate network of sheets and filaments known as the \emph{cosmic web} (CW). This structure has been predicted theoretically \citep{Zeldovich1970, bbks1986}, is found in cosmological simulations \citep[e.g.][]{Bond96, Springel05}, and can be seen in observational surveys in both the distribution of galaxies \citep[e.g.][]{Colless2001, tegmark04, Huchra05} and in narrowband Ly$\alpha$ emission at $z\gsim 3$ \citep{Umehata19, Martin23}. Combined, CW filaments and sheets are thought to comprise between $\sim (25-50)\%$ of the volume and $\sim (50-75)\%$ of the mass in the Universe, with sheets dominating the volume and filaments dominating the mass \citep[e.g.][and references therein]{Wang12, Cautun14, Libeskind18}. Filaments connect the most massive galaxies at any given epoch, which are located at the nodes of the CW, while more typical galaxies lie along individual filaments. The filaments are embedded within and fed by sheets or `walls' that surround large `voids'. At high redshift, $z>2$, CW filaments manifest as streams of cold, dense gas ($T\sim 10^4\,\Kel$, $n\sim 10^{-2}\cmc$), predicted to be the main mode of accretion onto galaxies, feeding them directly from the CW \citep[e.g.][]{Dekel.Birnboim.06, Dekel09a, Aung24}. This is thought to be true even in massive halos, $\Mv \gsim 10^{12}\msun$, where the CGM is hot, $T\gsim 10^6\Kel$, with very long cooling times \citep[][]{Rees77, White78, Birnboim.Dekel.03, Fielding17, Stern21}.

\smallskip
Gas in the CGM and the cosmic web is highly diffuse and difficult to directly observe. It has traditionally been traced using absorption line spectroscopy along lines of sight (los) to distant quasi-stellar objects (QSOs) or galaxies \citep[e.g.][]{Bergeron86, Hennawi06, Steidel10, Lehner22}, or narrow-band emission line studies; such as Ly$\alpha$ emission at redshifts $z\gsim3$ using integral field unit (IFU) spectographs such as KCWI on Keck and MUSE on the VLT \citep{Steidel00, Cantalupo14, Martin14a, Martin14b, Martin23, Umehata19, Tornotti24a, Tornotti24b}. These observations reveal that the gas in and around galaxy halos has a complex multiphase structure, with cool clouds embedded in hotter ambient gas (see \citealp{Tumlinson2017} for a recent review in the context of the CGM). The sizes of these cool clouds are difficult to constrain, and can range from sub-pc to a few $100\pc$ in the CGM. They have order unity area covering fractions, and appear to have very small volume filling fractions of $\fv\sim 10^{-3}$ and very large overdensities with respect to the hot gas of $\chi=\rho_{\rm c}/\rho_{\rm h}\sim (100-1000)$ \citep{Tumlinson2017, Cantalupo19, FG_Oh2023}. A similar multiphase structure is seen in cosmological simulations when care is taken to resolve the CGM with sub-kpc resolution \citep{vandeVoort19, Hummels19, Suresh19, Peeples19}.

\smallskip
Recently, novel cosmological simulations that `zoom-in' on massive CW sheets and filaments at high-$z$ have shown that these are surrounded by accretion shocks similar to massive halos and have a similar multiphase structure to the CGM (\citealp[][hereafter \citeta{m19} and \citeta{m21}]{m19,m21}; \citealp{lu23}). Multiphase CW sheets and filaments can explain puzzling observations of extremely metal-poor strong HI absorbers, with $Z\lsim 10^{-3}\zsol$ and $\Nhi>10^{17.2}\cms$ \citep[e.g.][]{Robert19, Lehner22}. However, the amount and extent of cold, dense gas increases with simulation resolution and is not converged at $\sim 300\pc$ resolution in the CGM and $\sim \kpc$ resolution in the CW. 

\smallskip
A theoretical model for the formation of multiphase gas consisting of small-scale cold clouds embedded in hot gas has been proposed \citep[][]{mccourt18}. According to the model, when the cooling time of a gas cloud becomes less than its sound-crossing time, it cannot cool isobarically and also does not cool monolithically as had been presumed \citep[][]{Field65, Burkert_Lin00}. The cloud instead `shatters' into numerous small fragments that lose sonic contact, causing them to contract independently and disperse, similar to a terrestrial fog \citep{mccourt18, Gronke_Oh20}. The typical size of the resulting cloudlets is predicted to be of order the \emph{minimal cooling length}, $\lcmin \sim \min(\cs \tcool)$, with $\cs$ and $\tcool$ the sound speed and cooling time, and the minimal value obtained at $T \sim 10^4 \Kel$. For typical CGM conditions at $z\sim (2-3)$, this is $\gsim 10\pc$ consistent with inferred cloud sizes, though it can reach $\sim\kpc$-scale in cosmic web sheets at these redshifts \citepa{m19,m21}. This model explains the vastly different area-covering and volume-filling factors \citep{FG_Oh2023}, as well as a number of additional observations in the CGM, High Velocity Clouds, quasar Broad Line Regions, and the interstellar medium \citep{Gronke17, mccourt18, Stanimirovic.Zweibel.2018, FG_Oh2023, Sameer24}. It also explains why cosmological simulations have not converged in terms of cold gas properties in the C/IGM. Some high-resolution simulations of this process suggest that the resulting clouds can be significantly smaller than $\lcmin$, with a minimal size set by cloud disruption due to external turbulence or thermal conduction \citep{Yao25}. 

\smallskip
The distribution of cloud sizes in the C/IGM is thus poorly constrained and very difficult to probe using either absorption or emission line studies. Both of these methods also require large and uncertain ionisation corrections to infer the total gas density, often making them difficult to interpret. Our understanding of the phase structure of both the CGM and CW is thus greatly limited.

\smallskip
In recent years, fast radio bursts (FRBs) have become an increasingly powerful tool to probe the C/IGM. 
FRBs are compact radio sources, emitting short-duration radio signals lasting a few tens of $\rm \mu s$ to a few $\rm ms$ in a frequency range of $\sim0.1-8\ghz$ \citep{Lorimer07, Lorimer24, Gajjar18, Pleunis21}. They are highly abundant and bright, with a detectable (all-sky equivalent) rate of $\gsim10^{3}\,\rm d^{-1}$ at a fluency of $1\jy\ms$, and increasing rates at lower fluences.
While the identity of FRB sources remains under debate, one primary candidate is magnetars \citep{Bochenek20, Good20, Kumar+17, Wadiasingh2019, beniamini25}. This suggests that they are likely to exist at high redshifts, potentially into the epoch of reionisation \citep{beniamini21, Hashimoto_2021, Heimersheim_2022, Ziegler24, Shaw2024}, making them an extremely powerful probe of intergalactic gas over cosmological timescales.

\smallskip
As the radio signal from an FRB travels through a plasma, its group velocity is hindered due to the plasma refractive index. This results in a frequency-dependent time delay, where lower-frequency photons arrive later. By measuring the time delay as a function of frequency, one can derive the integrated electron density along the los from the source to the observer, commonly referred to as the dispersion measure (DM), namely $\dm=\int {\n }/{\opzo[]} ds$, where $\n$ is the electron proper density and $ds$ is the proper distance element. 
% DM-z
For distant FRBs, the DM is dominated by the IGM, allowing it to serve as a rough estimate for the source redshift by assuming a given cosmology and modelling the average electron density along the los. 
% high-z FRBs
Over a hundred FRBs have been localised to a host galaxy, reaching up to $z\sim2$ \citep{Lorimer24, Connor24, gordon24, caleb25}. However, others are believed to originate from higher redshifts, due to their large measured DM \citep[e.g.,][]{zhang18}.

\smallskip
Unlike absorption line spectroscopy or narrowband emission studies, the DM of FRBs require no ionisation correction as they are sensitive to the total electron density along the line of sight. This makes them a sensitive probe of the cosmic baryon content, confirming that most baryons reside in the C/IGM \citep{macquart20}. Using large populations of localised FRBs, such as from the ongoing CHIME survey \citep{chime21}, one can constrain the baryon fraction in the CGM of intervening halos and in the IGM \citep[e.g.][]{prochaska19-halo, lee22, khrykin24, Connor24}. 

\smallskip
Beyond their DM, FRB scattering and scintillation can be used as a sensitive probe of the small-scale and turbulent structure of C/IGM gas, much like Galactic pulsars have been used to constrain the Milky-Way ISM \citep[e.g.][]{rickett77, rickett90, cordes91, Armstrong95, Cordes.lazio.02}. This is made possible by the FRB's compact nature and short duration, radio wave sensitivity to small electron density fluctuations, and the growing number of FRB detections \citep{Lorimer24}. A fog-like structure of $\sim\pc$-scale clouds in the CGM can thus have a large effect on FRB signals \citep{vedantham19, ocker22-horizons, jow24, ocker25, mas-ribas25}, in which case, scattering and scintillation can be powerful probes of the CGM and the cosmic web on scales much smaller than accessible using absorption line spectroscopy or simulations. 
Observational evidence of this remains mixed, with some localised FRBs showing no signs of excess scattering in the CGM of foreground galaxies \citep{prochaska19-sci, Connor20M33, Faber2024, shin24}, whereas a population synthesis study based on the first CHIME/FRB catalogue as a whole is marginally consistent with CGM scattering \citep{chawla22}. 
While FRB scattering in the diffuse IGM is expected to be negligible \citep[hereafter \citeta{macquart13} and \citeta{beniamini20}]{macquart13,beniamini20}, CW sheets and filaments can have a similar multiphase structure to the CGM, and contribute significantly to FRB scattering. If so, this can help constrain cold overdensities in the CW. It may also create an ``extragalactic scattering horizon'' beyond which FRB temporal broadening becomes so significant that they cease to be detectable, limiting their use as cosmological probes \citep{ocker22-horizons}. Some models suggest that if CGM scattering is significant, this may be the case for $\sim 20\%$ of FRBs at z > 5 \citep{ocker22-horizons}. If the likelihood of intersecting a cosmic sheet is much larger than intersecting a halo, the effect of sheets on the FRB population may be larger than that of the CGM, even if the effect of passing through a single sheet is smaller than passing through a single halo due to the higher density ($\n$) in the CGM. 

\smallskip
Several studies have examined the impact of cool cloudlets in the CGM on FRBs. 
\citet{mas-ribas25} focused on refractive scattering by discrete clouds, while \citet{ocker25} adopted an empirical, turbulence-based framework to predict diffractive scattering and scintillation, both considering a single halo. 
\citet[][hereafter \citeta{vedantham19}]{vedantham19} modelled the effect of cumulative scattering in the CGM of halos along the los to an FRB, assuming a multiphase `shattered' CGM of small cold clouds of size $\lcmin$ embedded in a hot medium. 
In this work, motivated by the presence of shattered cold gas cloudlets in CW sheets and filaments in simulations (\citeta{m19,m21}; \citealp{lu23, Yao25}), we consider these structures in addition to the CGM (hereafter collectively CW objects, or CWOs).
We investigate whether small-scale cold structure in CWOs measurably affects FRB signals, and whether diffractive scattering can constrain such structure.
Throughout, we adopt the shattering model as a physically motivated realisation of cold gas on small scales within the hot ambient medium of CWOs. Our conclusions do not rely on the correctness of the shattering mechanism itself, but rather on the presence (or absence) of turbulent density fluctuations at the diffractive scales probed by FRB scattering.

\smallskip
Radio wave propagation through turbulent inhomogeneous plasma introduces temporal broadening (`scattering'), as well as spectral interference (`scintillation'). While both are governed by the same underlying diffractive physics, they are sensitive to different regimes.
In this work, we focus on \emph{scattering} as an incoherent population-level probe, leveraging its sensitivity to cumulative scattering along the los.
We model the effects of small-scale density fluctuations in CWOs across cosmic time and identify their expected cumulative scattering signatures.
In a companion paper \scintpaperl (hereafter \scintpaper), we explore \emph{scintillation} as a coherence-based diagnostic of individual CW plasma screens. Crucially, this can remain a sensitive probe even when temporal broadening is negligible. Together, these approaches provide complementary constraints on small-scale structure in the CGM and IGM.

Our current work relies in part on results presented in another companion paper (\dmpaperl, hereafter \dmpaper). In that paper, we derive an analytic model for the virial properties of CW sheets as a function of redshift and the halo mass in the nearby CW node. This is analogous to similar models for filaments derived in \citet{m18} and \citet{lu23}, in that it assumes a hierarchy of CWOs, where halos are fed by filaments and filaments are fed by sheets. In this way, the virial properties of filaments are constrained by the total accretion rate onto massive haloes in CW nodes \citep[][hereafter \citeta{m18}]{m18}, while at the same time the virial properties of sheets are constrained by the accretion rate onto the filaments embedded within them. Combining our derived virial properties of CWOs with an excursion set approach for ellipsoidal collapse in the CW following \citet{shen06}, we evaluate the number of CWOs along an average los as a function of redshift. Finally, we use these results to predict the contribution of each CWO to the total DM of an FRB source as a function of source redshift.

\smallskip
The remainder of this paper is organised as follows. In \se{shatter}, we present our fiducial model for the formation of multiphase gas in CWOs, modelled as small-scale cold cloudlets within a hot medium.  
In \se{scatt}, we present a basic model of radio wave scattering in a CW plasma screen with a single cloud. In \se{cosmo}, we briefly summarise our model for the virial properties of CWOs (\dmpaper) and use it to estimate the number of clouds within different CWOs as a function of redshift, as well as the number of CWOs intercepted along an average los. In \se{mscatt}, we estimate the expected scattering time caused by multiple clouds in multiple CW plasma screens across cosmic time. Finally, in \se{dc_Sc} we summarise our results and conclude. 
Throughout, we assume a flat $\Lambda$CDM cosmology, with $H_0=70\kms\Mpc^{-1}$, $\omm=1-\oml=0.3$, and a universal baryon fraction of $\fb=0.17$.

% ************************************************
% @@@@@@@@@@@@@@@@@@@@@@@@@@@@@@@@@@@@@@@@@
\section[shattering]{The Small-Scale Cold Structure of CWOs}
\label{se:shatter}
% @@@@@@@@@@@@@@@@@@@@@@@@@@@@@@@@@@@@@@@@@
%

To make concrete predictions for FRB scattering, we require a model for the characteristic size of cold overdensities in CWOs. Several physical processes may generate such small-scale overdensities, including thermal instability, turbulent mixing layers, and radiative shocks. 
In this section, we describe the shattering model and adopt it as a fiducial and physically motivated example. This choice largely impacts the assumed cloud size, $\lc$, in our model (and consequently, also the number of cloudlets intercepted along a los through a CWO, $\fa\propto\lc^{-1}$). For alternative models, one could simply modify the assumed $\lc$. However, as we describe in Appendix \se{shatterApp}, additional processes (e.g., thermal conduction) yield typical cloud sizes within a factor of a few from those predicted by shattering. We thus use the shattering model throughout for convenience, discussing the effects of deviations from it in \se{disc}.

% =========================================
\subsection{Thermal Fragmentation and Cloud Sizes}
\label{se:shatter_1}
% =========================================

% _________________________________________
{
\begin{figure}[ht]%
    \centering%
    \includegraphics[width=0.85\columnwidth]{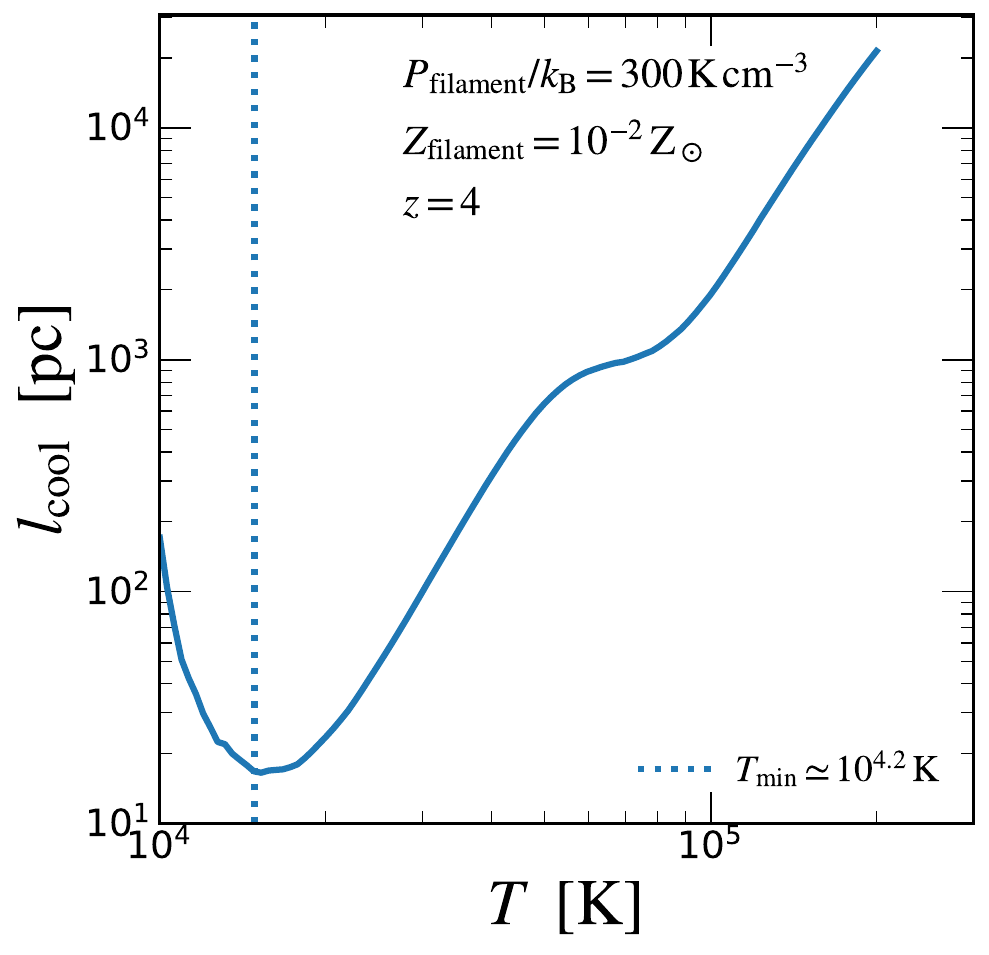}
   \caption{
   Example of the cooling length as a function of temperature calculated using \eq{lcool}, assuming isobaric cooling. This is shown for typical filament pressure and metallicity at $z=4$ \citepa{m21}.}
   \label{fig:lcool}
\end{figure}
}% fig:lcool
% _________________________________________

According to \citet{mccourt18}, when a thermally unstable cloud is large enough such that its cooling time is much shorter than its sound-crossing time, it does not cool monolithically, but rather shatters into many small fragments that proceed to cool isobarically. The size of these fragments is the local \emph{cooling length}, $\lcool=\cs\tcool$. At a given pressure, $\lcool$ is a strong function of temperature (\fig{lcool}). Therefore, this process was thought to be hierarchical, with clouds fragmenting continuously to smaller and smaller scales until reaching the minimal cooling length, $\lcmin=\min(\lcool)$, at $T\gsim 10^4\Kel$. This shattering scale depends primarily on the external pressure, the gas metallicity, and the presence of an ionising UV background. The cooling length as a function of temperature is
\be
\label{eq:lcool}
    \lcool = \cs \tcool = \pfrac{\gamma \kb^5}{\mu^3 \mpr (\gamma-1)^2 \xh^2}^{1/2}\frac{T^{5/2}}{P\, \Lamtz(T,Z,z)},
\ee
where $\mu$ is the mean molecular weight, $\kb$ is Boltzmann's constant, $\mpr$ is the proton mass, $\xh$ is the hydrogen fraction, $P=n \kb T$ is the gas thermal pressure, and $\nh^2\Lamtz(T,Z,z)$ is the net cooling rate per unit volume, accounting for heating from a background ionising radiation field, with $\nh=\mu \xh n$ the hydrogen number density. In \fig{lcool} we show $\lcool$ as a function of temperature assuming a confining pressure of $P/\kb=300\Kel\cmc$, a metallicity of $Z=10^{-2}\zsol$, and a $z=4$ \citet{haardt_madau96} UV background, typical of CW filaments at high-$z$ \citep{lu23}. We see that $\lcmin\sim 20\pc$ in this case. In general, one can write 
\be
    \label{eq:lshatter}
    \lcmin \simeq 15\pc~\frac{\Tx[min,2e4]^{5/2}}{P_3\,\Lamtz[min,-23]},
\ee
where $\Tmin[,2e4]=\Tmin/(2\tm 10^4\Kel)$ is the temperature where $\lcool$ is minimal at fixed pressure, and $\Lamtz[min,-23]=\Lamtz(\Tmin)/(10^{-23}{\ergs\cm^3)}$.%
\footnote{
We review some of the key physics behind the formation of multiphase gas via shattering in \se{shatterApp}.}

% =========================================
\subsection{Pressure in CWOs}
\label{se:P}
% =========================================
% 

To evaluate the properties of cloudlets in CWOs across cosmic time, the first question we must ask is which CWOs do we expect will shatter in the first place? 
We assume shattering to occur in CWOs where the virial temperature is at least $\Tv>10^5\Kel$ (see also \se{CW_shatter}), a factor $\sim (5-10)$ greater than the equilibrium temperature of $\sim (1-2)\tm 10^4\Kel$ \citep[see also][]{Gronke_Oh20, Gronke_Oh23, Yao25}. We envision an accretion shock near the outer edge of the CWO with a post-shock temperature of order $\Tv$ (\citeta{m19,m21}; \citealp{lu23}), with cloudlets forming in the post-shock gas as it cools. Note that we do not explicitly require the cooling time in the hot gas to be long compared to the crossing time of the CWO in order to maintain a fixed external pressure, which can be maintained by the ram pressure of infalling gas and turbulent pressure in the post-shock gas, both of which are comparable to the virial pressures estimated below. In the end, we only require that the background pressure does not vary much over the typical fragmentation timescale (e.g., the shattering timescale), comparable to the cooling time of non-linear density perturbations in the post-shock gas. This condition is almost always met for virial temperatures $\Tv\gsim (10^5-10^{5.5})\Kel$ (\citealp{mccourt18}; \citeta{m21}). 

\smallskip
The halo virial temperature is given by \citep[e.g.][]{dekel13},%
\be 
    \label{eq:Tvir_halo}
    \Tvh \sim 2.4\tm 10^6\Kel\:\Dvhf^{1/3}\opzo\Mhf^{2/3},
\ee 
where $\Mhf\sim \Mh/10^{12}\msun$ and $\opzo=\opzo[]/5$. 
The filament virial temperature can be expressed as a function of its mass per unit length, or alternatively as a function of redshift and the mass of the halo fed by the filament (\citeta{m18}; \citealt{lu23, Aung24}). 
In the EdS regime at $z>1$,%
\footnote{
See \dmpaper for corresponding expressions valid at $z<1$.}
the result is 
\be 
    \label{eq:Tvir_fil}
    \Tvf \sim 2.55\tm 10^5\Kel\:\Mhf^{0.77}\,\opzo^{2}\;\fff\,\Machf^{-1},
\ee
where $\fff$ is the fraction of total accretion onto a halo flowing along a given filament, normalised to a fiducial value of $1/3$ \citep{Dekel09a,danovich12}, and $\Machf$ is the velocity of filament gas normalised to the halo virial velocity. 
In our companion paper (\dmpaper), we present a detailed derivation of the virial properties of CW sheets, which yields 
\be 
\label{eq:Tvir_sheet}
\Tvsh
    \sim 2.4 \tm10^5\Kel \ &\fff^{}\fshf^{2}\Machff^{-1}\Machshf^{-2} \Dvshf^{-1} \Mhf^{0.77} \opzo^{3},
\ee 
where $\fshf$ is the fraction of total accretion onto the filament flowing from within the sheet normalised to a fiducial value of $1/4$ \citep{lu23}, $\Dvshf$ is the virial overdensity of the sheet normalised to a value of $6$ \citep{shen06}, and $\Machsh$ is the velocity of sheet gas normalised to the filament virial velocity. 

\smallskip 
When evaluating $\Tv$ in \eqsiti{Tvir_halo}{Tvir_sheet}, we assume a truncated isothermal profile and include the surface pressure term in the virial equation \citep[e.g.][]{krumholz2015notes,lu23}. Our condition for the formation of cloudlets, $\Tv>10^5\Kel$, thus occurs in CWOs spanning a wide range of redshifts and in the vicinity of a wide range of halo masses. 

\smallskip
The properties of the resulting clouds depend on the confining pressure of the hot gas. We present here a model for the expected `virial pressure' in a CWO at the virial temperature. We evaluate the gas pressure for sheets and filaments using the post-shock temperature and mean gas density, 
\be
    P_\mr{shock} \sim \frac{\rhob[,v] \kb \Tv}{\mu\mpr}
    \label{eq:P_shock}
\ee
where we take the temperature behind the shock to be the virial temperature of the CWO, $\Tv$, and the mean density to be $\rhob[,v]=\Dv\brho\fb$. $\Dv$ is the overdensity of the CWO above the mean density of the universe, which is given by $\brho=\omm\rhoc\opzo[]^3$, with a critical density of $\rhoc[,0]=3H_0^2/(8\pi G)\sim 9.2\tm 10^{-30}\gcmc$ at $z=0$. For the CGM, we multiply this density by an additional factor $\fcgm=0.5$ to account for the baryonic mass fraction in the hot CGM \citep[e.g.,][]{oren24,khrykin24}. For simplicity, we use an estimate of the mean gas density rather than introducing a detailed profile.  We hereafter assume a mean molecular weight of $\mu=0.59$ for the shocked background medium, typical of fully ionised gas of primordial composition, a Universal baryon fraction of $\fb=0.17$, and virial overdensities of $\Dv=6$, $36$, and $180$ for sheets, filaments and halos, respectively.

\smallskip
The pressure in sheets and filaments with virial temperature $\Tv=\Tvfid\Kel$ is, respectively,
\begin{align}
    &P_\mr{shock,sh}\sim 110 \Kel \cmc \Dvshf\Tvshf \opzo^3
\\
    &P_\mr{shock,f}\sim 650 \Kel \cmc \Dvff\Tvff \opzo^3.
\end{align}
These values are very similar to the thermal pressure measured in sheets \citepa{m19,m21} and filaments \citep{lu23} in cosmological simulations at $z=4$. The pressure in the  CGM near the outskirts of a halo with the same virial temperature is 
\be
    P_\mr{shock,h} \sim 1.9\tm 10^3 \Kel \cmc \Dvhf\Tvhf \opzo^3.
\ee
However, it is convenient to express this as a function of halo mass rather than virial temperature, yielding
\be
    P_\mr{shock,h} \sim 1.5\tm 10^4 \Kel \cmc \Dvhf^{4/3}\Mh[,12]^{2/3} \opzo^4.
\ee
At $z=0$ this translates to a pressure of $P\sim 20 \Kel \cmc$ for $\Mh[,12]\sim 1$.

% =========================================
\subsection{Cloud Properties in CWOs}
\label{se:CW_clumps}
% =========================================

% _________________________________________
{
\begin{figure*}[ht]%
    \centering%
    \includegraphics[width=0.8\textwidth]{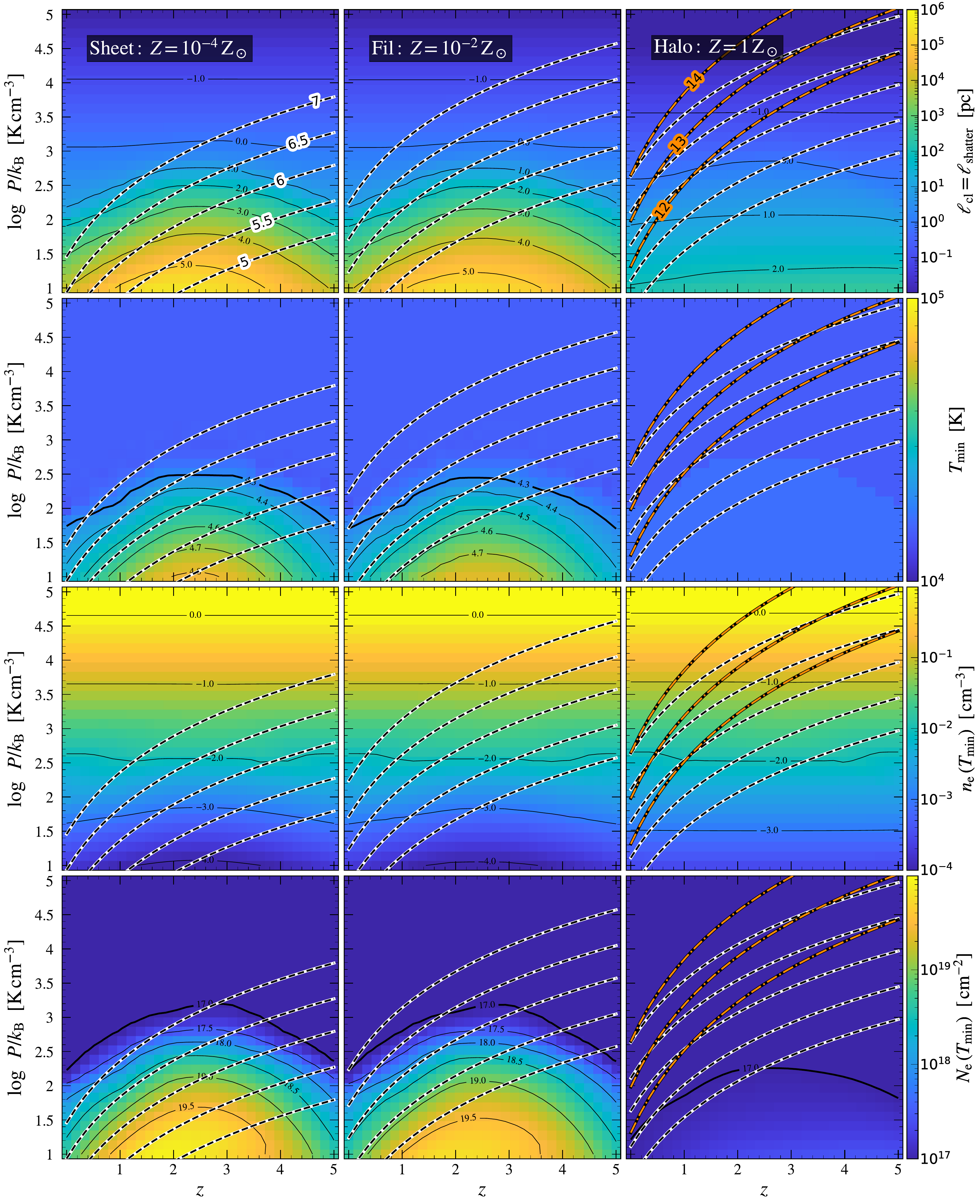}
    \caption{Properties of shattered cloudlets as a function of thermal pressure (y-axis), redshift (x-axis), and metallicity - $Z/\zsol=10^{-4}$ (left column), $10^{-2}$ (middle column), and $1.0$ (right column), crudely representing CW sheets, filaments, and halo CGM, respectively. From top to bottom, we show the cloud size (the minimal cooling length, $\lc=\lcmin$), the cloud temperature ($\Tmin$), the cloud electron density ($\n$), and the electron column density over a cloud diameter ($\N$).
    These are calculated assuming isobaric cooling at the given pressure, and account for a redshift-dependent \citet{haardt_madau96} UV background with self-shielding of dense gas \citep{rahmati13}. At a fixed pressure and redshift, metallicity is not important below $10^{-2}\zsol$. The dashed lines in each panel indicate the pressure near the outskirts of a CWO with virial (post-shock) temperature of $\Tv=10^{5}$, $10^{5.5}$, $10^{6}$, $10^{6.5}$, and $10^7\Kel$ from bottom to top, according to our model presented in \se{P}. 
    The black lines indicate the contours of the quantity shown in colour, and are labelled in log scale. 
    At a fixed virial temperature, sheets have the lowest pressures, largest cloud sizes, and smallest densities, while the CGM is at the opposite extreme with filaments in between. The orange dot-dashed lines in the right column represent halos with a fixed mass of $\Mh=10^{12}$, $10^{13}$, and $10^{14}\msun$ from bottom to top. 
    }
    \label{fig:clump_prop}
\end{figure*}
}% fig:clump_prop
% _________________________________________

In this section, we estimate the properties of shattered clumps in CWOs. Our fiducial model assumes that the cloud sizes are $\lc=\lcmin$, but as previously stated, the model can be applied to cloudlets of any size by adjusting $\lc$ (and $\fa$) accordingly.
Throughout, we assume a fixed metallicity for each CWO using typical values from cosmological simulations, $Z=(10^{-4},\,10^{-2},\,1)\zsol$ for sheets, filaments, and the CGM of halos, respectively. The Hydrogen mass fraction is fixed at $\xh=0.76$. We then use \eq{lcool} to evaluate the minimum cooling length, $\lcmin=\lcool(\Tmin)$, in each CWO as a function of redshift, assuming isobaric cooling at a given pressure which we allow to vary. We compute the mean molecular weight of the cloudlets, $\mu(\Tmin)$, their electron density, $\n(\Tmin)$, and $\Lamtz(\Tmin,Z,z)$ using the cooling module implemented in the \texttt{RAMSES} code \citep{Teyssier.02}, which accounts for atomic and fine-structure cooling for our assumed metallicity values. We include a redshift-dependent \citet{haardt_madau96} UV background (UVB) along with self-shielding of dense gas following \citet{rahmati13}.

\smallskip
An example of such a calculation showing $\lcool(T)$ for a filament at $z=4$ with $P/\kb=300\Kel\cmc$ is shown in \fig{lcool}, revealing $\lc\simeq 20\pc$ and $\Tmin\simeq10^{4.2}\Kel$. More generally, in the top row of \fig{clump_prop} we show $\lc$ as a function of thermal pressure on the y-axis and redshift on the x-axis. We show results for $z\le 5$, safely after the end of reionisation, where a spatially uniform UVB is a reasonable assumption. The three columns show results for metallicities $Z=10^{-4},\,10^{-2},$ and $1\zsol$ from left to right, crudely representing CW sheets, CW filaments, and the halo CGM, respectively. Adopting $0.1\zsol$ for the CGM yields very similar results. Overall, $\lc$ is nearly identical in sheets and filaments at a fixed $(P,z)$. In the CGM, $\lc$ can be up to 3 orders of magnitude smaller at $P\lsim 100\Kel\cmc$, though at pressures $P>1000\Kel\cmc$, more typical of realistic CGM gas, the difference is only $\sim 0.5\dex$. The large difference between low-$Z$ (left \& middle panels) and high $Z$ (right panel) at low pressures is due to the fact that at $Z/\zsol\lsim10^{-1}$ cooling is dominated by the hydrogen peak, which is very sensitive to both photo-ionisation and photo-heating by the UVB,%
\footnote{UVB heating can also influence the normalisation of the cooling function, although it is a minor effect.}
while at higher metallicities, cooling is dominated by metal lines, which are insensitive to the UVB.

\smallskip
While $\lc$ may be similar in sheets and filaments at a fixed $(P,z)$, realistic CW sheets and filaments occupy different regions in this plane. The black and white dashed lines in each panel represent the pressure as a function of redshift in the outskirts of a virialized CWO with a fixed $\Tv$ ranging from $10^7$ to $10^5\Kel$ from top to bottom in intervals of $0.5\dex$, based on our model described in \se{P}. For comparison, the orange and black dashed lines in the CGM panels represent the pressure as a function of redshift in the outskirts of a virialized halo with a fixed $\Mh=10^{14}$, $10^{13}$, and $10^{12}\msun$ from top to bottom. For a fixed $\Tv$, we have $P_\mr{v}\propto\Tv \nh[v]\propto\Tv\opzo[]^3$ in all CWOs, where $\nh[v]$ is the density inside the virialized CWO which scales as the Universal mean density. However, in the CGM, $P_\mr{v}\propto\opzo[]^4$ for a fixed halo mass. For a fixed $\Tv$, the pressure increases from sheets to filaments to the CGM, leading to a corresponding decrease in $\lc$. In cosmological simulations with high-resolution in sheets and filaments (the \texttt{IPM} simulations, \citeta{m19}), \citeta{m21} and \citet{lu23} found $P/\kb\sim 100$ and $\sim 300\Kel\cmc$ for sheets and filaments with $\Tv\sim 10^{5.5}\Kel$ at $z\sim 4$, respectively, in agreement with our model. \citeta{m21} measured cold cloud sizes of $\lsim 1\kpc$ in such sheets, also in agreement with our model. We predict $\lc\lsim 20\pc$ in such filaments, which is unresolved in the \texttt{IPM} simulations. For halos at the same redshift and $\Tv$, our model predicts $P/\kb\sim 2000 \Kel\cmc$ and $\lc\sim 0.5 \pc$. 
The main haloes in the \texttt{IPM} simulations have $\Mh\sim 10^{12}\msun$ at $z\sim 4$, with $\Tv\sim 10^{6.5}\Kel$, $P/\kb\sim 10^4\Kel\cmc$, and $\lc\sim 0.05 \pc$.

\smallskip
In the second row of \fig{clump_prop} we show $\Tmin$, the temperature where $\lcool$ is minimal. For solar metallicity (i.e. the CGM), we find that $\Tmin\sim (1-2)\tm10^4\Kel$ throughout the $(P,z)$ plane, as predicted by \citet{mccourt18}. However, for the very low metallicities in sheets and filaments, there is a region at low pressures where $\Tmin$ is a few times larger. This is primarily driven by UVB heating, which becomes more influential at low densities where self-shielding is negligible. For $Z/\zsol\gsim10^{-1}$, the effect of heating on $\Lamtz$ is minor compared to the metal cooling, keeping $\Tmin$ practically constant. 
At $z\sim 4$ and our fiducial virial temperature of $\Tv\sim3\tm10^5\Kel$, we expect sheets and filaments to host cold clouds with $T\sim 2.5\tm 10^4\Kel$ and $\sim 2\tm 10^4\Kel$, respectively, which matches the cold gas properties in the \texttt{IPM} simulations as found by \citeta{m21} and \citet{lu23}.

\smallskip
In the third and fourth rows, we show the electron density in the cold clouds, $\n$, and the electron column density of a single cloud, $\N=\n\lc$. This density is evaluated by assuming ionisation equilibrium at the external pressure and at temperature $\Tmin$. Unlike other properties, $\n$ is extremely similar at all metallicities at a given $(P,z)$. At $z\sim 4$ and $\Tv\sim3\tm10^{5}\Kel$, we have $\n\gsim 10^{-3}\cmc$ for sheets and $\gsim 10^{-2}\cmc$ for filaments, consistent with the cold gas density in \citeta{m21} and \citet{lu23}. In the CGM, on the other hand, we have $\n\gsim 0.05\cmc$ at $z\sim 4$ for the same virial temperature. The cloud column density in the CGM is $\N\sim 10^{17}\cms$ regardless of pressure or redshift, consistent with the original prediction of \citet{mccourt18}.
Generally, for sheets and filaments, we obtain larger column densities at low pressures, driven by the much larger $\lc$. At $z\sim 4$ and $\Tv\sim3\tm10^5\Kel$, we obtain $\N\sim 10^{18.5}\cms$ for sheets, while a filament at the same redshift and $\Tv$ enters the high-pressure zone, where self-shielding becomes substantial, resulting in a column density of $\lsim10^{17.5}\cms$.

% ************************************************
% @@@@@@@@@@@@@@@@@@@@@@@@@@@@@@@@@@@@@@@@@
\section{scattering -- single cloud}
% @@@@@@@@@@@@@@@@@@@@@@@@@@@@@@@@@@@@@@@@@
\label{se:scatt}

When a short-duration radio pulse passes through a turbulent plasma, density fluctuations in the medium can spread the intrinsic features of the burst, causing a temporal and angular broadening at a given observed central frequency. 
Measurements of this broadening can provide valuable information on small-scale substructure in the intervening plasma. Of particular interest for our purposes, this can shed light on the existence of cool and turbulent small-scale structure within different CW environments. The physics behind this type of diffractive scattering are well known and have been studied extensively using pulsars in the Milky Way for several decades \citep[see][for a review]{rickett90}.
In this section, we review some basic concepts and derive results in the context of CWOs. Specifically, we evaluate the diffractive length scale (also referred to as the coherence length scale) and the scattering time. A similar derivation in the context of FRB scattering in the diffuse IGM can be found in \citeta{beniamini20}. Here, we begin by studying the effect of temporal broadening caused by a single eddy/cloudlet embedded within a single screen. Later, in \se{mscatt}, we extend this to multiple CW screens across cosmic time, each hosting multiple cloudlets.

% -----------------------------------------
\subsection{Phase Shift and the Coherence Scale}
\label{se:lpi}
% -----------------------------------------

% The coherence scale
Density fluctuations in a plasma screen cause fluctuations in the refractive index, $\mu_i=[1-(\nu_\mr{p}/\nuz)^2]^{1/2}$, where $\nu_\mr{p}=[e^2\n/(\pi\me)]^{1/2}$ is the plasma frequency, with $\me$ and $e$ the electron mass and charge. We define $\nuo$ and $\lamo$ as the frequency and wavelength of the passing radio wave in the rest frame of the observer, so $\nuz=\nuo\opzo[]=c/\lamz$ is the frequency in the rest frame of the screen.%
\footnote{Hereafter, a tag notation, $\square'$, indicates the rest frame of the screen.} For typical densities of $\n \sim 0.01\cmc$ we have $\nu_\mr{p}\lsim 1\khz$ while observed frequencies for FRBs are in the range $\nu\sim (0.1-8.0)\ghz\gg \nu_\mr{p}$.
We assume the fluctuations to be comprised of turbulent eddies. As a radio wave passes through a single eddy of size $\ell$, it will suffer a phase shift of $\dphi=(2\pi/\lamz)\ell\delta\mu_i$. The phase shift of a single eddy as a function of the density fluctuations is thus
\be
    \dphi = \frac{2\pi\ell}{\lamz}~\frac{1}{2\mu_i}~\frac{e^2 \dn}{\pi \me {\nuz}^2} \approx ~ \re\ell\lamz\dn
    \label{eq:dphi}
\ee
with $\re=e^2/(c^2\me)$.

\smallskip
We assume a power law spectrum for the density fluctuations, 
\be
    \dngt \sim \n (\ell/\lo)^\alpha
    \label{eq:dn}
\ee
where $\ell$ is the size of an eddy, $\li$ is the inner dissipation scale,%
\footnote{We estimate the inner scale of turbulence following \citeta{beniamini20}, and briefly describe their derivation in appendix \se{li}. This is given by $\li=\max(l_\mr{vis}, l_\mr{B})$ where $l_\mr{vis}$ is the viscous scale length and $l_\mr{B}$ is the Larmor radius.
}
$\lo$ is the outer injection scale, transverse to the los, and \eq{dn} is valid for scales in the inertial subrange, $\li<\ell<\lo$. For Kolmogorov turbulence, we expect $\alpha=1/3$ \citep{Lithwick.Goldreich.01}.\footnote{For subsonic, weakly compressible, Kolmogorov turbulence, the eddy velocity scales as $\delta v\propto \ell^{1/3}$ because the energy dissipation rate is $\epsilon\propto \rho_0 \delta v^2\,\delta v/\ell\sim\const$. Meanwhile, since density fluctuations dampen on a sound crossing time, the continuity equation yields $\delta \rho\,\cs/\ell \sim \rho_0 \delta v/\ell$, so that $\delta \rho/\rho_0 \propto \delta v/\cs \propto \ell^{1/3}$.
}
We assume the thin screen approximation, where the total number of eddies of size $\ell$ within the scattering screen is $\dL/\ell$, with $\dL$ as the total width occupied by eddies of scale $\ell$ along a line of sight through the screen, which may be less than the total width of the screen, $\DL$.
The cumulative RMS phase-shift acquired through the screen due to eddies of size $\ell$ is $\Dphi\sim({\dL}/{\ell})^{1/2}\dphi$. Using \eqs{dphi}-\eqm{dn}, the total phase change of a wave passing through a turbulent ionised medium due to eddies of size $\ell$ is
\be
    \Dphigt  
    &\sim \re\lamz\,\dL^{1/2}\n\lo^{-\alpha} \ell^{\frac{1+2\alpha}{2}}
    \equiv
    \pfrac{\ell}{\lpigt}^{\frac{1+2\alpha}{2}},
    \label{eq:Dphi}
\ee
where $\lpi$ is the \emph{coherence} or \emph{diffractive length scale}, defined as the scale on which the RMS phase difference equals one radian, $\Dphi(\lpi)=1$. A substantial change to the wave phase, and thus a strong modulation of the radio wave flux through the medium, is achieved when $\Dphi\gsim1$, on scales $\ell\gsim \lpi$. However, as shown in Appendix \se{time_delay}, the scattering angle of the light ray increases with decreasing eddy size and becomes substantial at scales $\ell\lsim \lpi$. Therefore, the dominant eddies for diffractive scattering, yielding both a large scattering angle and large flux modulation, are eddies of scale $\ell\sim \lpi$ \citepa[see also][]{beniamini20}.

\smallskip
From \eq{Dphi}, the coherence length is 
\be
    \lpigt \sim \lt(\re\lamz\,\dL^{\frac{1}{2}}\n\lo^{-\alpha}\rt)^{-\frac{2}{1+2\alpha}}.
    \label{eq:lpigt}
\ee
However, this expression is only valid if the coherence length falls in the inertial range, $\li<\lpi<\lo$, where \eq{dn} is valid. If $\lpi<\li$, then \eqs{dn}-\eqm{lpigt} are invalid as the power-law scaling of density fluctuations breaks down. The derivation of the coherence length in this case is more complicated, and we do not present it here (see \citealp{lambert-rickett99, coles87, rickett90}). The end result is
\be
    \lpilt &\sim \lt(\re\lamz \dL^{\frac{1}{2}} \n \lo^{-\alpha} \li^{-\frac{1-2\alpha}{2}}\rt)^{-1}.
    \label{eq:lpilt}
\ee

% Area-covering fraction
\smallskip
Similar to \citeta{vedantham19}, we take the outer scale of turbulence to be the size of cold clouds, $\lo\sim\lc$. 
Due to the high density contrasts between the cold clouds and the surrounding hot gas, the turbulence in each media is decoupled and evolves differently. Since the cold clouds have much higher densities than their surroundings, we expect them to dominate the scattering (justified below, see also \citealp{ocker25}), and therefore take the cloud size as the outer scale of turbulence. However, if turbulence in the hot gas contributes as well, our results can be interpreted as a lower limit on the expected scattering. 

\smallskip
As can be seen in \eq{lpigt} and \eq{lpilt} for a Kolmogorov spectrum of turbulence, the dependence of the coherence-scale on the outer scale, $\lo$ (=$\lc$), is weak compared to the dependence on the cloud density, $\n$.
Therefore, despite the uncertainty in our assumed cloud size, $\lc=\lcmin$, the main cloud property affecting scattering is actually their density, $\n$. 
In \se{disc}, we discuss how potentially larger cloud sizes can affect our main results.

\smallskip
In a medium of size $\DL$, the volume filling fraction (\fv) of clouds with size $\lc\ll\DL$ is \citep{mccourt18,FG_Oh2023} $\fv\sim \num[cl]\lc^3/\DL^3$, where $\num[cl]$ is the total number of such clouds. The area covering fraction (\fa) of cold clouds is $\fa\sim \fv\DL/\lc \sim \num[cl]\lc^2/\DL^2 \sim {\Nt[,cold]}/({\n\lc})$, where $\n$ is the mean electron density in the clouds and $\Nt[,cold]$ is the average electron column density over the region assuming that this is dominated by the cold clouds which are much denser than their surroundings. It is straightforward to see that $\fa$ also represents the mean number of clouds encountered along a given los through the region. Therefore, $\fa\sim \dL/\lc$, where as before $\dL$ is the total width occupied by eddies, which we assume are confined to the cold clouds. Inserting this into \eqs{lpigt}-\eqm{lpilt}, along with $\lo=\lc$ and our assumed $\alpha=1/3$ for Kolmogorov turbulence,\footnote{Expressions for a general index $\alpha$ can be found in appendix \se{scatt_app} (see also \citeta{macquart13}; \citealt{xu16} and references therein). \label{foot:alpha}} we obtain for the coherence length
\be
    \lpi =
    \begin{cases}
        \Big(\frac{\cbeta\re \lamo}{1+z}\Big)^{-\frac{6}{5}} \fa^{-\frac{3}{5}} \lc^{-\frac{1}{5}} \n^{-\frac{6}{5}} 
        &  \li<\lpi<\lc, 
        \\
          \Big(\frac{\cbeta[,i]\re \lamo}{1+z}\Big)^{-1} \li^{\frac{1}{6}} \fa^{-\frac{1}{2}} \lc^{-\frac{1}{6}} \n^{-1}
        &   \lpi<\li.
    \end{cases}
    \label{eq:lpi}
\ee
where $\cbeta$ is a constant of order unity which depends on the slope of the power spectrum and can be found through a more detailed derivation using the structure-function (\citealt{rickett90}; \citeta{macquart13}). For a Kolmogorov power spectrum, $\cbeta\sim1.6$ in the inertial sub-range, and $\cbeta[,i]\sim1.4$ for $\lpi<\li$.

We present an estimate of the inner scale of turbulence for different CWOs following \citeta{beniamini20} in \se{li}, and choose which expression for $\lpi$ from \eq{lpi} to use accordingly. However, it turns out that $\lpi<\li$ only for a very small part of our parameter space for the CWOs considered here. This can occur for low $\nuo$ of a few $100\mhz$ or for high $\fa\gg1$, which may be relevant for the CGM in massive haloes at high redshift (corresponding to rare high $\sigma$-peak haloes). For filaments and sheets, we examined a wide range of parameters ($\fa=1-1000$ and $\nuo\sim0.4-1\ghz$), finding that the coherence length scale is within the inertial subrange.

\smallskip
For $\li<\lpi<\lc$,
\be
    \lpi 
    &\sim
        2.4\tm 10^{12}\cm\ \ 
        \nuf^{\frac{6}{5}} \opzo^{\frac{6}{5}}  \fa^{-\frac{3}{5}} \lc[,10]^{-\frac{1}{5}} \nf^{-\frac{6}{5}} 
\label{eq:lpi_num}
\ee% eq:lpi_num
where $\opzo=\opzo[]/5$, $\lc[,10]=\lc/(10\pc)$, $\nf=\n/(10^{-2}\cmc)$, 
and $\nuf=\nu/1\ghz$. 

\smallskip
In the top row of \fig{tau}, we show the diffractive scale $\lpi$ of CWOs on the $P-z$ plane for $\nuo=1\ghz$ and $\fa=1$, where
$\n$ and $\lc$ are taken from \fig{clump_prop}. The coherence length is roughly $\gsim 10^{12.5}\cm$ and $\gsim 10^{12}\cm$ for sheets and filaments, respectively, with $\Tv\sim 10^6\Kel$ in the redshift range $z\sim (2-5)$. 
For halos with $\Mh\sim 10^{12}\msun$ in the same redshift interval, $\lpi$ decreases from $\sim 10^{12}\cm$ to $\sim 10^{11}\cm$. 

As argued by \citet{mccourt18} and \citeta{vedantham19} and can be seen in \fig{clump_prop}, for typical pressures in a hot CGM at roughly solar metallicity, the electron column density of a single cloud is roughly constant at $\N\sim\n\lc\sim 10^{17}\cms$. It is thus convenient to express $\lpi$ in the CGM as a function of the column density of a single cloud,
\be 
    \lpi \sim 9.4\tm 10^{10}\cm \ \nuf^{\frac{6}{5}} \opzo^{\frac{6}{5}}  \fa^{-\frac{3}{5}} \lc[,0.1] \N[,17]^{-\frac{6}{5}}
    \label{eq:lpi_column}
\ee 
where $\lc[,0.1]=\lc/(0.1\pc)$ and $\N[,17]=\N/(10^{17}\cms)$, typical values for high-$z$ shattered clouds in the CGM. However, as seen in \fig{clump_prop}, the column densities in filaments and sheets can span a broad range, making this expression less useful than \eq{lpi_num}.

\smallskip
Another useful quantity which relates to the coherence length scale is the scattering measure (SM) of a screen. The SM is defined as $\sm=\int_s^{s+\dL} ds \cn$, where $ds$ is the path length element through the screen and $\cn$ is the \emph{spectral amplitude}, the amplitude of the density power spectrum. For a Kolmogorov power spectrum, it is defined as $\cn\equiv(54\pi)^{-1/3} \lo^{-2/3}\dn^2$. For gravitationally collapsed objects, it is common practice to assume the limit of a thin uniform screen where $\sm\sim\dL \cn$.%
\footnote{
This approximation is ill-suited for objects which are a part of the Hubble flow, e.g., when estimating the SM of the IGM.
}
The classic definition of the SM, originally used to quantify scattering and scintillation of pulsars passing through a MW-ISM plasma screen at $z=0$, depends only on the intrinsic properties of the screen, with $\sm=(54\pi)^{-1/3} \fa\lo^{1/3}\n^2$. The coherence length scale can then be expressed in terms of the SM, $\lpi\propto[\lamz^2\sm]^{-3/5}$. For cosmological distances, it is useful to absorb the factor of $\opzo[]^{-2}=(\lamz/\lamo)^2$ in $\lpi$ into the SM (see for example \citeta{macquart13}), giving
\be
    \label{eq:sm_def}
    \sm 
    &\sim \dL \cn \opzo[]^{-2}
    = (54\pi)^{-\frac{1}{3}} \fa \lo^{\frac{1}{3}} \n^2 \opzo[]^{-2}
    \\
    & \sim 5\tm 10^{11} \smunits\  \fa\, \lc[,10]^{\frac{1}{3}}\, \nf^2 \opzo^{-2}
\ee
Since we deal with cosmological distances, unless stated otherwise (\se{dm_cl_frac}), we use this definition of the SM.

% -----------------------------------------
\subsection{Temporal Broadening}
\label{se:tau}
% ----------------------------------------- 

% _________________________________________
{
\begin{figure*}[ht]%
    \centering%
    \includegraphics[width=0.85\textwidth]{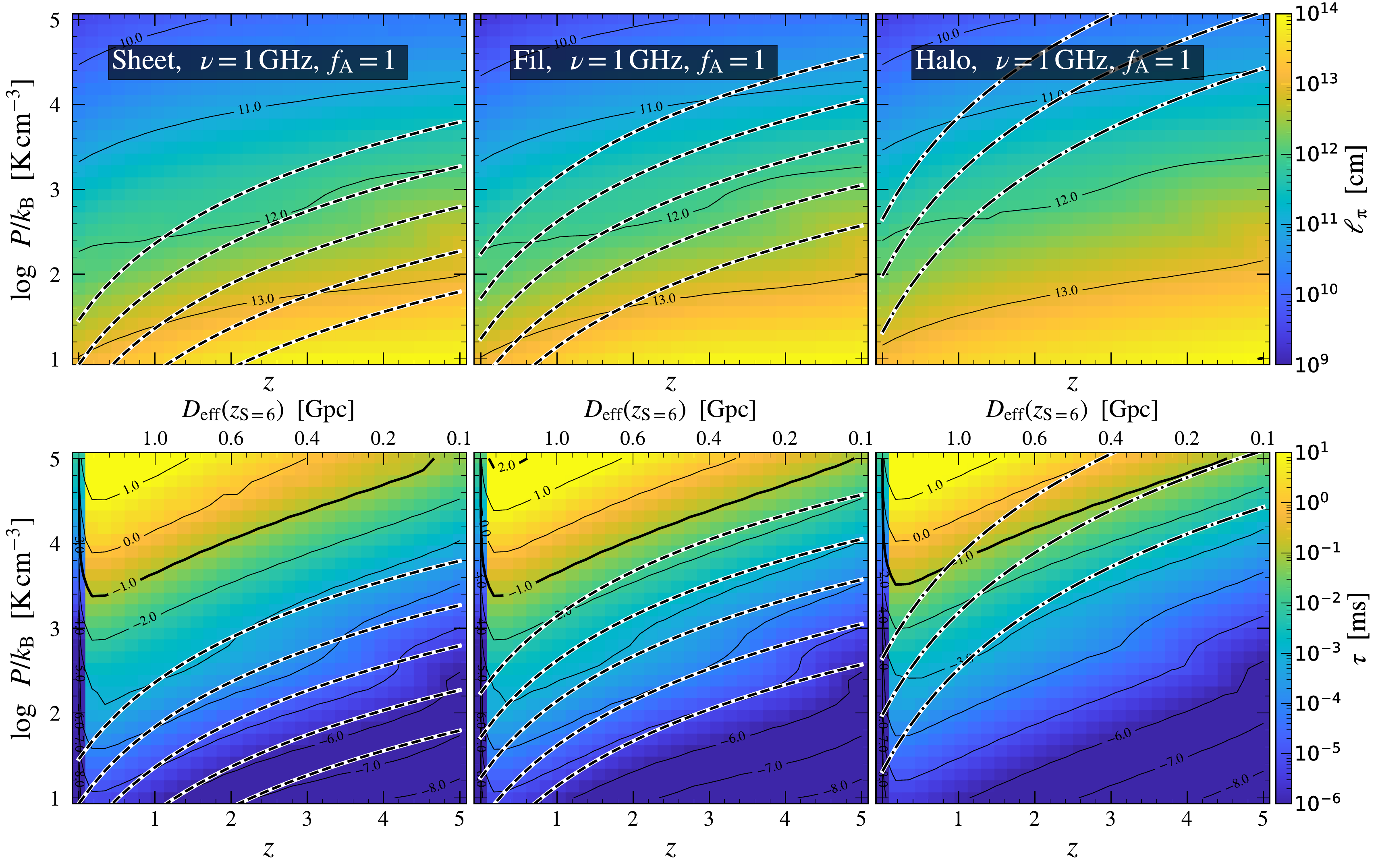}
    \caption{
    Similar to \fig{clump_prop}, but showing the diffractive scale ($\lpi$, top), and the temporal broadening ($\taus$, bottom) for CW screens as a function of thermal pressure and redshift, assuming an observed frequency of $\nuo=1\ghz$, $\fa=1$ clump per los across the CWO, and Kolmogorov turbulence with $\alpha=1/3$ in the clump. The redshift along the bottom $x$-axis indicates the CW screen redshift. For the temporal broadening, the source is assumed to be at redshift $\zs=6$ and $\Deff(z,\zs=6)$ is marked on the top $x$-axis. 
    For filaments with $\Tv\sim (10^{5.5}-10^{6})\Kel$ we estimate a temporal broadening of $\taus\sim(10^{-4}-10^{-3})\ms$, while for sheets $\taus$ is two orders of magnitude lower. For the CGM, our model predicts a temporal broadening of $\taus\gsim0.1\ms$ for $10^{14}\msun$ halos at $z>0.5$, or $10^{13}\msun$ halos at $z>4$, scaling approximately as halo mass at a fixed redshift.}
    \label{fig:tau}
\end{figure*}
}% fig:tau
% _________________________________________

Scattered rays travel longer path lengths, leading to a delay in their arrival time and thus to a temporal broadening of the observed signal. We offer a detailed description of the geometric time delay due to scattering in Appendix \se{time_delay}, though the final expression is well known.
It is convenient to express it as, 
\be 
    \label{eq:tau_obs_0}
    \taus \simeq \frac{\lamo}{2\pi c} \pfrac{\rf}{\lpi}^2,
\ee 
where $\rf$ is the Fresnel scale, the transverse scale on the screen over which the path length difference is of the order of $\lam$, 
\be
    \rf 
    = \pfrac{\Deff\lamz}{2\pi}^{\half} \sim 5.4\tm 10^{13}\cm \pfrac{\Deff[,9]}{\nuf\opzo}^{\half}.
    \label{eq:rF}
\ee% eq:rF
where $\Deff[,9]=\Deff/(1\Gpc)$ is the effective angular diameter distance to the source. The strength of the scattering regime can be defined as $u\equiv\rf/\lpi$, such that screens with $u>1$ ($u<1$) are in the strong (weak) regime.

\smallskip
Using \eqs{lpi}, \eqm{tau_obs_0} and \eqm{rF}, the scattering time is
\be
    \taus = 
    \begin{cases}
        \frac{\lt(\cbeta\re\rt)^{\frac{12}{5}}  }{4\pi^2 c} 
         \D \lamo^{\frac{22}{5}} \opzo[]^{-\frac{17}{5}} \fa^{\frac{6}{5}} \lc^{\frac{2}{5}} \n^{\frac{12}{5}}
         &  \li<\lpi<\lo 
        \\
        \frac{\lt(\cbeta[,i]\re\rt)^2}{4\pi^2 c}  \D \lamo^4 \li^{-\frac{1}{3}} \opzo[]^{-3}   \fa \lc^{\frac{1}{3}} \n^2
         &   \lpi<\li
    \end{cases} 
    \label{eq:taus}
\ee % eq:tau
For $\li<\lpi<\lc$ we obtain
\be
    \taus 
    &\sim
        8\tm10^{-5} \ms\, 
        \Deff[,9]\nuf^{-\frac{22}{5}} \opzo^{-\frac{17}{5}} \fa^{\frac{6}{5}}  \lc[,10]^{\frac{2}{5}} \nf^{\frac{12}{5}}.
    \label{eq:tau_num}
\ee% eq:tau_num
Note that \eq{taus} suggests somewhat counter-intuitively that the temporal broadening increases with cloudlet size. However, this is for a fixed $\fa$, namely a fixed number of cloudlets along the line of sight. If we consider instead a fixed volume filling fraction for cold gas, $\fv\simeq \fa\lc/\DL$, which is related to the mass-fraction of cold gas, $\fm\propto \fv\rho_{\rm cold}/\rho_{\rm hot}$, the temporal broadening increases with decreasing cloudlet size \citepa[see also][]{vedantham19}. We address constraints on both $\fa$ and $\fv$ in different CWOs in \se{fa_fv}. In either case, $\taus$ depends much more strongly on cloud density than on cloud size.

\smallskip
In the bottom row of \fig{tau}, we show the observed temporal broadening at $\nuo=1\ghz$ induced by one CW screen with a single cloud in it ($\fa=1$) for sheets (left), filaments (middle) and halos (right), as a function of thermal pressure and the redshift of the CWO, assuming an FRB source at a fixed redshift of $\zs=6$. 
In each case, we estimated the dissipation scale, $\li$, following \citeta{beniamini20} assuming a fixed magnetic field, $B=10^{-8},10^{-7}$, and $10^{-6}\G$ for sheets, filaments, and haloes, respectively (see \se{li}). This determines whether $\lpi<\li$ or $\lpi>\li$ in \eq{lpi} and \eq{taus}. Overall, the impact of $B$ on $\li$ is small, and we find that for all values considered, we rarely have $\lpi<\li$.

\smallskip
Under these assumptions, only clouds in the CGM of massive haloes with $\Mh>10^{14}\msun$ at $z>0.5$ or $\Mh>10^{13}\msun$ at $z>3.5$ result in a detectable temporal broadening of $\taus\gsim 0.1\ms$, which is the threshold robustly measurable by the CHIME/FRB instrument \citep{chawla22}. For sheets and filaments with $\Tv\lsim10^6\Kel$, individual clouds respectively induce $\taus<10^{-4}$ and $<10^{-3}\ms$ at $\nuo[9]=1$. This large difference in temporal broadening is due to the different pressures in each CWO at a given redshift, as shown by the dashed lines in each panel. However, we stress that for a given pressure and redshift, our model predicts very similar temporal broadening for clouds in different CWOs, despite the different values assumed for metallicity and magnetic fields. This is despite the different values of $\lc$ in different CWOs (\fig{clump_prop}, top row), and highlights the relatively weak dependence of $\taus$ and $\lpi$ on cloud size for a given $\fa$. These quantities do have a strong dependence on cold gas density, which at a given pressure and redshift is similar in all CWOs in our model (\fig{clump_prop}, third row).

% Deff, z_max
\smallskip
The top x-axis of the lower panels in \fig{tau} shows the effective angular diameter distance $\D$ corresponding to the redshift of the CWO in the bottom x-axis, assuming an FRB source at $\zs=6$. However, $\D$ is not very sensitive to this choice. A source at higher $\zs$ will slightly increase $\taus$, though the difference is not substantial. For example, for a screen at $z=1$, the distance $\D$ is roughly constant at $\zs\gsim 2$. Similar behaviour is seen for screens at different redshifts (see \fig{Deff} below and fig.~5 in \citeta{macquart13}). 

% ************************************************
% @@@@@@@@@@@@@@@@@@@@@@@@@@@@@@@@@@@@@@@@@
\section{Counting CWOs and cloudlets}
\label{se:cosmo}
% @@@@@@@@@@@@@@@@@@@@@@@@@@@@@@@@@@@@@@@@@
%

We now wish to account for multiple CW screens along a given los, and the probability that a given los intersects a given CWO.  To do this, we estimate the number density, $\nc(\mobj,z)$, and projected area, $\Ax(\mobj,z)$, of different CWOs, and obtain the average number of intercepted CWOs, $\num(\mobj, \zs)$, along a los from an FRB source at redshift $\zs$.

% projected area
\smallskip
For simplicity, we assume that the los is perpendicular to the filament axis and the sheet plane. 
Since these orientations maximise the cross section but minimise the path length through the CWO, they represent an upper limit on the probability for the CWO to be intercepted by a given los, and a lower limit on the total scattering contributed by the CWO. For the CGM, we assume that the cross section is the projected area of the virial sphere, $\Ax[h]=\pi\Rvh^2$, with this being dominated by the outer halo.

\smallskip
Below, we evaluate the mass, comoving number density, and cross section of CW sheets and filaments as a function of redshift and their virial temperature, which in our model determines whether or not a CWO will form cold cloudlets (\se{P}).

% =========================================
\subsection{Average Number of Intercepted CWOs}
\label{se:Nobj}
% =========================================
%

% _________________________________________
{
\begin{figure*}[ht]%
    \centering%
    \includegraphics[width=0.85\textwidth]{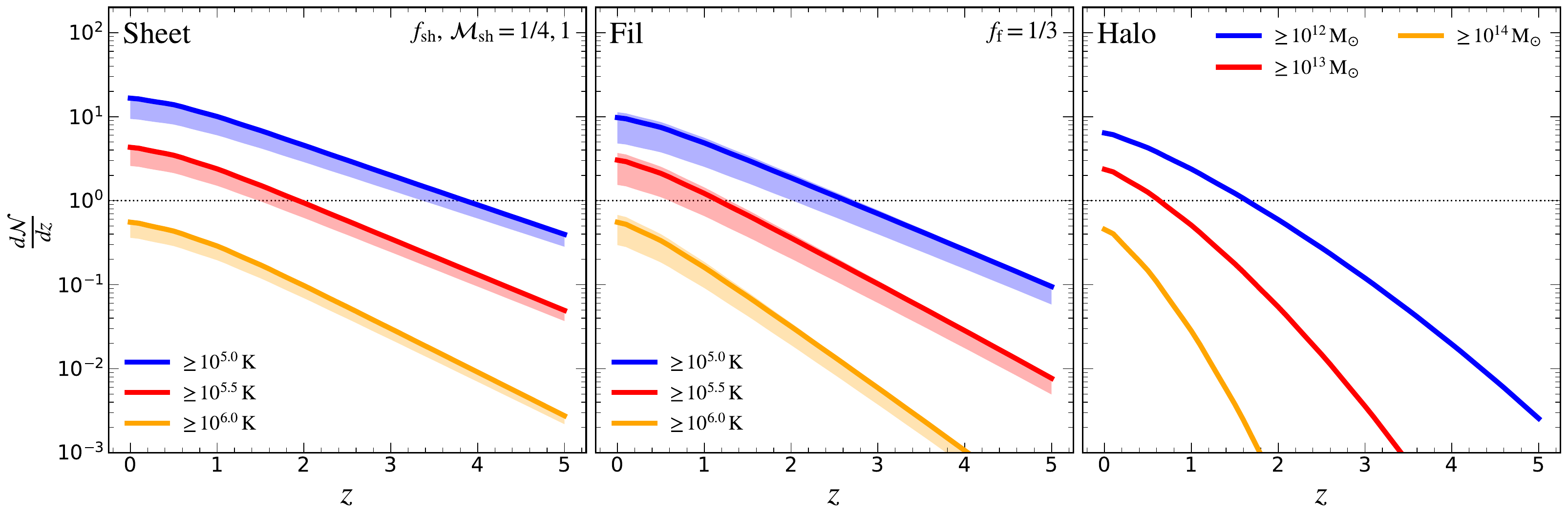}
    \caption{
    The average number per unit redshift of CW sheets (left) and filaments (middle) with virial temperature above a given threshold, and halos with virial mass above a given threshold (right), intercepted by a random los as a function of the CWO redshift, $z$. 
    For filaments and sheets, the solid lines correspond to the median mass using the conditional mass distribution function, and the shaded regions indicate the scatter (see text).}
    \label{fig:Nobj}
\end{figure*}
}% fig:Nobj
% _________________________________________

The average number of systems with total mass of $M=\mobj$ intersecting the los to a source at a redshift of $\zs$ is given by \citep{bahcall69, padmanabhan02}, % ~ total optical depth for the intersection of objects
\be
    \num(\zs,M) 
    &= \int_0^s ds\, \Ax[p]\,n_\mr{p} 
    \label{eq:Nobj}
\ee
where $ds$ is the proper los distance travelled by a ray during the redshift interval of $dz$, $ds= c H(z)^{-1} \opzo[]^{-1}dz$, $n_\mr{p}(z,M)\sim \nc(z,M)\opzo[]^3$ is the proper number density of such objects, $\nc$ is the comoving number density, and $\Ax[p]$ is the projected proper cross-section of the object (e.g., for halos $\Ax[p,halo]=\pi \Rvh^2$). The total mass of a CWO includes all hierarchical substructures (e.g., the total mass of the sheet includes the embedded filaments and halos).  % $d_\mr{H}(z)=c H_0^{-1}[\Oml+\omm\opzo[]^3]^{-1/2}=c H(z)^{-1}$. 

\smallskip
In our companion paper (\dmpaper), we evaluate the abundance and projected area of CWOs as a function of their virial temperature, calculate the derivative of \eq{Nobj} with respect to redshift, and integrate over masses above $\mobj$. The average number of CWOs with masses $\geq \mobj$ intercepted in the interval $dz$ about redshift $z$ is given by
\be
\label{eq:dNdzdM}
\frac{d\num(z, >\mobj)}{dz} 
    = \frac{c\opzo[]^2}{H(z)} \int_{\mobj}^\infty dM\, \Ax[p](z,M)\,\frac{d\nc(z, M)}{dM}.
\ee
In \fig{Nobj}, we show $\dNdz$ as a function of $z$ for different CWOs. For sheets and filaments, our model evaluates the relation between the total mass of a CWO and its virial temperature by combining the virial properties of CWOs developed in \dmpaper and \citeta{m18} respectively, with the excursion set model for the CW developed by \citet{shen06}. This model provides the conditional mass function of sheets/filaments connecting to halos of a given mass. 
For a given halo mass, we use the conditional mass function distribution to find the median sheet/filament masses, and the $(16, 84)$ percentile scatter about this mass. These are marked as $\kappa=0, -1, +1$, respectively.
Using our model, we then convert the mass of the CWO into a virial temperature (see \dmpaper for details). Each line in the left and middle panels represents $\dNdz$ of objects with $\Tv$ above the virial temperature threshold quoted in the legend. 
For filaments and sheets, the solid lines represent the median masses ($\kappa=0$), and the scatter about them is derived from the scatter in filament/sheet mass ($\kappa=\pm 1$). However, note that the $\dNdz$ of the $84$th percentile filament mass is often similar to or lower than that of the median mass. While more massive filaments cover a larger area across the sky, their number density can be significantly lower and may overall decrease $\dNdz$ below that of the median mass. We set the scatter in \fig{Nobj} by finding the minimal/maximal $\dNdz$ among $\kappa=0,\pm1$. 
For halos (right panel), each line indicates $\dndz$ of halos with $\Mh$ above the threshold mass quoted in the legend.

% =========================================
\subsection{Average Electron column density of CWOs}
\label{se:Nt}
% =========================================

We now wish to evaluate the electron column density across a virialized CWO. 
In \dmpaper, we develop a model for the virial properties of sheets, motivated by the model by \citeta{m18} for the virial properties of filaments which feed massive haloes. 
Here, we briefly summarise a few relevant properties for CWOs, and refer the reader to \citeta{m18} and \dmpaper for further details.

Given a sheet with virial temperature $\Tvsh=\Tvfid\Kel$, the virial scale length ($\Xv$) and density ($\rhovsh$) are
\begin{align}
    &\Xv 
    \sim 170 \kpc \ 
    \Tvshf^{0.5}\Dvshf^{-0.5} \opzo^{-1.5}
    \label{eq:Xv_Tvsh}
    \\
    &\rhovsh 
    \sim 2.1\tm10^{-27} \gcc\; \Dvshf \opzo^3
    \label{eq:rhov_sh}
\end{align}
where $\Dvshf$ is the virial overdensity of the sheet, normalised to a value of $6$ \citep{shen06}.

The virial radius and density of a filament with a temperature of $\Tvf=\Tvfid\Kel$ are
\begin{align}
    &\Rvf  
    \sim 63\kpc\ \Dvff^{-0.5} \ \Tvff^{0.5} \opzo^{-1.5} 
    \label{eq:Rv_fil_T}
    \\
    &\rhovf
    \sim 1.24\tm10^{-26} \gcc\; \Dvff \opzo^3
    \label{eq:rhov_fil}
\end{align}
where $\Dvf$ is the virial overdensity of the filament, with $\Dvff=\Dvf/36$ (\citealt{shen06}; \citeta{m18}). 

For haloes, the virial radius and density of a halo as a function of halo mass \citep[e.g.,][]{dekel13} are 
\begin{align}
    &\Rvh\sim 64\kpc \, \Dvhf^{-1/3}\Mhf^{1/3} \opzo^{-1}
    \label{eq:Rv_h}
    \\
    &\rhov[,h]
    \sim 6.2\tm10^{-26} \gcc\; \Dvhf \opzo^3
    \label{eq:rhov_h}
\end{align}

\smallskip
To estimate the contribution of each CWO to scattering, we must first estimate its electron column density. We estimate the average electron column density of a single CWO as 
\be 
    \Nt
    \sim \frac{\xie}{\mpr} \rho_\mr{g} \DL
    \label{eq:Nt}
\ee% eq:Nt
where $\rho_\mr{g}\sim\fb\rhov$ is the average gas density in the screen, $\xie\sim0.88$ is the free electron fraction per nucleon assuming a primordial composition of fully ionised hydrogen and helium,%
\footnote{At $z\gsim 2$, prior to helium reionisation, one obtains $\xie\sim 0.82$. However, we use a fixed $\xie=0.88$ at all redshifts.} 
and $\DL$ is the path length within the screen. For all objects, we ignore the internal density profile,%
\footnote{ 
For haloes, the profile is potentially more influential, and we therefore repeat some of our calculations with several density profiles in \se{profiles}. However, this does not change any of our main conclusions.}
and consider a smooth gas distribution rather than clumpy substructure. 

Using \eq{Rv_fil_T}, \eq{rhov_fil} and \eq{Nt}, the average $\Nt$ of a single filament for a given $\Tvf$ is 
\be
    \Nt[,f] 
    \sim \frac{\xie \fb}{\mpr} 2\eta\Rvf \rhovf
    \label{eq:Nt_f}
\ee
where $\eta=\sqrt{1-b^2}$, $b$ is the impact parameter normalised by the filament virial radius ($b=R_\mr{\perp}/\Rvf$), and the path length through the filament is $2\eta \Rv[,f]$. Since we are ignoring the filament density profile, the interception at different impact parameters affects only the path length through the plasma screen.

For a single sheet, using \eq{Xv_Tvsh}, \eq{rhov_sh}, and \eq{Nt}, the average $\Nt$ is
\be
    \Nt[,sh] 
    \sim \frac{\xie\fb}{\mpr} 2\Xv\rhovsh.
    \label{eq:Nt_sh}
\ee% eq:Nt_sh
Note that for a face-on sheet, as assumed, there is no dependence on the impact parameter. For the CGM in a single halo, using \eq{Rv_h}, \eq{rhov_h}, and \eq{Nt}
\be
    \Nt[,h]
    &\sim \frac{\xie\fb\fcgm}{\mpr} 2\eta\Rvh\rhovh
    \label{eq:Nt_h}
\ee
where $\fcgm$ is the fraction of baryonic mass in the halo that is in the CGM, with a fiducial value of $1/2$ \citep[e.g.,][]{oren24,khrykin24}. Similarly to the filament, $\eta=\sqrt{1-b^2}$, with $b=R_{\perp}/\Rvh$, and the path length through the halo is $L=2\eta \Rvh$.

% =========================================
\subsection[tcool/tff]{$\tcool/\tff$ -- Constraints on fragmentation at Low-$z$}
\label{se:tcool_tff}
% =========================================
% 

% _________________________________________
{
\begin{figure*}[ht]%
    \centering%
    \includegraphics[width=0.85\textwidth]{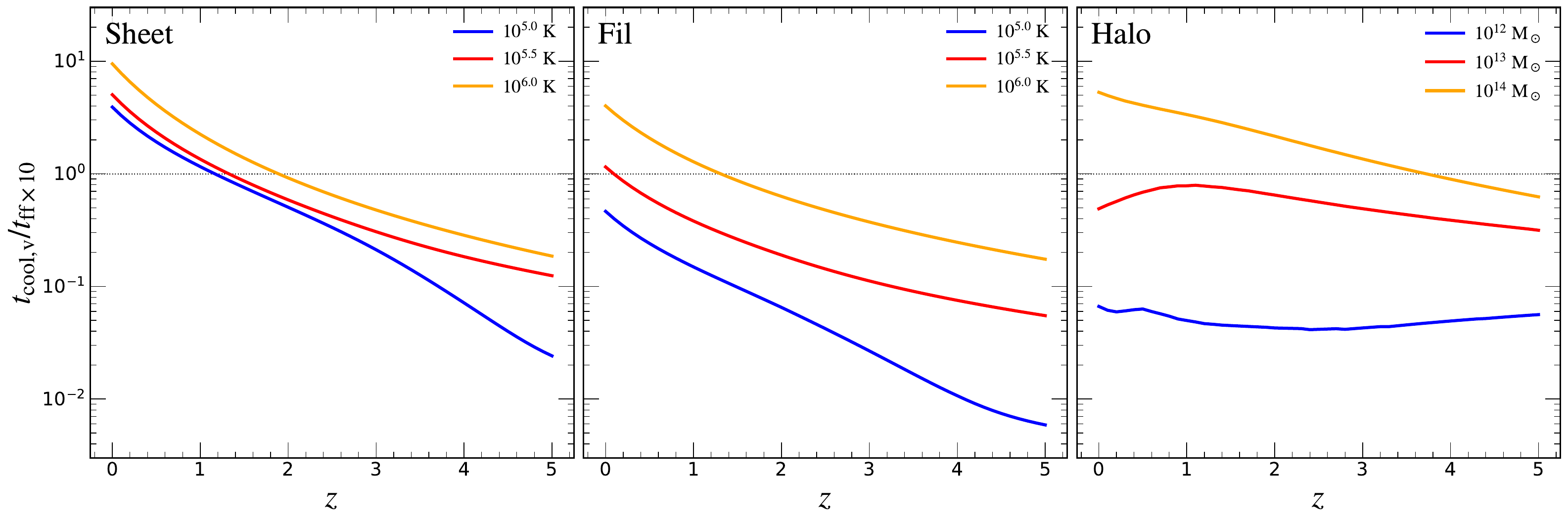}
    \caption{
    The ratio of cooling time to ten times the free-fall time for virialized gas in CW sheets (left), filaments (middle), and halos (right), with different virial temperatures or masses as indicated in the legend. According to our simple model, CWOs are unstable to thermal condensation and fragmentation if this ratio is $\lsim 1$. This is always the case for low and intermediate mass halos and filaments, though the most massive halos are stable at $z\lsim4$, and the most massive filaments are stable at $z\lsim 1$. Low and intermediate mass sheets are stable at $z\lsim 1$ while high mass sheets are stable at $z\lsim 2$.
    }
    \label{fig:tcool_tff}
\end{figure*}
}% fig:tcool_tff
% _________________________________________

To estimate whether the post-shock medium in a virialized CWO is susceptible to local thermal instabilities that can lead to fragmentation, we estimate the ratio of the cooling time to the free-fall time close to the virial radius. Studies of the intracluster medium (ICM) in galaxy clusters suggest that hot gas is subject to local thermal instabilities when $\tcool/\tff\lsim 10$ \citep{mccourt12, sharma12, voit15, voit-donahue15}. These instabilities form overdense clouds that proceed to cool and condense and can eventually fragment into small cloudlets. 
Systems with plane-parallel and cylindrical geometries have been shown to exhibit similar thresholds in $\tcool/\tff$ for condensation \citep{meece15, choudhury16}. We hereafter adopt a threshold of $\tcool/\tff=10$ for the presence of small-scale structure in all CWOs. We consider only the mean density and post-shock temperature in each CWO, neglecting the profiles in $\tcool$ and $\tff$ which, depending on assumptions, can make condensation more likely closer to the centre or at the outskirts. 
Furthermore, it is worth noting that condensation has also been found to occur at much larger values of $\tcool/\tff\lsim 20$ for non-linear perturbations \citep{choudhury16,choudhury19}. 
We thus stress that this is only a crude approximation of a fragmentation criterion in CWOs, meant for illustrative purposes. 

\smallskip 
We account for different geometries when evaluating the free-fall time. For spherical halos, the free-fall time is $\tff[,h]=\sqrt{3\pi/(32 G \rhov)}$, where $\rhov$ is the average density in the sphere. For filaments and sheets, the free-fall time is given by $\tff[,f]=\sqrt{1/(4 G \rhov)}$ and $\tff[,sh]=\sqrt{1/(2\pi G \rhov)}$, where $\rhov$ is the mean density in the filament and sheet, respectively. The cooling time for gas in the vicinity of the virial radius, with temperature $\Tv$ and density $\rhov[,g]\sim \omb\Dv\rhoc[,0]\opzo[]^3$, is given by 
{
\be
    \tcool[,v] 
    &= 
    \frac{\kb\mpr}{\mu \xh^2 (\gamma-1) \rhoc[,0]\omb}
    \frac{\Tv}{\opzo[]^3 \Dv\, \Lamtz(\Tv,Z,z)}\\
    \label{eq:tcoolv}
\ee
}% eq:tcoolv
where for haloes we include a factor of $\fcgm$ in the gas density.

\smallskip
In \fig{tcool_tff} we show the ratio $\tcool[,v] /(10\tff)$ for each CWO, to constrain the redshift range where one would expect to find multiphase substructure within the shocked medium. According to our simple model, halos with $\Mv\lsim 10^{13}\msun$ and filaments with $\Tv\lsim 10^{5.5}\Kel$ are susceptible to thermal instabilities and condensation at all redshifts. More massive halos, $\Mv\sim 10^{14}\msun$, are only likely to form cloudlets at $z\gsim4$, where such massive halos are unlikely to be found anyway. Filaments with $\Tv\sim 10^6\Kel$ are unstable to condensation at $z\gsim 1$. 
Sheets, on the other hand, are only unstable at $z\gsim 1$ for $\Tv\sim 10^{5-5.5}\Kel$, and at $z\gsim2$ for $\Tv\sim 10^6\Kel$.

% =========================================
\subsection[Constraints on fa]{Constraints on $\fa$}
\label{se:fa_fv}
% =========================================

% _________________________________________
{
\begin{figure*}[ht]%
    \centering%
    \includegraphics[width=0.85\textwidth]{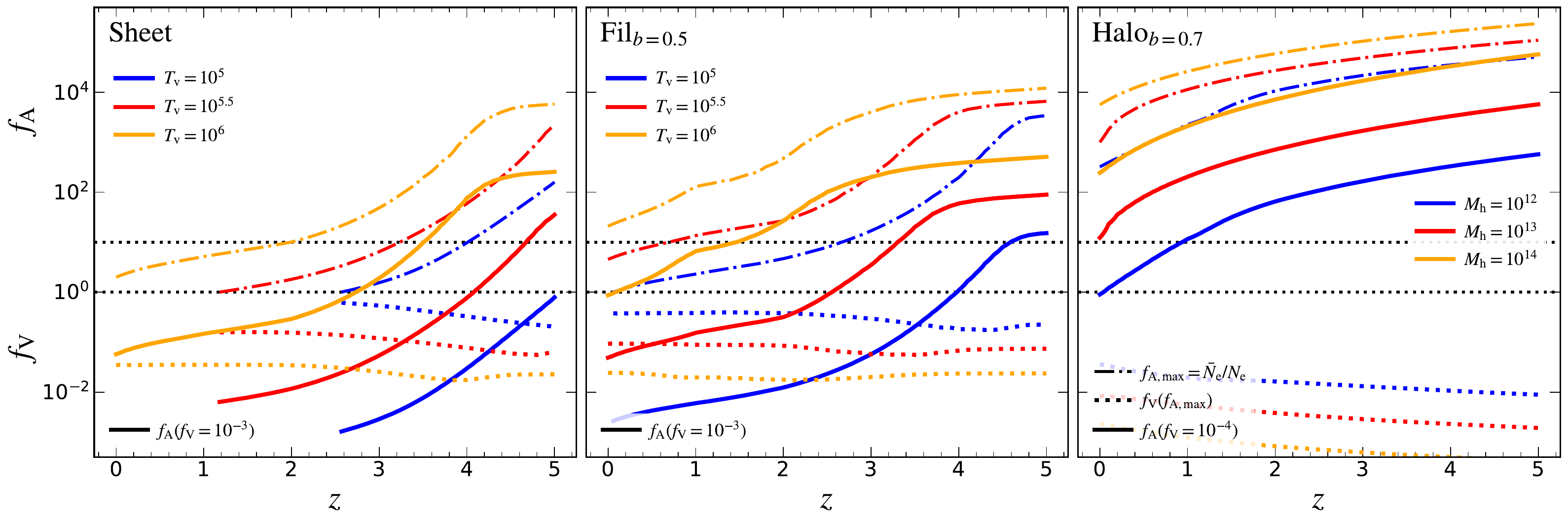}
    \caption{
    Areal-covering fractions, $\fa$, and volume filling fractions, $\fv$, for CW sheets (left), filaments (middle) and haloes (right), with different values of $\Tv$ for sheets and filaments and different values of $\Mh$ for haloes, as indicated in the legend. Dot-dashed lines represent the maximal covering fraction, $\famax\sim \Nt/\N$, defined as the ratio of the mean electron column density through the CWO (\se{Nt}) to the typical column density of a cloudlet (\se{shatter}). The existence of small-scale cloudlets of size $\lc$ is only possible when this is $>1$, which is always the case for filaments, haloes, and massive sheets with $\Tv\gsim 10^6\Kel$, though for lower mass sheets this is only true at high redshift. The dotted lines represent the volume filling fractions corresponding to $\famax$, $\fvmax=\fv(\famax)$. This is similar for sheets and filaments, though smaller for haloes, and in all cases, there is no noticeable redshift dependence. The solid lines represent the \fa obtained for a fixed $\fv=10^{-3}$ for sheets and filaments and $10^{-4}$ for haloes. 
    }
    \label{fig:fa_fv}
\end{figure*}
}% fig:fa_fv
% _________________________________________

In this section, we estimate upper limits on the covering fraction ($\fa$) and the corresponding volume filling fraction (\fv) of cool gas in CWOs, and use these to draw conclusions on scattering.

\smallskip 
The maximal covering fraction for each CWO, $\famax$, is estimated as the ratio of the total average electron column density, $\Nt$ (\eqnp{Nt}), and the column density of a single cloud, $\N$ (\se{shatter}), assuming an impact parameter of $b=0.7$ and $0.5$ for haloes and filaments, respectively.%
\footnote{
$b=0.5$ ($b=0.7$) is the median impact parameter for a random distribution of sightlines through an edge-on cylinder (projected sphere).}
Throughout this work, $\Nt$ is defined as the total electron column density, including both the hot and cold components. Here, we estimate $\famax$ by considering the extreme limit, where the entire column density is in the cold phase and in the form of dense cloudlets. Namely, $\Nt=\Nt[,cold]=\famax\N$, where $\Nt[,hot]=0$.
We show $\famax$ as dash-dotted lines in \fig{fa_fv} for sheets (left), filaments (middle), and halos (right), with different colours representing different virial temperatures (sheets and filaments) and masses (halos). We only show cases where $\Nt\geq\N$ so that $\fa[,max]\geq1$. 
For sheets with $\Tv\sim 10^{5.5}\Kel$ and $\Tv\sim 10^{5}\Kel$, $\fa[,max]<1$ at $z\lsim 1$ and $z\lsim 2.5$, respectively. This suggests that low- and intermediate-mass sheets at lower $z$ may not form cold cloudlets and thus not contribute to scattering. On the other hand, more massive sheets with $\Tv\sim 10^6\Kel$, as well as all filaments and halos in the range of virial temperatures and masses considered, maintain $\fa[,max]>1$ at all redshifts, suggesting that these objects can potentially contribute to scattering of FRB sources at all redshifts. However, we caution that in massive CWOs, the formation of multiphase gas may be suppressed, limiting their potential contribution to scattering (see \se{tcool_tff} and \fig{tcool_tff}).

\smallskip 
The dotted lines are the corresponding upper limits on \fv calculated as $\fvmax=\famax \lc/\DL=\nt/\n$, where $\lc$ is the cloud size
and $\DL$ is the total width of the CWO calculated from its virial scale length, $\DL=2\eta\Rv[,obj]$ (see \se{Nt}). Interestingly, while $\famax$ for filaments can be an order of magnitude larger than for sheets with the same $\Tv$, the corresponding $\fvmax$ are rather similar. On the other hand, $\fvmax$ for haloes is roughly two orders of magnitude smaller; however, this is not an apples-to-apples comparison because we are looking at haloes of fixed $\Mh$ rather than fixed $\Tv$. In all cases, $\fvmax$ is roughly independent of redshift, despite both $\famax$ and $\fa$ for a fixed $\fv$ (solid lines, described below) declining towards lower redshift. This suggests that $\fv$ may be a more stable parameter than $\fa$ with less variation over time, and also that $\fv$ may be similar in sheets and filaments, though smaller in the CGM.

\smallskip
$\famax$ provides an upper limit on \fa and helps place crude limits on the redshift range where fragmentation to cloudlets is possible. However, it represents an extreme upper limit relying on the unrealistic assumption that the entire electron column density of the CWO consists of cloudlets, with no hot component, $\Nt[,hot]=0$. Therefore, we show as solid lines in \fig{fa_fv} the \fa resulting from assuming a fixed $\fv=10^{-3}$ for sheets and filaments and $\fv=10^{-4}$ for halos, where for a given $\fv$, it is evaluated as $\fa=\fv\DL/\lo$. The similar values adopted for sheets and filaments and the smaller value adopted for the CGM are motivated by the different values of $\fvmax$ discussed above. Recalling that the mass fraction of cold gas is $\fm \sim \chi\fv$ with $\chi$ the density contrast between the cold and hot components, these values span the range typically inferred for shock-heated CWOs and the CGM in both simulations (e.g. \citeta{m21}; \citealp{lu23}) and observations \citep[e.g.][]{Cantalupo19,prochaska19-sci}. 

\smallskip
In summary, we consider the following criteria for fragmentation into $\lc$-sized cloudlets:
\be
    \label{eq:frag_cond}
    1.\quad &\famax =\frac{\Nt}{\N}\geq 1\\
    2.\quad &\frac{\tcool}{\tff} < 10
\ee
The first condition is necessary - the column density of a single cloudlet cannot exceed the total available column density within a CWO (nor can its size exceed the total path length through the CWO).%
\footnote{The first condition limits only the maximal covering fraction, $\famax\geq1$, while $\fa$ is allowed to be lower than one. 
}
However, the second condition is a rough guideline, as discussed in \se{tcool_tff}.

% =========================================
\subsection{Clumpy DM fraction}
\label{se:dm_cl_frac}
% =========================================

% _________________________________________
{
\begin{figure*}[ht]%
    \centering%
    \includegraphics[width=0.85\textwidth]{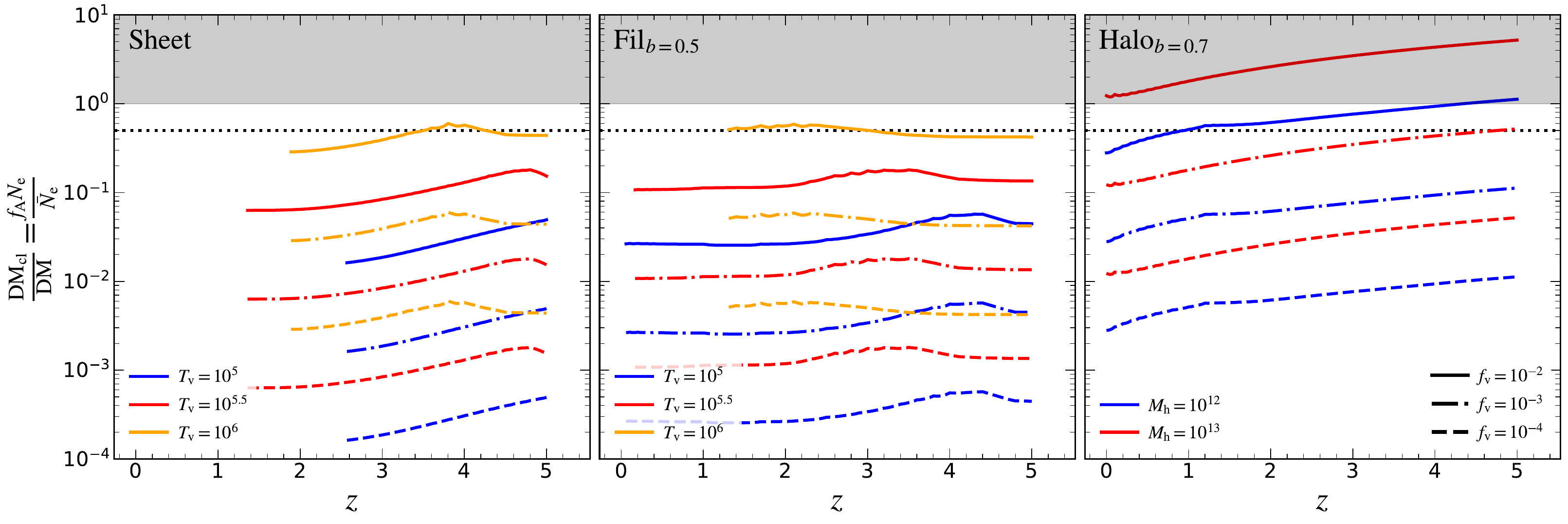}
    \caption{
    Fraction of total $\dms$ contributed by cloudlets in CW sheets (left), filaments (middle), and CGM (right). Different colours represent different values of $\Tv$ for sheets and filaments and different values of $\Mv$ for the CGM. Different line styles represent different assumed \fv for the cloudlets. The horizontal dotted line marks a clumpy DM fraction of 0.5, while the shaded region represents a forbidden region where the DM from cloudlets is greater than the total $\dms$, and can be used to rule out too large values of $\fv$ in halos. In sheets and filaments, cloudlets typically contribute only a small fraction to the total $\dms$, except in the most massive systems with the largest \fv.
    }
    \label{fig:dm_cl_frac}
\end{figure*}
}% fig:dm_cl_frac
% _________________________________________

In this section, we ask whether the ionised regions that dominate FRB scattering in CWOs are the same as those that dominate their $\dms$. More specifically, we ask whether the volume filling fractions of these two gaseous components are similar. This is important for determining the relation between the $\dms$ and the $\sm$ of CWOs. 
Following \citet{cordes91}, we write $\dms=\DL\mean{\fvdm\nt}$ and $\sm=\DL \mean{\fvsc \cn}$ (note that this is the classic definition of the SM without a redshift dependence). The SM of the screen can be expressed as a function of the DM,
\be
    \label{eq:sm_dm_c91}
    \sm 
    &=
    \csm\frac{\fvsc}{\fvdm^2}
    \frac{\epsilon^2\zeta}{\lo^{\frac{2}{3}}} \frac{\dms^2}{\DL}
    = 
    \csm F_\mr{c}  \frac{\dms^2}{\DL}
\ee
where $\fvsc$ and $\fvdm$ are respectively the \fv for the regions dominating scattering and the DM, $\mean{.}$ denotes averaging along the los, $\mean{\n}_\mr{los}=\mean{\fvdm\nt}=\dms/\DL$, $\cn=\csm\epsilon^2\nt^2\lo^{-2/3}$ is the spectral amplitude, $\csm$ is a constant of order unity which depends on the power spectrum index of density fluctuations, $\epsilon=\dn[,rms]/\nt$, $\dn(x)=\n(x)-\nt(x)$ at position $x$ in the screen, $\zeta=\mean{\nt^2}/\mean{\nt}^2$ is the cloud-to-cloud internal density variations, and we assume a Kolmogorov spectrum of turbulence. 

\smallskip
If the ionised regions which dominate the scattering are the same as those dominating the DM, then \eq{sm_dm_c91} is simplified by taking $\fvsc\sim\fvdm$, yielding $\sm/\dms\propto\fvsc^{-1}\epsilon^2\zeta\lo^{-2/3}\mean{\n}_\mr{los}$. This is found to be the case in the MW-ISM (as discussed in our companion paper, \scintpaper). 
However, in the limit where small-scale cloudlets dominate scattering while the $\dms$ is dominated by the volume-filling hot phase, we have $\fvsc\ll\fvdm$ and $\fvdm\ra 1$. In this case, 
$\sm/\dms=\csm{\fvsc\epsilon^2\zeta}{\lo^{-2/3}} \mean{\n}_\mr{los}$, proportional to $\fvsc$ rather than inversely proportional to it. If CWOs are in the latter limit, yet we use observationally motivated relations between the temporal broadening, $\taus$, and the total DM through the screen based on ISM studies which are in the former limit, this will lead to erroneous estimates of the \fv for the cold component. 

\smallskip
Since the conditions in the ISM are very different from those in CWOs, in terms of densities and temperatures, turbulence-driving mechanisms and scales, radiation fields, and magnetic field strengths, among others, we wish to examine what limit we might expect to be valid in CWOs. If the same regions dominate both $\taus$ and DM, then the $\dmsc$ due to dense clumps in a CW screen should be a significant fraction of the total DM. In \fig{dm_cl_frac}, we show the fraction of the total DM contributed by cloudlets in CW sheets (left), filaments (middle), and the CGM of haloes (right), as a function of the screen redshift. The clumpy DM fraction is ${\dmsc}/{\dms}={\fa\N}/{\Nt}$ (y-axis), and can also be expressed as $\fv\DL\n/\Nt$. 
Different coloured lines indicate different $\Tv$ for sheets and filaments, and different $\Mh$ for haloes. Different line styles represent $\fv=10^{-2},\ 10^{-3}$ and $10^{-4}$ (see legends). As before, we assume an impact parameter of $b=0.7,\ 0.5$ for haloes and filaments, respectively, and we show only systems which obey our conditions for fragmentation (\eqnp{frag_cond}). The grey shaded region marks the forbidden region ($\dmsc/\dms>1$), where the DM from cloudlets exceeds the total DM. 

\smallskip
% sheets + filaments dm_cl
From the two left panels in \fig{dm_cl_frac}, we find that the only sheets and filaments where the $\dms$ has a sizeable clumpy fraction ($\sim0.5$) are those with a high $\fv\gsim10^{-2}$, and high temperatures of $\Tv[,sh|f]\gsim10^{6}\Kel$. For all other sheets and filaments, the DM clumpy fraction is very small (typically $\lsim10^{-1}$). This suggests that the regions which dominate the DM are likely not the clumpy regions that dominate the scattering, and different volume filling fractions should be assigned to these components. 

% cgm dm_cl
\smallskip
For the CGM (right panel), the clumpy DM fraction of haloes with $\fv\le 10^{-3}$ is almost always very small, except for $10^{13}\msun$ haloes with $\fv=10^{-3}$ that reach a fraction of $\sim 0.5$ at $z\sim 4-5$ where such massive halos are unlikely to be found. 
If $\fv=10^{-2}$, then $10^{12}\msun$ halos reach a clumpy DM fraction of $\sim0.5$ at $z\gsim1$, and enter the forbidden zone, where $\dmsc/\dms>1$, at $z>4$. Halos of $\Mh=10^{13}\msun$ with $\fv=10^{-2}$ are in the forbidden zone at all redshifts. This suggests that the \fv of cloudlets of such haloes should be lower, $\fv<10^{-2}$. In \se{mscatt} we obtain even stronger constraints on $\fv$ using different arguments.

% ************************************************
% @@@@@@@@@@@@@@@@@@@@@@@@@@@@@@@@@@@@@@@@@
\section{Scattering}
\label{se:mscatt}
% @@@@@@@@@@@@@@@@@@@@@@@@@@@@@@@@@@@@@@@@@ 

In the current section, we combine our results from \se{scatt} for scattering by a single cloud within a CWO with our model for the distribution of and degree of fragmentation in CWOs over cosmic time from \se{cosmo} to explore the effect of scattering by multiple clouds within a single CWO, as well as scattering by multiple CWOs across cosmic time.

% =========================================
\subsection{Scattering by Multiple Clouds Within a Single CWO}
\label{se:mscatt_1cwo}
% =========================================

% _________________________________________
{
\begin{figure*}[ht]%
    \centering%
    \includegraphics[width=0.85\textwidth]{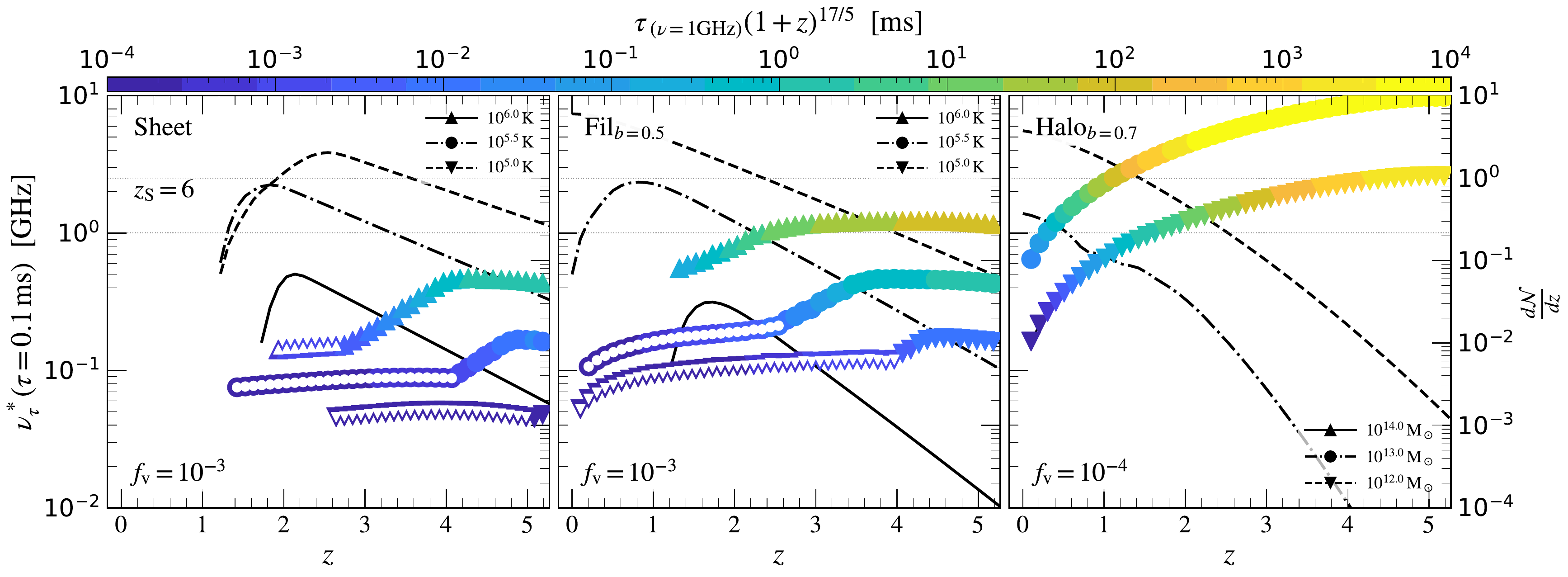}
    \caption{
    Detectability of temporal broadening from CWOs. Coloured symbols show, as a function of redshift, the maximal frequency below which the temporal broadening induced by scattering in CWOs is $\taus\geq0.1\ms$, $\nuot$ (left y-axis, see \eqnp{nuot}). This is taken as a rough detectability threshold for $\ms$ FRBs. The symbol colour indicates the intrinsic scattering time expected passing through a single CWO at $\nuo=1\ghz$. The assumed $\fa$ used in \eq{nuot} is based on \se{fa_fv} (\fig{fa_fv}), where $\fa=\DL\fv/\lc$, and we assume a fixed $\fv$ of $10^{-3}$ for sheets and filaments, and $\fv=10^{-4}$ for haloes. CWOs with $\fa<1$ are set to one, and are marked with open symbols.
    We show CW sheets (left), filaments (middle), and halos (right). Triangles, circles, and reverse triangles represent sheets and filaments with $\Tv=10^5$, $10^{5.5}$, and $10^6\Kel$, and halos with $\Mv=10^{12}$, $10^{13}$, and $10^{14}\msun$, respectively. Dashed, dash-dotted, and solid black lines show the number of CWOs per unit redshift interval, $\dNdz$ (right y-axis), where we include only shattered CWOs. For filaments and sheets, $\dNdz$ is shown for the temperature ranges of $10^{5-5.5}, 10^{5.5-6}$ and $10^{6-6.5}\Kel$; and mass ranges of $10^{12-13}, 10^{13-14}$ and $10^{14-15}\msun$ for haloes.
    }
    \label{fig:nuot}
\end{figure*}
}% fig:nuot
% _________________________________________

We have seen that the temporal broadening caused by scattering, $\taus$, decreases with increasing frequency (\eqnp{tau_num}). 
We define the transition frequency ($\nuot$) between detectable ($\taus[det]$) and non-detectable temporal broadening, where $\taus/\taus[det]=1$. Using \eq{tau_num}, $\nuot$ is given by
\be
    \nuot 
    &= \nuo \pfrac{\taus(\nuo)}{\taus[det]}^{\frac{5}{22}}
    \\
    &\sim 0.2 \ghz\ \taus[det,-1]^{-\frac{5}{22}}
        \Deff[,9]^{\frac{5}{22}}
        \opzo^{-\frac{17}{22}} 
        \fa^{\frac{3}{11}}
        \lc[,10]^{\frac{1}{11}} 
        \nf^{\frac{6}{11}}
    \label{eq:nuot}
\ee
where $\taus[det,-1]=\taus[det]/(0.1\ms)$.
In \fig{nuot}, we show using coloured symbols the frequency where $\taus/(0.1\ms)=1$, $\nuot(\taus=0.1\ms)$ (left y-axis), as a function of redshift for CW sheets (left), filaments (middle), and halos (right). Different symbols (triangles, circles, and reverse triangles) represent different values of $\Tv$ for sheets and filaments and different $\Mh$ for halos, as indicated in the legend. The symbol colour indicates the expected scattering time at $1\ghz$. We show the results for the fiducial \fv of cold cloudlets, $\fv=10^{-3}$ for sheets and filaments and $10^{-4}$ for haloes (\se{fa_fv}). The source is assumed to be at $\zs\sim 6$. 
We take $\taus[det]=0.1\ms$ as a rough detectability threshold for temporal broadening, meant to serve only as a rough guideline for a typical FRB with an intrinsic burst duration of a few $\!\ms$. Recently, however, a growing number of FRBs with durations of a few $\!\mus$ have been observed \citep{hewitt23, snelders23}. For such FRBs, a significantly lower scattering time may be detectable.
Note that for each CWO, we show $\nuot(\taus=0.1\ms)$ only above a minimal redshift, determined by the criteria for fragmentation, according to \eq{frag_cond}, assuming values for $\Nt$ and $\DL$ corresponding to an impact parameter of $b=0.7$ for haloes, and $b=0.5$ for filaments.

\smallskip
The three black lines in each panel show the number of CWOs per redshift interval along a los, $\dNdz$ (right y-axis). Unlike \eq{dNdzdM}, which expresses $d\num/dz$ above a given virial mass/temperature, we here impose upper limits of $10^{0.5}\Tv$ for sheets and filaments, and $10\Mh$ for haloes, namely the range corresponds to the separation between two adjacent $\Tv$ or $\Mh$ values shown in the figure. 
$\dNdz$ here differs from the quantity shown in \fig{Nobj} in that we show only the medians ($\kappa=0$), include only shattered CWOs (\eqnp{frag_cond}), and impose upper limits on the temperatures and masses in each bin.

% frb rate
\smallskip
$\dNdz$ can also be thought of as the covering fraction of the sky per unit redshift for a given type of CWO, or alternatively, for $\dNdz<1$, the fraction of FRBs which will intercept a given type of CWO per unit redshift.%
\footnote{We often refer to $\dNdz$ as the fraction of the sky covered by a CWO; however, it is actually the covering fraction of the sky per unit redshift.
}
In order to estimate the rate at which a los to an FRB intersects a given CWO, we first need to estimate the rate of FRBs with $\zs>z_\mr{CWO}$. \citet[][hereafter \citeta{beniamini21}]{beniamini21} used results from \citet{lu-piro19} to evaluate the rate of FRBs during the reionisation epoch, $z>6$, with an observed specific fluence ($e_{\nuo}$) above a given threshold. 
They conservatively estimated a rate of $\dotNum(\zs>6)\sim 10^4\yr^{-1}$ for FRBs during reionisation with an observed fluence of $e_{\nuo}\geq 1 \jy\ms$ at $\nuo=1\ghz$. This rate is over the entire sky, so if a given survey has a sky covering fraction of $f_\mr{\rm sky}$, the corresponding rate becomes $\dotNum(\zs>6)\sim f_\mr{sky} 10^4\yr^{-1}$. The rate increases substantially for FRBs at lower redshifts \citepa[see their fig.~3]{beniamini21}.

\smallskip
The high rate of FRBs across the sky may allow the detection of what would otherwise seem like a rather rare occurrence. 
For example, filaments with $\Tvf=10^{6-6.5}\Kel$ at $z\sim2.5\pm0.5$ are expected by our model to cover $\sim 10^{-2}$ of the sky. Considering only sources at $\zs>6$, with an FRB rate of $\sim 10^4\yr^{-1}$, this results in $\sim 100$ FRB sight lines passing through such filaments per year over the entire sky. 
From \fig{nuot}, turbulent cloudlets in such filaments could cause significant scattering with $\taus\gsim0.1\ms$ at $\nuo\leq\nuot\sim1\ghz$.%
\footnote{
The detectability rate of FRBs scales roughly linearly with frequency. 
}
We note that although an encounter with such a filament is expected to cause substantial temporal broadening, it will likely be a challenge to distinguish its contribution from scattering in the host. As we show in \se{mscatt_multi} below, the signature of CWOs on observed temporal broadening is, in general, better suited for stronger and more abundant screens, such as massive haloes at $z\lsim1$, using their cumulative signature. 
For an encounter with a single CWO, scintillation may be a better method of detection, as we discuss in our companion paper (\scintpaper).
Similarly, the expected number of $10^6\Kel$ sheets at $z\sim 4\pm0.5$ is similar, leading to $\sim 100$ sight lines to FRBs at $\zs>6$ passing through such sheets per year over the entire sky. However, the maximal observed frequency below which cloudlets in such sheets can cause temporal broadening of $\taus\gsim0.1\ms$, is rather low, $\nuo\leq\nuot\sim0.4\ghz$. 
While scattering by the CGM of intervening halos seems very promising \citepa[see also][]{vedantham19}, in \se{mscatt_multi}, we explore the effect of multiple scattering screens, and show that observational constraints introduce significant limitations on such scattering in the redshift range $z\sim 0-1$.

% @@@@@@@@@@@@@@@@@@@@@@@@@@@@@@@@@@@@@@@@@
\subsection{Scattering by multiple CW screens}
\label{se:mscatt_multi}
% @@@@@@@@@@@@@@@@@@@@@@@@@@@@@@@@@@@@@@@@@ 

% _________________________________________
{
\begin{figure}[ht]%
    \centering%
    \includegraphics[width=0.85\columnwidth]{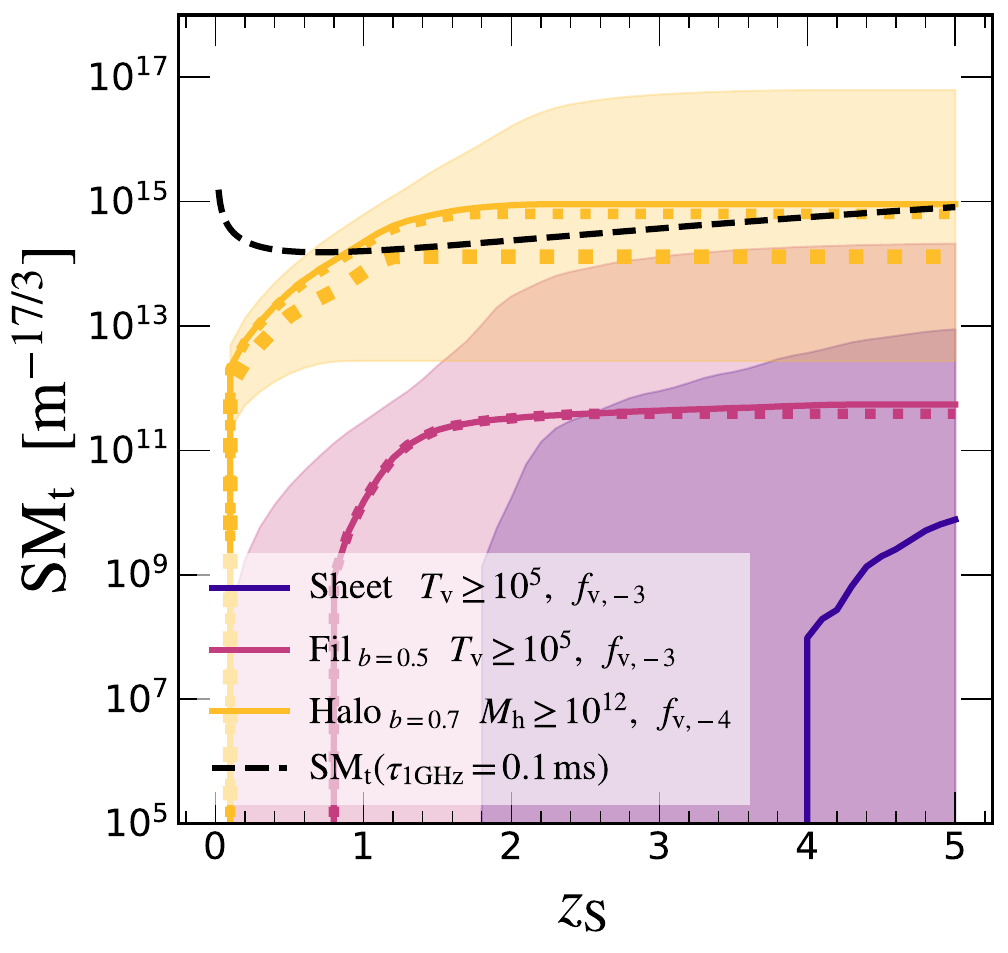}
    \caption{
    The cumulative $\smt$ of CWOs as a function of source redshift, $\zs$.
    The solid blue and purple lines indicate the cumulative median $\sm$ (y-axis), for sheets and filaments, respectively, with $\Tv[,sh|f]\geq10^{5}\Kel$ and $\fv=10^{-3}$. Solid orange lines represent haloes with masses $\Mh\geq10^{12}\msun$ and $\fv=10^{-4}$. 
    The dotted lines with corresponding colours indicate filaments with $\Tvf\geq10^{5.5}\Kel$, and haloes with $\Mh\geq10^{12.5}\msun$ and $\geq10^{13}\msun$ (thicker orange lines correspond to higher mass). 
    The shaded regions show the $(5,95)$ percentile on  $\dndz$ using the Poisson distribution. 
    Haloes and filaments are assumed to have an impact parameter of $b=0.7$ and $0.5$, respectively. The dashed black line is a rough estimate for the SM required to achieve a temporal broadening of $\taus(\nuo=1\ghz)=0.1\ms$, where we treat the total $\smt$ due to all CWOs along the los as one thin scattering screen, placed at the median redshift between the source and the observer.
    For an average los, the contribution of filaments and sheets to the cumulative scattering is expected to be negligible. For haloes, the scattering is dominated by haloes of masses $\Mh\geq10^{12.5}\msun$.
    }
    \label{fig:smt}
\end{figure}
}% fig:smt
% _________________________________________

To address scattering by multiple CW screens, we estimate the contribution of each screen to the total scattering measure ($\smt$). 
As described in \se{scatt}, the SM of a uniform thin screen is given by $\sm[\ii]\sim\dL[\ii]^{}\cn[,\ii] (1+z_i^{})^{-2}$, where $\cn$ is the amplitude of the density power spectrum, often called the spectral amplitude. The diffractive scale can be expressed as a function of the SM, $\lpi[,\ii]\propto [\lam^2 \sm[\ii](z)]^{-3/5}$, where the redshift dependence of $\lpi$, namely the factor of $\opzo[]^{-2}=(\lamz/\lam)^2$, has been absorbed into the definition of the SM.
For a Kolmogorov spectrum, $\cn=(54\pi)^{-1/3} \lo^{-2/3}\dn^2$. Using our notations from \se{scatt} (\eqnp{sm_def}), the effective SM along a sightline to a source at $\zs$, with a total of $\num[obj]$ intervening scattering screens at redshifts $\{z_i\}$, is 
\be
    \smt(\zs)
    = \sum_i^{\num[obj]} \sm[\ii] 
    \sim \frac{1}{(54\pi)^{1/3}}  \sum_i^{\num[obj]} \fv\DL[\ii]\lo[,\ii]^{-\frac{2}{3}}\n[,\ii]^2 (1\!+\!z_i^{})^{-2}
\label{eq:SM}
\ee% eq:SM
where we have used $\fa[,\ii]=\fv\DL[\ii]/\lo[,\ii]$. For simplicity, for a given type of CWO, we assume a fixed $\fv$.
% 

% fig:SM
\smallskip
In \fig{smt}, we show the cumulative SM of each CWO along a typical los, as a function of the source redshift. Coloured solid lines and shaded regions show the median and $(5,95)$ percentile ranges for sheets/filaments with $\Tv[,sh|f]\geq 10^5\Kel$ and for haloes of $\Mh\geq10^{12}\msun$. The coloured dotted lines indicate higher thresholds, $10^{5.5}\Kel$ for filaments, and $\Mh\geq10^{12.5}$ and $\geq10^{13}\msun$ for haloes, where the thicker dotted lines correspond to a higher mass. In general, we find that sheets have the smallest contribution to the $\smt$, followed by filaments, with the CGM of halos contributing the most.

\smallskip
In order to assess the SM required for detectable scattering ($\taut\gsim 0.1\ms$), we treat the total $\smt$ as one thin scattering screen, placed at the median redshift ($z_\mr{id}$) between the source and the observer, with the corresponding distance, $\Deff[,id]$. Using \eqs{sm_def} and \eqm{taus}, we then write the SM as a function of the scattering time, $\smt(\taut) \propto [\taut (1+z_\mr{id}) /(\Deff[,id] \lambda^{22/5})]^{5/6}$. 
The dashed black line in \fig{smt} represents $\smt(\taut=0.1\ms)$, the SM corresponding to our fiducial value for a detectable scattering time at $1\ghz$. 
As before, we constrain the redshifts where CWOs contribute to scattering by the conditions necessary for fragmentation (\eqnp{frag_cond}), adopting values of $\Nt$ and $\DL$ corresponding to impact parameters of $b=0.5$ and $0.7$ for filaments and haloes, respectively.

\smallskip
We find that the cumulative SM for filaments approaches $\smt(\taut=0.1\ms)$ only for a source at $\zs\gsim 2$ and only considering the 95th percentile of $\dNdz$. 
Comparing the purple solid and dotted lines, we see that the cumulative SM due to filaments with $\Tv\ge 10^5\Kel$ is nearly identical to that of filaments with $\Tv\ge 10^{5.5}\Kel$, suggesting that the latter dominate the scattering. 
The same is true for haloes with $\Mh\ge10^{12}\msun$ (solid orange line) and $\Mh\ge10^{12.5}\msun$ (thinner dotted orange line), suggesting that the latter dominate halo-scattering. 
At $\zs\lsim1$, halos of mass $\Mh\gsim10^{13}\msun$ (thicker orange-dotted) significantly contribute to the total SM from haloes, though their contribution diminishes at higher redshifts. 
At $\zs>1.5$, an average los is not expected to encounter even haloes with $\Mv>10^{12}\msun$, causing the median cumulative SM to saturate, with scattering limited to less common sightlines. 
Overall, under the assumptions of our fiducial model, the contribution of haloes to $\smt$ far outweighs that of other CWOs.

% _________________________________________
{
\begin{figure*}[ht]%
    \centering%
    \includegraphics[width=0.43\textwidth]{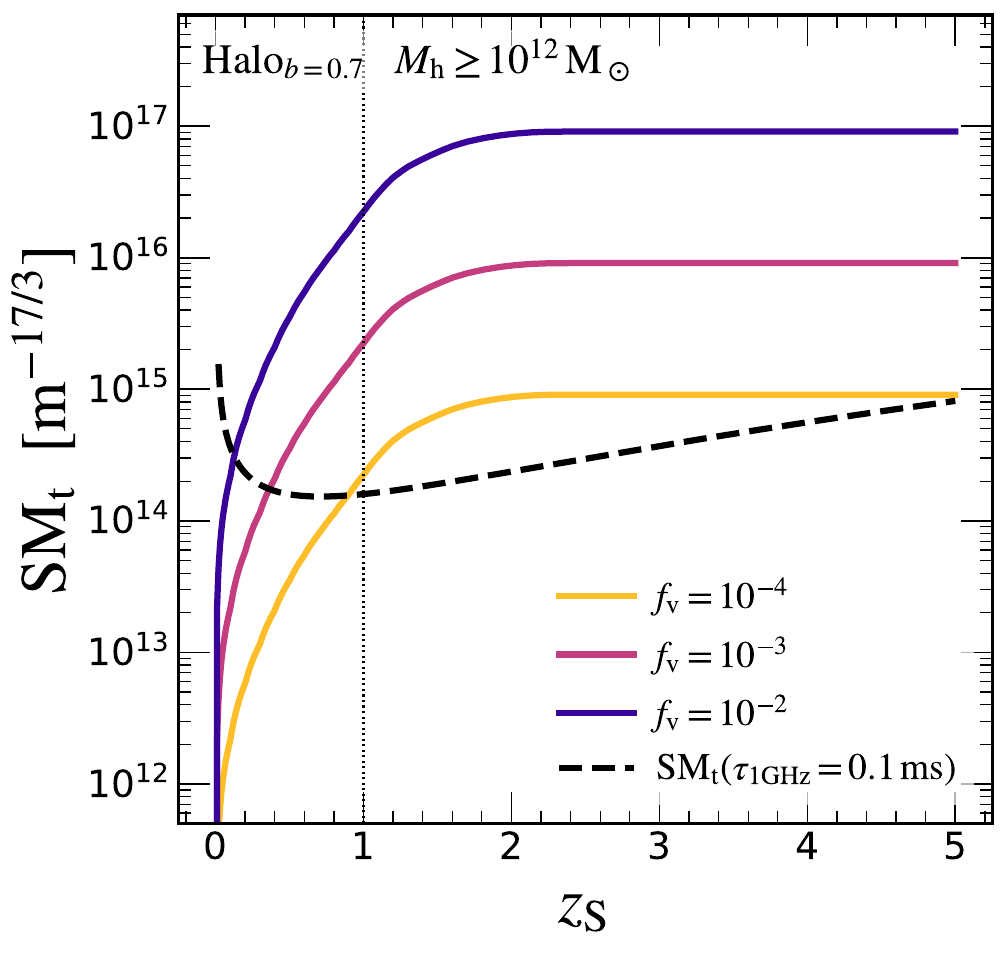}
    \includegraphics[width=0.43\textwidth]{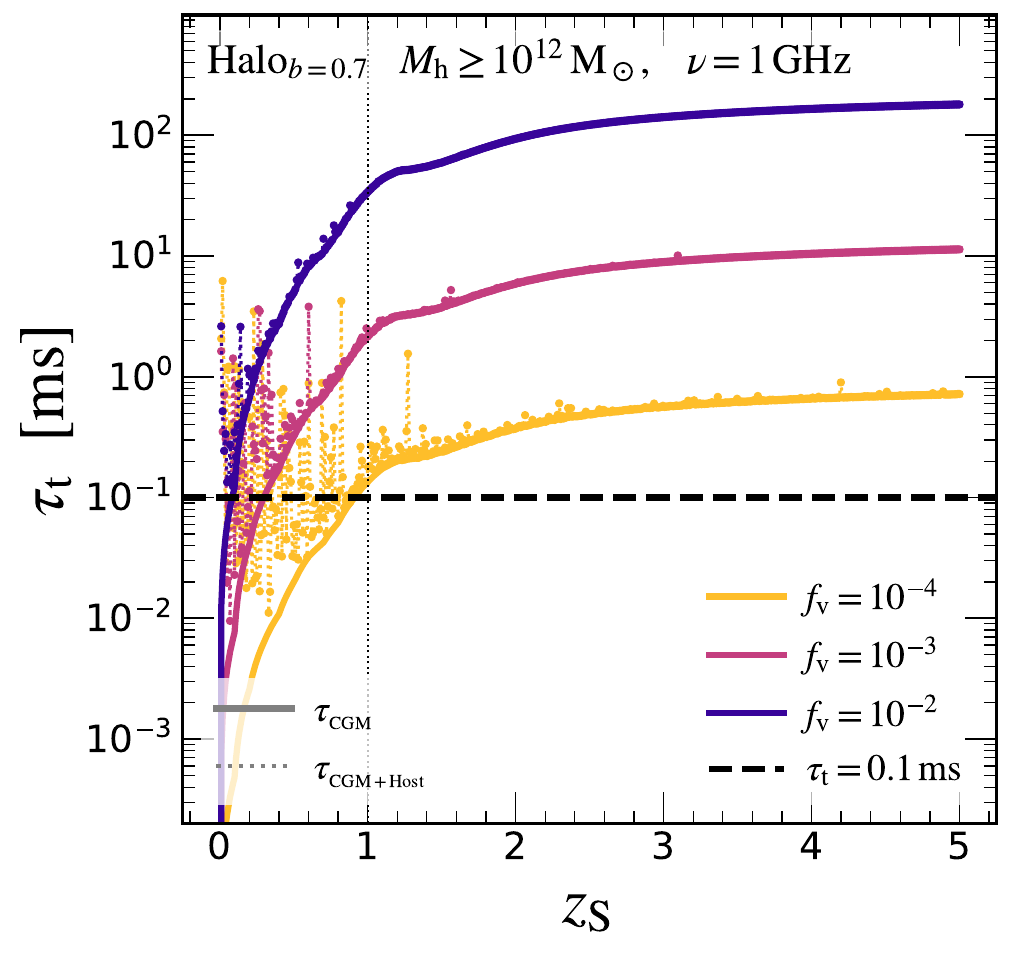}
    \caption{
    Constraints on the volume filling fraction of turbulent, cold cloudlets in the CGM of massive haloes, resulting from scattering observations of FRBs. 
    {\bf Left:} We show the median values of the cumulative $\smt$ caused by the CGM in haloes with $\Mh\geq 10^{12}\msun$ along the los to an FRB at redshift $\zs$, shown on the x-axis. We assume an impact parameter of $b=0.7$ for each halo. Different coloured lines show results for different $\fv$ as indicated in the legend. The dashed black line represents an estimate for the minimal $\smt$ which allows detection (see \fig{smt}). 
    If cool gas in the CGM is in the form of small-scale turbulent cloudlets, in pressure equilibrium with the hot phase and with sizes of order the minimal cooling length, then our model predicts a correlation of the $\smt$ with source redshift for $\zs\lsim1$, which should be observable if $\fv\gsim10^{-3}$. 
    {\bf Right:} Here the $y$-axis shows the median values (using Poisson distribution) of the cumulative scattering time, $\taut$, of the same halos, at the same impact parameter and with the same \fv as shown in the left panel. The dashed black line indicated a crude threshold for detection of $\taut=0.1\ms$. 
    In thin dotted lines with symbols, we introduce at each $\zs$ a random scattering time for a screen in the FRB host, drawn from a log-normal distribution \citep{chime21}, and add it to the cumulative scattering time from intervening CGM (described in the text). 
    If cool gas in the CGM is in the form of small-scale turbulent cloudlets, in pressure equilibrium with the hot phase and with sizes of order the minimal cooling length, then our fiducial model predicts a correlation of $\taus[CGM]$ with source redshift at $\zs\lsim1.2$, which is likely to be observable if $\fv>10^{-3}$. 
    Including a random $\taus[Host]$ introduces scatter, but this does not significantly mask the correlation for the higher $\fv$. 
    }
    \label{fig:taut}
\end{figure*}
}% fig:taut
% _________________________________________

\smallskip
% fig - SM_halo_fv
The strong scattering predicted by our model due to small-scale turbulent cloudlets in the CGM of haloes with $\Mh\gsim 10^{12}\msun$, combined with the large number of such haloes intercepted by a typical los, allows us to place constraints on the prevalence of turbulent cold cloudlets in such haloes and their corresponding \fv.  
In the left panel of \fig{taut}, we show the median SM contributed by intervening haloes with masses $\Mh\geq10^{12}\msun$ along a los from a source at redshift $\zs$ assuming our fiducial impact parameter for halos of $b=0.7$. Different coloured lines represent different volume filling fractions $\fv=(10^{-4}-10^{-2})$. The black dashed line indicates $\smt(\taut\sim0.1\ms)$ (see \fig{smt}), namely an estimate for the minimal SM required for detectable temporal broadening of $\taut\geq0.1\ms$.

\smallskip
In the right panel of \fig{taut}, we estimate the observed scattering time at $\nuo=1\ghz$ from the CGM for a los passing through multiple halos. 
To estimate the scattering time for a collection of screens between the source and the observer, we take into account that each screen perceives the previous screen as its source. As a result, the Fresnel scale of each screen is diminished through the effective angular diameter distance, $\Deff[,\ii]=d_\mr{\ii,o} d_{i-1,i}/d_\mr{\ii-1,o}$.%
\footnote{
$d_{i,j}$ is the angular diameter distance between $z_{i}$ and $z_{j}$, with $z_{i}>z_{j}$.}
Namely, the $i$-th screen receives rays from the previous screen ($i-1$), if such exists, or from the source (see \fig{time_delay}). 
To find $\Deff[,\ii]$ for the $i$-th screen, we roughly estimate the location of the previous screen ($i-1$) by searching for the previous integer increment in the cumulative number of haloes above $\geq10^{12}\msun$ (see \eq{Nobj}). 
This allows us to crudely estimate, for a given $\zs$, the scattering time contribution of every screen with $z_{i}<\zs$. Namely, we obtain the scattering time of each CWO as a function of its redshift and $\Mh$ (haloes) or $\Tv$ (filaments \& sheets). 
Finally, we integrate to find the cumulative $\taut$ as a function of $\zs$,
\be
    \taut(\zs, >\mathcal{Q}_\mr{min}) = \int_0^{\zs} \int_{\mathcal{Q}_\mr{min}}^{\infty} \taus(z,\mathcal{Q}) \frac{d^2\num}{dzd\mathcal{Q}} dz d\mathcal{Q}
\ee
where $\mathcal{Q}=\Mh$ for haloes, and $\mathcal{Q}=\Tv$ for filaments \& sheets.
The result is shown in the right panel of \fig{taut}, where the thick solid coloured lines indicate the medians (using Poisson statistics) of the cumulative $\taut$ for haloes, assuming an impact parameter factor of $b=0.7$ for each halo. 
The different colours represent different $\fv=(10^{-4}-10^{-2})$, and we exclude any haloes where the conditions for fragmentation (\eqnp{frag_cond}) are not met. 
The thin dotted lines indicate the cumulative scattering time of intervening haloes with the addition of a randomly drawn \citep{chime21} $\taus[Host]$ at each $\zs$ bin (indicated by symbols), as described in \se{disc} below. 
The black dashed line indicates $\taut\sim0.1\ms$.

\smallskip
Focusing on the redshift range of $z\sim0-1$, our results in \fig{taut} suggest that for $\fv>10^{-4}$ an average los is expected to undergo significant scattering due to the CGM of haloes with $\Mh\geq10^{12}\msun$. Moreover, the total $\smt$ and $\taut$ increase with redshift, which is detectable if the volume filling fraction of shattered, turbulent cloudlets in the CGM of such haloes exceeds $10^{-3}$. For $\fv\sim 10^{-3}$, this correlation may be partially masked by noise; however, the baseline normalisation of the cumulative scattering may help indicate the typical values of $\fv$ in haloes (although, unlike the $\taut-\zs$ correlation, an increased baseline will be harder to distinguish from scattering by the host). 
Given the dominance of $\sim10^{12.5-13}\msun$ haloes on the scattering in \fig{taut}, and the $\fv$ limitation on such haloes (see \fig{dm_cl_frac}), our model predicts that the strongest $\taut-\zs$ correlation is roughly limited to $\fv\sim10^{-2}$ (blue line).

\smallskip
Unfortunately, since our model predicts such small contributions to the total scattering at $z\lsim1$ from cosmic sheets and filaments, we cannot place similar constraints on $\fv$ in these CWOs using this method. However, we refer the reader to our companion paper, \scintpaper, where we find constraints using scintillation.

% =========================================
\subsection{Discussion and Comparison to Observations}
\label{se:disc}
% =========================================

\Fig{taut} shows that our model predicts an observed correlation between $\taut$ and $\zs$, resulting in a correlation between $\taut$ and the total $\dmt$, including the contribution of $\dmigm$. However, to date, most observations do not find significant correlations in either relation. 
\citet{sand25} used a sample of 137 FRBs and found no significant correlation between the scattering time and the extragalactic DM. 
While some studies previously found hints for a possible correlation between $\taut$ and the extragalactic DM \citep[e.g.,][]{ravi19, gupta22}, these included considerably smaller samples, and given the uncertainties, stated moderate or no evidence for correlation.
About a hundred observed FRBs are localised with known redshifts, and the highest redshift detection is at $\zs\sim 2$ \citep{caleb25}. Of these, a few tens have scattering-time measurements.
\citet{acharya25} used a sample of 24 localised FRBs with scattering time measurements, and initially found a hint of correlation between the scattering time and the observed DM, but found no significant correlation when they considered a much larger sample from \citet{chime21}. Using the sample of 24 localised FRBs, they found no significant correlation between the source rest frame scattering time, $\taut(1+\zs)^{17/5}$, and the redshift of the source.
\citet{glowacki25} showed consistent results with no significant correlation between the source rest frame scattering time and $\zs$, with a sample of 28 localised FRBs.

\smallskip
The lack of an observed correlation between $\taut$ and $\zs$ is still uncertain, due to the relatively small number of localised FRBs, particularly at $\zs\gsim0.6$. While the lack of correlation between $\taut$ and $\dmt$ in larger samples of unlocalised FRBs provides support for this claim, inferring the associated source redshifts from the DM introduces uncertainties of a different nature. The rapidly increasing number of localised observations may soon help reduce these uncertainties, allowing us to ascertain whether such a correlation exists and place constraints on our model. 

\smallskip 
Additionally, there are observations of specific FRB sightlines which are in tension with our model of scattering in the CGM. A compelling example is FRB~181112, first studied by \citet{prochaska19-sci}. The source was localised to $\zs\sim0.5$, measured to have a very low $\taus\lsim 40\mus$ (at $\nuo\sim1.3\ghz$), and confirmed to pass through a massive halo at $z\sim0.4$. \citet{cho20} later showed that $\taus$ is even lower - $\sim20\mus$, from the scintillation bandwidth. Moreover, they found no evidence for strong frequency dependence, implying that this broadening is not necessarily due to propagation through a plasma screen. 
FRB~181112 is a rather unique sightline with a small impact parameter ($\sim30\kpc$) through the halo. Assuming a hydrostatic CGM, the pressure is higher for small $b$, implying higher $\n$ and smaller $\lc$ (\fig{clump_prop}), and thus higher $\taus$ (\fig{tau}, bottom right). \citeta{vedantham19} argued that self-shielding by a fog of cloudlets is expected to reduce $\taus$ for small $b$. Nevertheless, their model still predicts significant scattering for an encounter with such a halo. 

\smallskip
Below, we discuss several possible effects and modifications to our fiducial model that may reduce scattering in the CGM, and/or hide an underlying correlation between $\taut$ and $\zs$. We begin by noting how $\taus$ scales with the properties of cloudlets in CWOs. Using \eq{taus} with $\fa=\fv\DL/\lo$, we obtain $\taus\propto\lc^{-4/5}\n^{12/5}\fv^{6/5}$, for $\lpi$ within the inertial subrange. We begin by discussing the effect of varying these parameters.

% -----------------------------------------
\par\smallskip\emph{\bud Larger cloud size, $\lc>\lcmin$:}
{
Our fiducial model assumes that all cloudlets have the same size of $\lc=\lcmin$. One may consider the effect of a spectrum of cloud sizes; however, since the smallest cloud size is expected to dominate the scattering, we may simplify the discussion by focusing on the effect of a larger threshold for the minimum cloud size (e.g., due to coagulation, incomplete fragmentation, or turbulent disruption of the smallest clouds \citealt{Gronke22}).
For a fixed $\fv$ and $\n$, the scattering dependence on the cloud size is $\taus\propto\lc^{-4/5}$ (\eqnp{taus}), implying that a threshold of $\lc=10\lcmin$ will reduce the scattering by a factor of $\sim6$. This may hide the correlation shown in \fig{taut} for $\fv\sim10^{-3}$, but not for higher $\fv$. 
Assuming a fixed $\fv=10^{-2}$, and all other model parameters at their fiducial values, the three lines in \fig{taut} correspond to $\lc/\lcmin=1$, $10^{1.5}$, and $10^3$ from top to bottom. We conclude that for $\lc/\lcmin \sim 10^{3}$ scattering in the CGM will be undetectable, while for $\lc/\lcmin \lsim 10^{1.5}$, $\taut$ is expected to show a correlation with $\zs$. 

\smallskip
A related possibility is anisotropically shaped cloudlets. 
Scattering depends on the \emph{transverse} length scale of the density fluctuations, rather than the global cloudlet geometry. 
Several studies of scattering by both pulsars \citep{higdon84, romani87, rickett90} and FRBs
\citep{xu16, jow24} suggest that scattering could be caused by filamentary or sheet-like density fluctuations. Such geometries would likely involve additional processes such as strong magnetic fields \citep{wang26} or strong compression (e.g., shocks). 
We have assumed a Kolmogorov spectrum of turbulence, motivated by `the big power-law in the sky' \citep{Armstrong95}. However, if shocks are involved, supersonic conditions may yield a different slope \citep[see e.g.,][and references therein]{xu16}, which will change the strength of scattering. If magnetic fields are sufficient to induce filamentary or sheet-like structures, then in addition to cloud geometry, one must also consider the magnetic field contribution to the overall pressure in the cloud, and the impact on $\n$ (as discussed below). 
}

% -----------------------------------------
\par\smallskip\emph{\bud Substantial non-thermal pressure, lower $\n$:}
{ 
$\taus$ is very sensitive to density fluctuations, with a dependence of $\taus\propto\n^{12/5}$. Our fiducial model assumes negligible non-thermal pressure, so the density ratio between cold and hot gas is roughly the inverse temperature ratio. However, if there is a significant non-thermal contribution to the pressure, such as magnetic fields or cosmic rays, this could result in lower-density cloudlets, which may reduce the scattering or erase the $\taus-\zs$ correlation shown in \fig{taut}. The strong dependence on $\n$ implies that a relatively small variation can significantly alter $\taus$.
Assuming a fixed $\fv=10^{-2}$, and all other model parameters at their fiducial values, the three lines in \fig{taut} correspond to $\n, 10^{-0.5}\n$, and $10^{-1}\n$, from top to bottom, where $\n$ is the cloudlets density in our fiducial model (\fig{clump_prop}).
Here, we isolated the effect of $\n$ by keeping $\lc=\lcmin$ fixed for clarity. However, if non-thermal pressure is important, both $\n$ and the cooling length (\eqnp{lcool}) will be affected, with $\n\propto P_\mr{th}$ and $\lcool\propto P_\mr{th}^{-1}$.%
\footnote{
$\n$ and $\lcool$ also depend on $\mu$, which at lower pressures could decrease, and slightly influence them beyond the effect of $P_\mr{th}$.}
For instance, if the thermal pressure fraction is $f_{P_\mr{th}}\sim 0.5$, the effect on $\taus$ is roughly a factor of $f_{P_\mr{th}}^{16/5}\sim0.1$, rather than a factor of $\sim0.2$ considering only the effect on $\n$.

\smallskip
Demonstrating the above point, \citet{werk14} inferred remarkably low cold-phase densities in haloes of $\sim10^{12}\msun$ at $z_\mr{med}\sim0.2$, comparable to the expected hot phase density, possibly indicating significant non-thermal pressure support \citep{butsky_quinn18, ji20}. These results were obtained using quasar absorption-line spectra to study cold gas in the CGM of $z\lsim 1$ galaxies as part of the COS-Halos survey \citep{werk14, prochaska17, faerman23}. 
}

% -----------------------------------------
\par\smallskip\emph{\bud Small $\fv\lsim10^{-4}$:} 
{
As discussed in \se{mscatt_multi}, given the dependence of $\taus\propto\fv^{6/5}$, the correlation of $\taut-\zs$ in \fig{taut} can be hidden if $\fv\lsim10^{-4}$. However, such a small $\fv$ may present tension with low-$z$ observations of $\sim 10^{12}\msun$ haloes in the COS-Halos survey \citep{werk14, prochaska17, faerman23}, finding a median $\fv\sim10^{-1}$ \citep{werk14}. 
While our results in \fig{taut} are shown for haloes of $\geq10^{12}\msun$, higher mass haloes, $\Mh\gsim10^{12.5-13}\msun$ (\fig{smt}), dominate this correlation. Observations of higher mass haloes, $\Mh\sim10^{13}\msun$ at a median redshift of $z\sim0.4$ \citep[e.g. COS-LRG][]{zahedy19} indicate substantial density ratios between cold and hot gas, and infer a volume filling fraction in the outer halo of $\fv\sim{\rm few}\times10^{-4}$. 

}

% -----------------------------------------
\par\smallskip\emph{\bud Noise due to scattering by the host FRB:}
{
Scattering from the host screen may introduce variations which could hide the $\taut-\zs$ (or $\taut-\dmt$) relation.
\citet{chime21} used simulated-event injections to correct for selection effects in the data. They modelled the intrinsic broadening time, finding that this can (marginally) be modelled as a log-normal distribution,\footnote{
\citet{scott25} argued that a log-uniform distribution is at least as consistent with the histogram shown in \citet[][fig.~17]{chime21}. We choose a log-normal distribution, which has a greater potential to smear the $\taut-\zs$ relation if such exists.
}
$\log[\taus(\nuo=0.6\ghz)/\ms]=\mathit{N}(\mu_{\taus}, \sigma_{\taus})$,  with $\mu_{\taus}=\ln(2.02)$ and $\sigma_{\taus}=1.72$ (see their Table~4). 
We randomly draw an intrinsic host scattering time, $\tausin[Host]$, from this distribution, and scale it to find the observed scattering due to the host, $\taus[Host]=\tausin[Host](1+\zs)^{-17/5}$ at $\nuo=1\ghz$.
In the right panel of \fig{taut}, the thin coloured dotted lines indicate the cumulative scattering time of intervening haloes with the addition of a randomly drawn $\taus[Host]$ at each $\zs$ bin (indicated by symbols).%
\footnote{
Our main focus in \fig{taut} is at relatively low-$z$, $z\lsim1$, where to date, there is no observational evidence for a redshift evolution in the distribution of $\tausin[Host]$. Future observations at higher redshifts may prove otherwise.}
For $\fv\sim10^{-2}$, the median scattering time by haloes shows correlation with $\zs$ even when we include a random $\taus[Host]$. For $\fv=10^{-3}$, the variations are somewhat stronger with respect to the normalisation, but overall $\taus[Host]$ does not appear to smear out the correlation.
}

% -----------------------------------------
\par\smallskip\emph{\bud Turbulence cascade in the clouds:} 
{
We have assumed throughout strong turbulence in cold cloudlets that cascades down to the coherence scale.
This implies that:
(i) cloudlets survive long enough to allow this cascade ($t_\mr{survival}\geq t_\mr{eddy-turnover}$),
(ii) our estimates of $\li$ do not severely underestimate damping mechanisms (e.g., see \citealt{Lithwick.Goldreich.01} and the appendix in \citealt{prochaska19-sci}). 
In practice, the former may also be treated as a cascade interrupted by a damping mechanism, at a scale length corresponding to $v_\mr{eddy}t_\mr{survival}$. 
Exploring specific damping mechanisms is beyond the scope of this work; however, we can assess the effect of damping at a larger inner scale, using \eq{taus} in the regime of $\lpi<\li$, where $\taus\propto\li^{-1/3}\lc^{-2/3}$. 
As before, assuming a fixed $\fv=10^{-2}$ with all other parameters at their fiducial values, the three coloured lines in \fig{taut} correspond to $\li/\lpi\sim 1$, $10^{3.6}$, and $10^{7.2}$, from top to bottom. For comparison, at $\zs\lsim1$ for $\Mh\sim10^{12.5-13}\msun$, $\lo/\lpi\sim10^9$. 

}

% ************************************************
% @@@@@@@@@@@@@@@@@@@@@@@@@@@@@@@@@@@@@@@@@
\section[Conclusions]{Conclusions}
\label{se:dc_Sc}
% @@@@@@@@@@@@@@@@@@@@@@@@@@@@@@@@@@@@@@@@@

% {
In this paper, we considered the formation of multiphase gas in the hot medium of cosmic web sheets and filaments and in the circumgalactic medium (CGM) of dark matter halos (collectively cosmic web objects, or CWOs), leading to the formation of small-scale cold ($T\gsim 10^4\Kel$) and dense cloudlets of size $\lc$. 
We then considered how this small-scale multiphase structure would impact scattering of FRBs at cosmological distances, and whether detections of such scattering could constrain the small-scale structure of CWOs. 
Our main results can be summarised as follows:

\begin{enumerate}

\item \underline{\emph{Fiducial Model for Small-Scale Cold Clouds in CWOs (\se{shatter}):}}

\smallskip
$\bullet$ 
    We use the shattering model \citep{mccourt18} as a fiducial model to translate ambient conditions into cloud properties. However, it is important to note that our results should be interpreted more generally as constraints on the presence of small-scale density fluctuations in CWOs, rather than as constraints on a specific cloud model. To evaluate the properties of cloudlets, we find the cooling length $\lcool$ (\eqnp{lcool}) as a function of $T$ for a given pressure, redshift and metallicity, and identify its minimum as the cloud scale $\lc=\lcmin$ (\figsiti{lcool}{clump_prop}). We then associate the corresponding temperature, $\Tmin$, and electron density, $\n$, with the cloudlet.  

\smallskip

\item \underline{\emph{Scattering -- single cloud (\se{scatt}):}}

\smallskip
$\bullet$ 
    When a radio wave passes through a small-scale turbulent electron density fluctuation in a plasma screen, it will suffer a phase shift and deflect the ray. 
    For a given turbulent power spectrum, one can find the scale that will govern the observed scattering, which is the \emph{diffractive scale} length ($\lpi$, \eqnp{lpi}, and top row in \fig{tau}).
    We assume throughout that cloudlets are turbulent with a Kolmogorov power spectrum, and an outer scale of turbulence $\lo=\lc$. 

\smallskip
$\bullet$ 
    A deflected ray travels a longer path length, leading to a delay in its time of arrival (\se{tau}, \se{time_delay}, \fig{time_delay}), manifested as \emph{temporal broadening} $\taus$ at a given central frequency $\nuo$; with a strong dependence on frequency ($\propto\nuo^{-22/5}$, \eqnp{tau_num}). We find that a ray emitted from an FRB source and intercepting with a single cloudlet ($\fa=1$) within a $10^{12}\msun$ halo or $10^{6}\Kel$ sheet|filament, results in $\taus$ well below our crude threshold for detection, $\taus[det]\gsim0.1\ms$ at $\nuo=1\ghz$ (\fig{tau}, bottom row).

\smallskip

\item \underline{\emph{Counting CWOs (\se{Nobj}):}}

\smallskip
$\bullet$ 
    To account for multiple screens along a los from an FRB, we evaluate in our companion paper (\dmpaper) the probability that an average los will intersect with a given CWO. 
    Here, we confine our attention to either massive haloes or to high-temperature filaments and sheets, as these systems are likely to form small-scale cool cloudlets in their hot medium.

\smallskip
$\bullet$ 
    One may naively presume that sheets and filaments can be more abundant than haloes and have a larger covering fraction over the sky. 
    However, we find that once they are limited to high temperatures, the number of systems and their sky coverage become comparable to those of the massive haloes (\fig{Nobj}).
    Therefore, compared to haloes, the abundance of sheets \& filaments cannot compensate for their relatively low $\n$ and large $\lc$.

\smallskip

\item \underline{\emph{Counting cloudlets (\se{Nt}-\se{dm_cl_frac}):}}

\smallskip
$\bullet$ 
    We use the extreme upper limit on $\fa$, where the entire column density is comprised of cold clouds, $\famax=\Nt/\N$ (and corresponding $\fvmax$), where $\Nt$ is the column density through the CWO, to find rough constraints on the redshift ranges where fragmentation is possible (\eqnp{frag_cond}).

\smallskip
$\bullet$ 
    We find that $\fvmax$ has only a small dependence on redshift for all CWOs, for filaments \& sheets $\fvmax$ is rather similar, with $\fvmax\gsim10^{-2}$.
    For $\gsim10^{13}\msun$ haloes, $\fvmax$ is lower than $10^{-2}$ at all redshifts (\fig{fa_fv})

\smallskip
$\bullet$
    We quantify the ratio between the DM due to all encountered cloudlets ($\dmsc$) and the overall $\dms$ through a CW screen (\se{dm_cl_frac}, \fig{dm_cl_frac}). We find a significant $\dmsc/\dms$ fraction only for CWOs with $\fv\gsim10^{-2}$. $\Mh\gsim10^{13}\msun$ haloes with such high $\fv$ are in the `forbidden zone' ($\dmsc>\dms$), indicating that $\fv$ must be lower than $10^{-2}$ in these haloes for $\lc\sim\lcmin$.

\smallskip
$\bullet$ 
    Most CWOs have $\dmsc/\dms<0.1$ for $\fv\lsim10^{-3}$; in such cases, regions which dominate scattering are not comparable to those dominating the $\dms$. We argue that the $\taus-\dms$ relation, often used for the MW-ISM or the FRB host screens, should be used with care when applied to CWOs; and emphasize the need to use different $\fv$ for the regions which dominate the scattering ($\fvsc$) and the $\dms$ ($\fvdm$) of CWOs (\eqnp{sm_dm_c91})

\smallskip

\item \underline{\emph{Scattering by a single CWO (\se{mscatt_1cwo}):}}

\smallskip
$\bullet$ 
    Using our estimates for the number of cloudlets in a CWO with a given $\fv$, we evaluate the expected scattering from multiple clouds in a single CWO (\se{mscatt_1cwo}), and find the transition frequency $\nuot$ which allows a detectable $\taus$, namely is the maximal frequency below which $\taus\geq\taus[det]$ (\fig{nuot}). 

\smallskip
$\bullet$ 
    A single filament at $z\sim 2.5-5$ with $\Tv\sim10^{6}\Kel$ and $\fv\sim10^{-3}$ is expected to cause a temporal broadening of $\taus\gsim0.1\ms$ at $\nuo\lsim1\ghz$ for a source at $\zs=6$ with an impact parameter of $b=0.5$ (\eqnp{nuot}, \fig{nuot}). 
    Our model predicts that such filaments cover a small $\dNdz\sim10^{-2}-10^{-4}$ of the sky (\fig{nuot}).
    However, the high rates of FRBs may still permit detections of relatively rare sightlines. 
    Using conservative estimates of FRB rates \citepa{beniamini21}, considering only sources at $\zs\gsim6$, and filaments at $z\sim2.5\pm0.5$, this results in $\sim 10^{2}\yr^{-1}\sky$ FRB sightlines passing through such systems (\se{mscatt_1cwo}). 

\smallskip
$\bullet$ 
    We note that the detection of small-scale cloudlets in a single CW screen via temporal broadening is challenging, and the distinction from the host screen is bound to be difficult. 
    For encounters with rare high-$z$ systems (or with low $\fv$ haloes), we suggest better-suited methods based on \emph{scintillation} measurements, which we explore in our companion paper, \scintpaper.

\smallskip

\item \underline{\emph{Scattering by multiple CWOs (\se{mscatt_multi})}}

\smallskip
$\bullet$ 
    Conversely, using temporal broadening for encounters with multiple CW screens is possible in some cases, as the cumulative effect may result in a correlation between the observed $\taut$, and the source redshift, $\zs$ (or the total observed $\dmt$).
    In such a case, distinguishing the host screen could be easier, as host scattering is not expected to cause such a correlation.

\smallskip
$\bullet$ 
    We find that for an average los, filaments and sheets are expected to have a negligible contribution to the cumulative scattering measure, $\smt$ (\eqnp{SM}, \fig{smt}). While the intrinsic scattering time of individual high-$z$ filaments could be substantial (\fig{nuot}), the strong redshift dependence of $\taus$ ($\propto\opzo[]^{-17/5}$), along with the relatively low number of expected encounters with high $\Tv$ objects at these epochs, results in a negligible contribution to the cumulative scattering (see more optimistic prospects in \scintpaper). 

\smallskip
$\bullet$ 
    On the other hand, our fiducial model predicts that if turbulent shattered cloudlets exist in the CGM with $\fv\gsim10^{-3}$, the observed $\taut$ due to all haloes above $10^{12}\msun$ along the los, is expected to be strongly correlated with the source redshift at $\zs\lsim1$ (\fig{taut}). 

\smallskip
$\bullet$ 
    At present, most observations find no significant correlation between the total observed scattering time ($\taut$) and either the redshift of the source, $\zs$, or the total observed $\dmt$. While more observations of localised FRBs are needed to reduce the uncertainties, the rapidly increasing number of such observations allows us to place constraints and predict the permitted parameter ranges and the validity of some of our model assumptions. 

\smallskip
$\bullet$ 
    The expected correlation in our model can become too subtle for detection if $\fv\lsim10^{-4}$, which drives the overall normalisation of the $\taut-\zs$ relation below our rough detectability threshold $\taus[det]=0.1\ms$. Furthermore, a lower $\taus[det]$ is unlikely to help in this context, as the scattering time of screens in the FRBs host galaxies ($\taus[Host]$) introduces variabilities which are likely to erase the expected correlation in our model for $\fv\lsim10^{-4}$ (\fig{taut}).

\smallskip
$\bullet$ 
    Although we use the shattering model to translate the conditions in CWOs to cloudlet properties, the results can be treated more generally by exploring the effects of variations from the fiducial model assumptions. We discuss the effects of varying $\lc$, $\n$, and the assumption of substantial turbulence in the clouds (\se{disc}). With large forthcoming FRB samples, we could perform hierarchical inference of the effective scattering amplitude and its redshift dependence, thereby constraining combinations of $\fv$, $\n$, $\lc$, and turbulence amplitude, with degeneracies that can be partially broken using frequency dependence, sightline-to-sightline variance, and external priors.
    
\end{enumerate}

% ************************************************
% @@@@@@@@@@@@@@@@@@@@@@@@@@@@@@@@@@@@@@@@@
\begin{acknowledgements}
  We thank Yakov Faerman, Wenbin Lu, Matt McQuinn, Stella Ocker, Xavier Prochaska, and Nadav Shoval for very helpful discussions. 
  SL and NM acknowledge support from BSF grant 2022281 and NSF-BSF grant 2022736. 
  PB's work was funded by grants (no. 2020747, 2024788) from the United States-Israel Binational Science Foundation (BSF), Jerusalem, Israel, by a grant (no. 1649/23) from the Israel Science Foundation and by a NASA grant (80NSSC24K0770). 
  SPO acknowledges NSF grant AST240752 for support.
\end{acknowledgements}
% @@@@@@@@@@@@@@@@@@@@@@@@@@@@@@@@@@@@@@@@@

%%%%%%%%%%%%%%%%%%%%%%%%%%%%%%%%%%%%%%%%%%%%%%%%%%%%%%%%%%%%%%
\smallskip\noindent
    {\tiny
    {\bf Abbreviations}\\
    \citeta{beniamini20}:\; \citealt{beniamini20}\\
    \citeta{beniamini21}:\; \citealt{beniamini21}\\
    \citeta{macquart13}:\; \citealt{macquart13}\\
    \citeta{m18}:\; \citealt{m18}\\
    \citeta{m19}:\; \citealt{m19}\\
    \citeta{m21}:\; \citealt{m21}\\
    \citeta{vedantham19}:\; \citealt{vedantham19}
    }
%%%%%%%%%%%%%%%%%%%%%%%%%%%%%%%%%%%%%%%%%%%%%%%%%%%%%%%%%%%%%%
\bibliographystyle{aa}
\bibliography{bib}
%%%%%%%%%%%%%%%%%%%%%%%%%%%%%%%%%%%%%%%%%%%%%%%%%%%%%%%%%%%%%%
%%%%%%%%%%%%%%%%%%%%%%%%%%%%%%%%%%%%%%%%%%%%%%%%%%%%%%%%%%%%%%%
\begin{appendix}
%%%%%%%%%%%%%%%%%%%%%%%%%%%%%%%%%%%%%%%%%%%%%%%%%%%%%%%%%%%%%%%

%%%%%%%%%%%%%%%%%%%%%%%%%%%%%%%%%%%%%%%%%%%%%%%%%%%%%%%%%%%%%%%

% @@@@@@@@@@@@@@@@@@@@@@@@@@@@@@@@@@@@@@@@@
\section{shattering}
\label{se:shatterApp}
% @@@@@@@@@@@@@@@@@@@@@@@@@@@@@@@@@@@@@@@@@

Our fiducial model for multiphase gas described in \se{shatter} relies on the original shattering picture of \citet{mccourt18}. This model naturally explained the small cloud sizes, large \fa, and small \fv observed in the CGM at both $z\sim 0$ and $z\gsim 2$, as well as many additional observations in different environments \citep{mccourt18, FG_Oh2023}. However, subsequent work has revealed that the picture is not so simple. 
There is still some debate over the conditions for a large cooling cloud to `shatter', or otherwise fragment, with some suggesting this depends on the final overdensity between the cold and hot gas \citep{Gronke_Oh20, Yao25} and others that it depends on the thermal stability conditions in the initial cloud (\citealp{Waters.Proga.19a, Das21}; \citeta{m21}). Furthermore, 3D simulations of cloud shattering reveal that this does not proceed hierarchically as proposed by \citet{mccourt18}. Rather, initially large ($r_{\rm cl}\gg \lcmin$) and non-linear ($\delta \rho/\rho \gg 1$) clouds begin cooling isochorically when they first lose sonic contact, becoming strongly compressed by their surroundings until their central pressure overshoots and they explode into many small fragments \citep{Gronke_Oh20, Yao25}. The fragmentation appears due to Richtmyer-Meshkov instabilities \citep[RMI;][]{Richtmyer1960TaylorII, Meshkov1969InstabilityOT, Zhou2017a, Zhou2017b}, and is thus not seen in 1D simulations \citep{Waters.Proga.19a,Das21}.

Once the initial fragmentation occurs, the resulting small cloudlets either disperse throughout the medium or coagulate to form larger clouds \citep{Waters.Proga.19b, Gronke_Oh20, Gronke_Oh23, Yao25}. 
While several coagulation mechanisms have been discussed in the literature \citep[see summary in][]{FG_Oh2023}, the most relevant for the turbulent environments prevalent in the CGM and the CW is coagulation due to the advective flow generated by hot gas condensing onto cold clouds through turbulent mixing layers at their interfaces. 
In this case, the coagulation can be modelled as an effective force between two clouds which behaves similarly to gravity, with the cloud area playing the role of mass and the force scaling as $r^{-n}$, with $r$ the distance between the clouds and $n=2$, $1$, or $0$ for systems with spherical, cylindrical, or planar symmetry \citep{Gronke_Oh23, Yao25}. Clouds tend to coagulate if the final density ratio of cold to hot gas after they regain pressure equilibrium, $\chi_{\rm f}\equiv \rho_{\rm cl,f}/\rho_{\rm h}$, is less than a critical value, $\chi_{\rm f}<\chi_{\rm f, crit}$, while larger density clouds remain shattered \citep{Gronke_Oh20, Gronke_Oh23, Yao25}. For typical clouds, the critical overdensity is of order $\sim (200-300)$, with a weak dependence on cloud size, scaling as $(r_{\rm cl}/\lcmin)^{1/4}$, as well as on the symmetry of the system, with the coagulation efficiency increasing from spheres to cylinders (filaments) to planar systems (sheets).

In practice, unless coagulation is very efficient (i.e. for low density and/or initially small clouds), this process results in a scale-free cloud mass function, $dN/dm\propto m^{-2}$ \citep{Gronke22, Yao25}. Contrary to earlier expectations \citep{mccourt18, Liang_Ian20}, this distribution continues until the resolution scale in numerical simulations, even if this is much smaller than $\lcmin$ \citep{Yao25}. Thus, the shattering process itself does not seem to impose a minimal cloud size. However, additional processes such as thermal conduction or hydrodynamical disruption in turbulent environments \citep{Gronke22} result in a minimal size for cold clouds. For instance, \citet{Gronke22} find that for a cloud to survive in a turbulent environment with turbulent Mach number $\Mach$, it must be larger than 
\be
    \label{eq:lturb}
    \ell_{\rm turb} \simeq 80\pc~\frac{\Tx[cl,4]^{5/2}\,\chi_{100}\, \Mach}{P_3\,\Lamtz[mix,-22.4]} \pfrac{f(\Mach)}{0.25}^{-1},
\ee
where $\Tx[cl,4]$ is the cloud temperature in units of $10^4\Kel$, $P_3=nT/1000\Kel\cmc$ represents the cloud pressure, $\Lamtz[mix,-22.4]=\Lamtz(\Tx[mix])/(10^{-22.4}{\rm erg\,s^{-1}\,cm^3)}$ is the net cooling function in the mixing layer at temperature $\Tx[mix]\sim (\Tx[cl]\Tx[h])^{1/2}$, $\chi_{100}=\chi/100$ represents the density ratio between the cloud and the background, and $f(\Mach)\sim 10^{-0.6\Mach}$ is an empirical Mach dependent fudge-factor from simulations. Comparing \eqs{lshatter} and \eqm{lturb}, we see that for typical conditions and mildly subsonic turbulence, $\Mach\lsim 1$, $\ell_{\rm turb}$ is larger than $\lcmin$. 
On the other hand, the minimal cloud size set by thermal conduction is the Field Length, which for typical conditions in the high-$z$ CGM and $\sim 10\%$ Spitzer conductivity, is \citep{Begelman_Mckee90, Armillotta16, m20} 
\be
    \label{eq:lfield}
\ell_\mr{cond} \simeq 24\pc~\frac{f_\mr{suppress,0.1}\Tx[cl,4]^{11/4}\,\chi_{100}^{7/4}}{P_3\,\Lamtz[min,-23]^{1/2}},
\ee
usually comparable to $\lcmin$.

To summarise, despite modifications to the original shattering picture of \citet{mccourt18}, the end result of a large and thermally unstable cloud is still expected to be a mist of small cold cloudlets with a minimal size of order $\lcmin=\min(\lcool)$, surrounding larger clouds with a scale-free mass spectrum of $dN/dm\propto m^{-2}$.

% =========================================
\subsection{Shattering in Different CW Environments}
\label{se:CW_shatter}
% =========================================

% _________________________________________
{
\begin{figure}[ht]%
    \centering%
    \includegraphics[width=0.85\columnwidth]{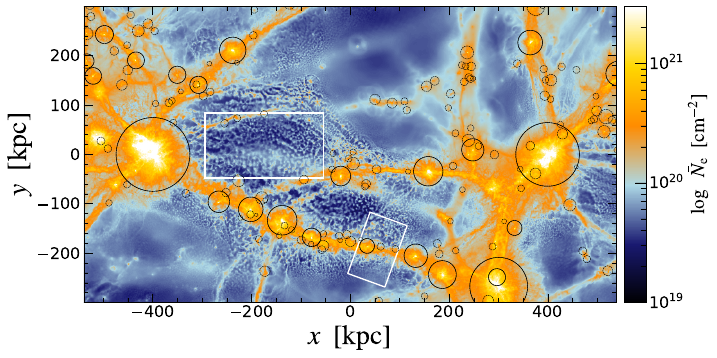}
    \includegraphics[width=0.85\columnwidth]{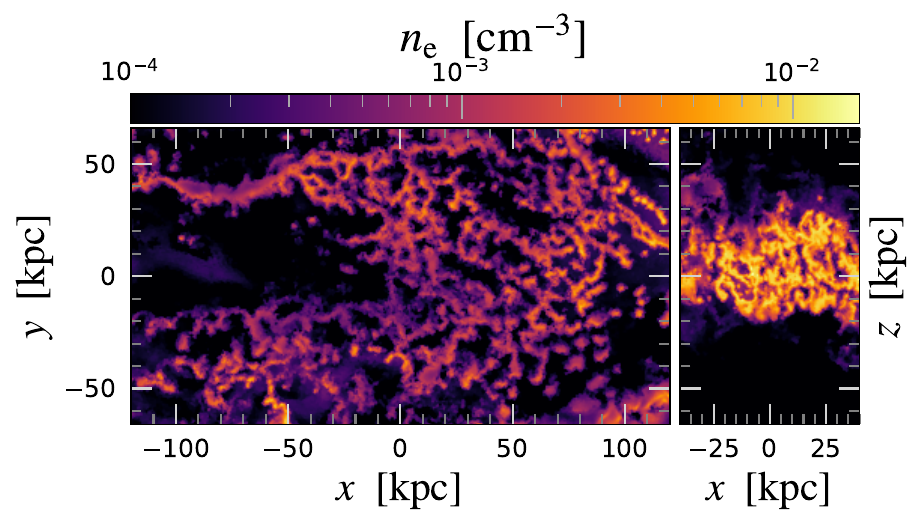}
    \caption{
    {\bf Top:} 
    Electron column density, $\Nt$, of the cosmic sheet from the highest resolution \texttt{IPM} simulation at $z\sim 4$, integrated over $\pm 200\kpc$. The virial radii of halos with masses $\Mh\geq 10^9\msun$ are marked with black circles. 
    {\bf Bottom:}
    Shattering in the cosmic web. We show the electron density in a thin slice through the sheet (left) and filament (right) regions, highlighted with white rectangles in the top panel. In the sheet region, the minimum cooling length is marginally resolved, and we find a shattered cold substructure with cloud sizes of $\lcmin\sim 700\pc$ \citepa{m19, m21}. For the filament, despite the fact that the minimum cooling length is not resolved in this region ($\sim20\pc$ at $z\sim4$), one can see indications for a shattered structure of dense clumps inside the shock radius of the filament.
    }
    \label{fig:N_e_map2d_sheetAll}
    \label{fig:ne_sheet_fil_map2d}
\end{figure}
}% fig:smt
% _________________________________________

The theoretical framework described above is most commonly considered in reference to the CGM around massive halos, where the volume is filled with hot gas at the virial temperature. However, it applies equally well to CW sheets and filaments, which are massive enough to sustain a strong gravitational accretion shock at their boundary. This was demonstrated in cosmological simulations by \citeta{m19,m21}. Using the moving-mesh code \texttt{AREPO} \citep{Springel.10}, these authors zoomed-in on a large patch of the IGM between two massive haloes, which at $z\sim 2$ are $\Mh\sim 5\tm 10^{12}\msun$ each and connected by a $\sim$ Mpc-scale cosmic filament. They simulated this system at five different resolutions, the highest having a gas-cell mass of $m_{\rm gas}\sim 1.5\tm 10^4 \msun$, roughly 10 times better than the Illustris-TNG50 simulation \citep{Nelson19}. This large filament forms at $z\sim 2.5$ from the collision of several co-planar filaments embedded in a large-scale CW sheet, which itself formed from the collision of two smaller sheets at $z\sim 5.5$. These simulations are referred to as the \texttt{IPM} simulations, which refers to the \emph{intra-pancake medium} of multiphase gas within cosmological sheets (or `Zel'dovich pancakes') found in these simulations \citep[see also][]{Pasha23}. In \fig{N_e_map2d_sheetAll} we show a face-on projection of the electron column density, $\Nt$, in the sheet at $z\sim 4$. We see multiple co-planar filaments connecting three massive, $\Mh\sim 10^{12}\msun$ halos, with multiple halos of $\Mh>10^{10}\msun$ aligned along the filaments.

\smallskip
These authors found that, following the collision of the two smaller sheets, the resulting post-merger sheet was surrounded by a stable accretion shock with hot gas at $T\sim 10^{5.5}\Kel$ surrounding a cold component near the sheet midplane. At high resolution, the cold gas morphology resembled the shattered structures predicted by \citet{mccourt18}, with cloud temperatures $T\sim (2-3)\tm 10^4\Kel$, densities $\nh\sim 10^{-2.5}\cmc$, and sizes comparable to $\lcmin\lsim 1\kpc$, which was marginally resolved in the sheet at their best resolution (\fig{ne_sheet_fil_map2d}, bottom-left panel). The authors found that this shattered structure only appeared when the cooling length at $T\sim 10^5\Kel$ was resolved,%
\footnote{The origin of this convergence criterion is unclear, but it is potentially related to the fact that at these temperatures, isochoric thermal instabilities are stable while unstable isobaric instabilities can only operate on scales smaller than $\lcool$. See discussion in \citeta{m21} and \citet{Das21}.} and is thus not seen in most cosmological simulations. The densities, metallicities, and spatial distribution of these cold clouds resembled observations of extremely metal-poor $(Z\lsim 10^{-3}\zsol)$ Lyman-Limit Systems ($\Nhi>10^{17.2}\cms$) which are otherwise difficult to explain \citep{Robert19, Lehner22}, lending observational support to the picture of shattering in the IPM.

\smallskip
Using the same \texttt{IPM} simulations, \citet{lu23} studied the internal structure of CW filaments at $z\sim 4$, finding a robust three-zone radial structure for the filament cross section. The outer filament is a virialized zone surrounded by a strong cylindrical accretion shock and filled with hot gas at the `virial temperature' of the dark matter filament. Interior to this is a multiphase zone with high turbulence and vorticity and small-scale cold clouds embedded in warm gas. The centre of the filament is a dense, isothermal core representing the cold streams that feed galaxies. While $\lcmin$ in this region was not resolved, cold clouds were present down to the resolution scale (bottom-right panel of \fig{ne_sheet_fil_map2d}).

% @@@@@@@@@@@@@@@@@@@@@@@@@@@@@@@@@@@@@@@@@
\section{scattering}
\label{se:scatt_app}
% @@@@@@@@@@@@@@@@@@@@@@@@@@@@@@@@@@@@@@@@@

% _________________________________________
{
\begin{figure}
    \centering
    \includegraphics[width=0.7\columnwidth, trim={1.8cm 0.8cm 2cm 0.8cm},clip]{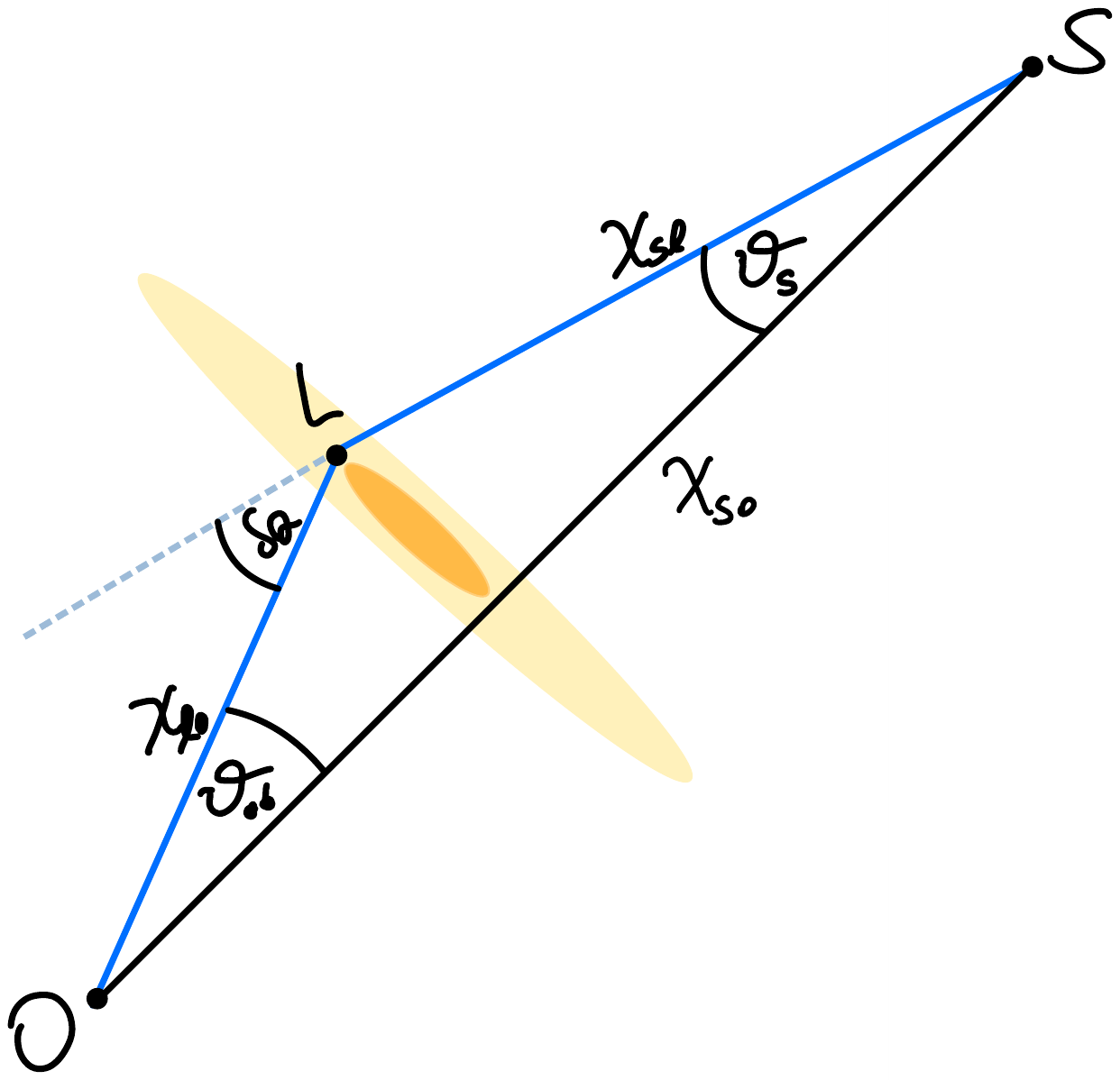}
    \includegraphics[width=0.85\columnwidth, trim={0.4cm 3.4cm 0.3cm 4cm},clip]{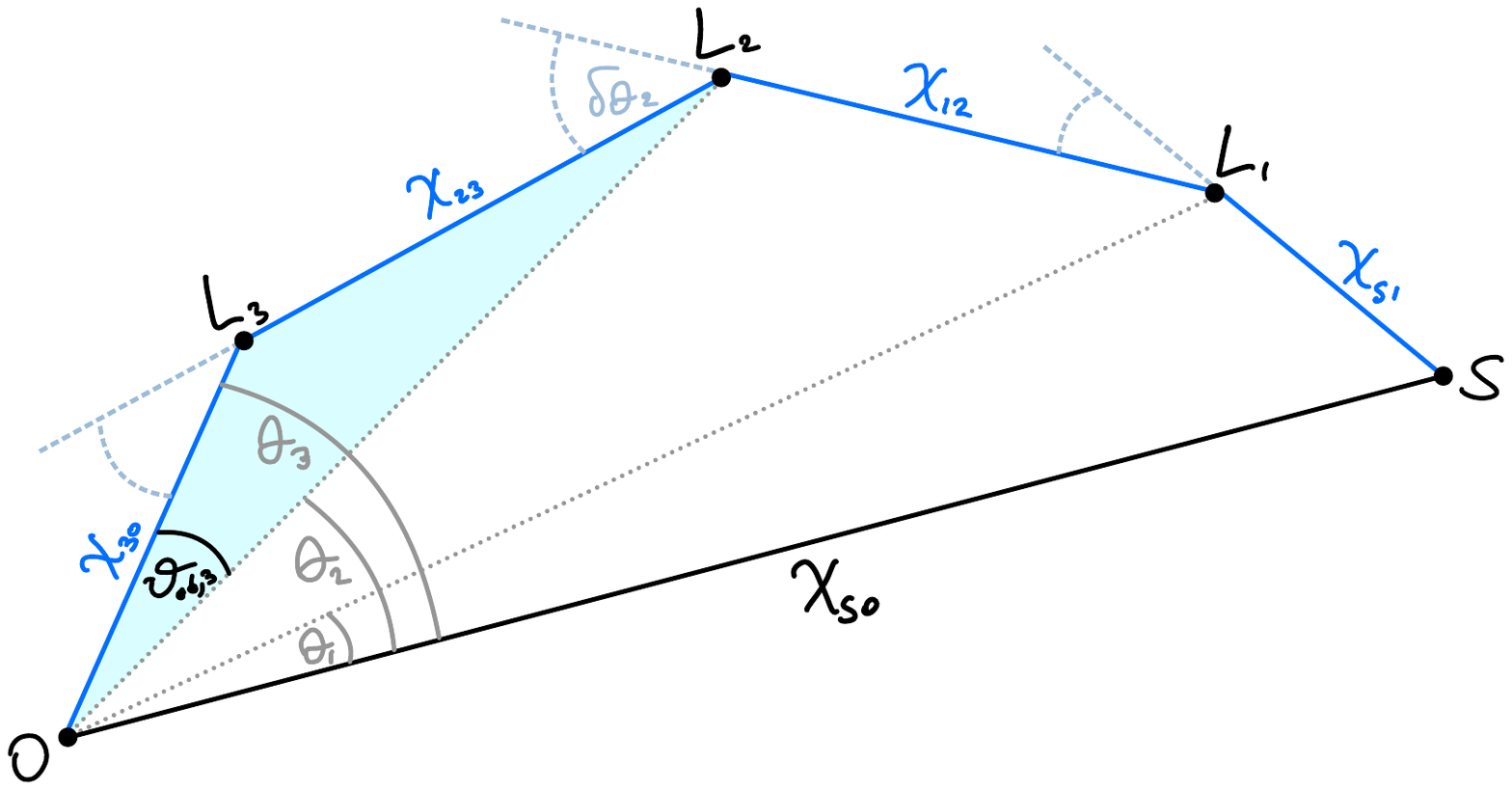}
    \caption{
    Geometric ray path (blue solid lines). 
    {\bf Top:} Geometric ray path with a single screen (located at $L$) between the source ($S$) and an observer ($O$). $\dth$ marks the scattering angle caused by the screen, and $\tho$ marks the observed angle. The scattering disc is indicated by the darker orange ellipse.
    {\bf Bottom:} Geometric ray path with a three-screen set-up, located at $L_{1,\, 2,\, 3}$, ordered respectively from the source ($S$) to the observer ($O$). 
    One can find the observed scattering angle of the $i$-th screen by $\tho[,\ii]=\theta_\mr{\ii}-\theta_\mr{\ii+1}$. Note that the source in the top panel is analogous to $L_2$ in the bottom panel, and the cyan-shaded triangle in the bottom panel highlights the geometry of the top panel.
    None of the quantities illustrated in these cartoons is to scale. 
    }
    \label{fig:time_delay}
\end{figure}
}% fig:time_delay - cartoon
% _________________________________________

% =========================================
\subsection[Inner scale of turbulence, li]{Inner scale of turbulence, $\li$}
\label{se:li}
% ----------------------------------------- 

As described in \se{scatt}, the expressions for the coherence scale and temporal broadening are different depending on whether the coherence scale is in the inertial range of turbulence or not, namely whether it is larger or smaller than the inner (dissipation) scale of turbulence. We here estimate this scale following \citeta{beniamini20}. We include here only a brief description of their derivation and refer the reader to section \S3.1 in \citeta{beniamini20} for further details.

\smallskip
In low-density astrophysical plasmas, characterised by a large mean free path ($l_\mr{mfp}$), the smallest eddies are unlikely to be smaller than the Larmor radius ($l_\mr{B}$). 
Assuming the proton Larmor radius satisfies $l_\mr{B}\ll l_\mr{mfp}$, the viscous length scale can be estimated by equating the dissipation time to the eddy turnover time. For a Kolmogorov spectrum of turbulence, the viscous scale is
\be
    l_\mr{vis} 
    \sim \frac{l_\mr{B}^{3/2}\lo^{1/4}}{(\Mach \,l_\mr{mfp})^{3/4}} 
    \sim 3.4\tm10^{7}\cm\ \pfrac{ \n\, \lof^{1/3}\, \B[\Bfv]^{-2} }{\Mach\,\Tf}^{3/4}
    \label{eq:l_vis}
\ee% eq - l_vis
where the Larmor radius is $l_\mr{B}={(3c^2\mpr \kb T)^{1/2}}{(e B)^{-1}}$, $B$ is the magnetic field, the mean free path of protons at temperature $T$ is $l_\mr{mfp}\sim {\kb T^2}/(\n e^4\ln\Lambda)$, the Coulomb logarithm is $\ln \Lambda\sim 20$, $\Mach$ is the Mach number of the outer scale,%
\footnote{
For an eddy of size $\ell$ with velocity $u_\mr{\ell}$, we assume a Kolmogorov spectrum for velocity fluctuations, giving in the inertial subrange $u_\mr{\ell}=u_\mr{o}(\ell/\lo)^{1/3}$, where the speed of the outer scale eddy is $u_\mr{o}$. 
The Mach number of the outer scale of turbulence is then defined as 
$\Mach=u_\mr{o}/u_\mr{p,rms}$, where $u_\mr{p,rms}={\mpr}/{(3\kb T)}$ is the average thermal velocity of the protons.
} 
$\Tf=T/10^{\Tfv}\Kel$, and $\Bf=B/10^{\Bfv}\G$.
Finally, the smallest eddy size is given by $\li=\max(l_\mr{vis}, l_\mr{B})$. For these values $\li \sim 10^{\lifv} \cm$.

% =========================================
\subsection{The coherence scale and temporal broadening}
\label{se:lpi_tau_alpha}
% =========================================

The general forms of \eqs{lpi} and \eqm{taus} for the coherence scale and temporal broadening for an arbitrary turbulent spectral slope, $\alpha$, are 
\be
    \lpi \sim 
    \begin{cases}
        \Big(\cbeta\re\lamz\,\fa^{\frac{1}{2}} \lo^{\frac{1-2\alpha}{2}} \n \Big)^{-\frac{2}{1+2\alpha}} &  \li<\lpi<\lo 
        \\
         \,\Big(\cbeta[,i]\re\lamz \li^{-\frac{1-2\alpha}{2}} \fa^{\frac{1}{2}} \lo^{\frac{1-2\alpha}{2}} \n \Big)^{-1} &   \lpi<\li
    \end{cases} 
    \label{eq:lpi_fa_0}
\ee% eq:lpi_fa_0

{\be
    \tau \sim 
    \begin{cases}
        \frac{\lamo \D}{4\pi^2 c} \Big( \cbeta^2\, \re^2\, \lamz^{\frac{5+2\alpha}{2}}\,\fa\,\lo^{1-2\alpha}\,\n^2\Big)^{\frac{2}{1+2\alpha}}
         &  \li<\lpi<\lo 
        \\ \\
        \frac{\lamo\D}{4\pi^2 c}\, \cbeta[,i]^2 \re^2 \lamz^3  \li^{-(1-2\alpha)} \fa \lo^{1-2\alpha} \n^2
         &   \lpi<\li
    \end{cases} 
    \label{eq:tau_obs_alpha}
\ee}% eq:tau_obs_alpha
where
{
\be
&\cbeta^2 = 
    \widetilde{\cbeta}^2 2^{2-\beta}\pi\beta\frac{\Gamma(-\beta/2)}{\Gamma(\beta/2)}
    &&  \li<\lpi<\lo, 
        \\
&\cbeta[,i]^2 = \widetilde{\cbeta}^2 
    \frac{\pi\beta}{4}\Gamma(-\beta/2)
    &&  \lpi<\li
\ee
$\beta=2\alpha+3$, and $\widetilde{\cbeta}^2=C_n^2 \lo^{\beta-3} \n^{-2}=({\beta-3})/({2 (2\pi)^{4-\beta})}$, assuming $\beta>3$.
}

% _________________________________________
{
\begin{figure}[ht]%
    \centering%
    \includegraphics[width=0.85\columnwidth]{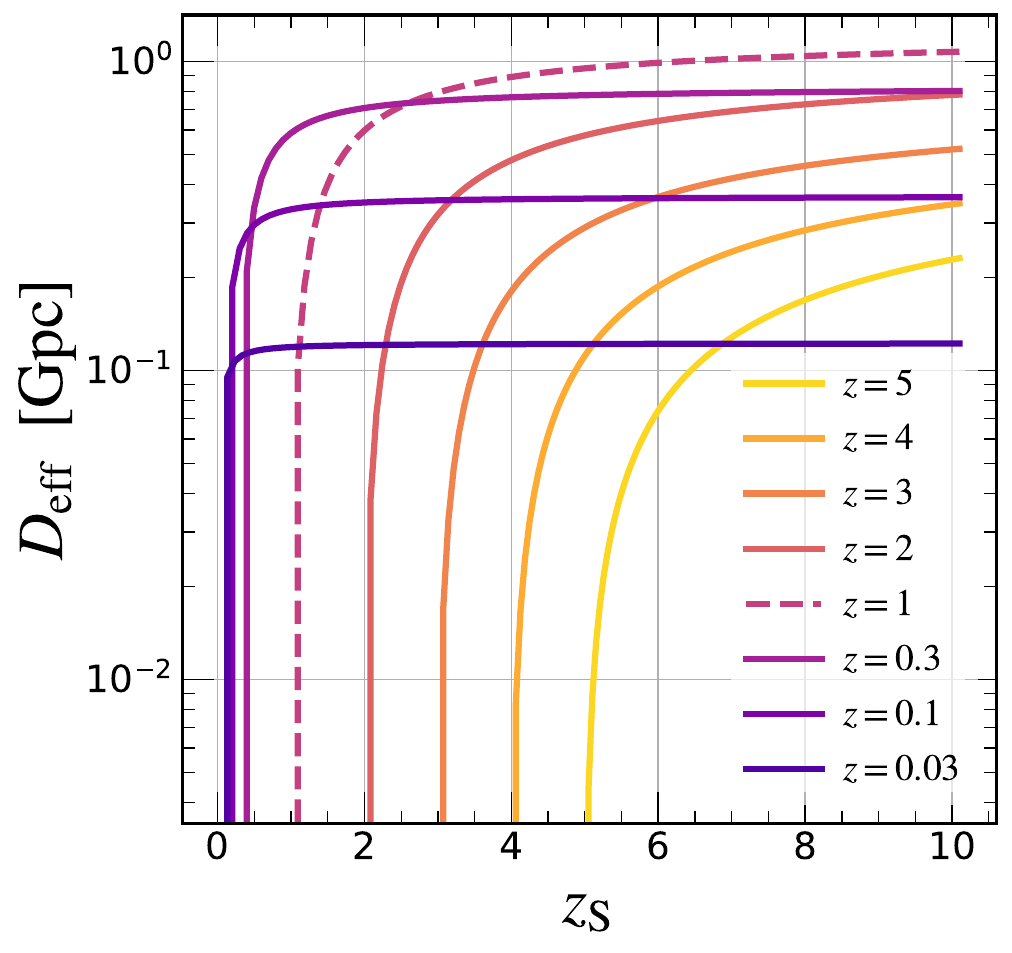}
    \caption{
    The effect of screen location. We show here the effective angular diameter distance, $\D$, as a function of the source redshift, $\zs$, for screens at different $z$. 
    For example, for a screen at $z=1$ (highlighted as a dashed line), the distance $\D(\zs>1)$ is roughly constant at $\zs\gsim 2$. Similar behaviour is seen for screens at different redshifts.
    }
    \label{fig:Deff}
\end{figure}
}% fig:Deff
% _________________________________________

% -----------------------------------------
\subsubsection[The effective angular diameter distance, Deff]{The effective angular diameter distance, $\Deff$:}
% -----------------------------------------

In \fig{Deff}, we illustrate the effect of the source redshift, $\zs$, on the effective angular diameter distance, $\Deff$, for a screen at a given $z$. We show $\Deff$ as a function of $\zs$, and the different coloured lines indicate different screen redshifts. For a screen at a given $z$, $\Deff$ saturates to a constant value at a somewhat higher $\zs$, rendering $\Deff$ not very sensitive to the choice of $\zs$.

% =========================================
\subsection{Temporal broadening from geometrical optics}
\label{se:time_delay}
% =========================================

When a ray is scattered by an angle $\dth$, it travels a longer path length, thus delaying the arrival time to the observer (=temporal broadening). Here, we describe the broadening of a signal through simple geometric considerations.

% -----------------------------------------
\subsubsection{Deflection Angle}
\label{se:theta}
% -----------------------------------------

Consider a wave with wavevector $\bf{k'}$ in the rest frame of the plasma screen. In the absence of perturbations, the wavefront, representing a surface of constant phase, is perpendicular to the direction of propagation, namely to $\bf{k'}$. However, if the wave is propagating through a turbulent medium, random perturbations induce random phase shifts. If the phase shift is constant along the wavefront, then this does not change the direction of propagation, since the original wavefront remains at constant phase. However, changes in the phase shift along surfaces perpendicular to $\bf{k'}$ result in a new surface of constant phase which has an angle with respect to the original wavefront. This is the deflection angle. For example, if ${\bf{k'}}=k'{\bf{\hat{z}}}$ then the original wave is proportional to $\exp(-ik'{\rm z})$ while the perturbed wave is proportional to $\exp[-i(k'{\rm z}+\Delta \phi(x,y))]$. It is thus straightforward to see that the deflection angle of the wave is
\be 
    \label{eq:theta}
    \dth = \frac{\abs{{\bf{\nabla}}_{\perp}\Delta \phi}}{k'},
\ee 
where ${\bf{\nabla}}_{\perp}\Delta \phi$ is the gradient of the phase shift in the plane perpendicular to $\bf{k'}$. 

\smallskip
Inserting \eq{Dphi} into \eq{theta}, we see that $\dth \propto (\ell/\lpi)^{\alpha-1/2}$. For $\alpha<0.5$ as expected based on observational grounds \citep{rickett77, romani86}, and in particular for our fiducial choice of $\alpha\sim 1/3$, the angle grows larger with decreasing eddy size. Thus, as stated previously, the dominant eddies for diffractive scattering, yielding both a large scattering angle and large flux modulation, are eddies of scale $\ell\sim \lpi$, assuming that $\lpi>\li$. For such eddies, one can approximate $\abs{{\bf{\nabla}}_{\perp}\Delta \phi} \sim \sqrt{2}/\lpi$, where we have assumed a phase shift of $\sim 1$ radian on scales of $\lpi$, and the factor $\sqrt{2}$ comes from the sum in quadrature $[(\partial \Delta \phi / \partial x)^2+(\partial \Delta \phi / \partial y)^2]^{1/2}$. If $\lpi<\li$, then eddies at scales $\lpi$ are not available, and we instead have $\abs{{\bf{\nabla}}_{\perp}\Delta \phi} \sim \sqrt{2}\Delta \phi(\li)/\li$. Altogether, 
\be
    \dth \sim
    \begin{cases}
        \frac{\lamz}{\sqrt{2}\pi \lpi} 
        &  \li<\lpi<\lo, 
        \\
          \frac{\lamz}{\sqrt{2}\pi \li}\pfrac{\li}{\lpi}^{\frac{2\alpha+1}{2}}
        &   \lpi<\li,
    \end{cases}
    \label{eq:theta_2}
\ee%
where $\lamz = 30\cm\,\nuf^{-1} \opzo[]^{-1}$ is the wavelength and $\lpi$ in each regime is given by \eq{lpi}.

% -----------------------------------------
\subsubsection{Temporal broadening due to a single screen}
% -----------------------------------------

The relation between the deflection angle and the temporal broadening is best understood by first considering Euclidean geometry. Consider a source at distance $\dso$ from the observer and $\dsl$ from a scattering screen. The distance of the screen to the observer is $\dlo=\dso-\dsl$. Consider two rays travelling from the source to the observer, one in a straight line and the other deflected by an angle $\dth\ll 1$ by the screen. Straightforward geometrical considerations show that in order for the deflected ray to reach the observer, it must have left the source at an angle $\ths=(\dlo/\dso)\dth$. The difference in arrival time between the two rays is thus 
\be 
\label{eq:time_delay_1}
    \taus[_{Euc}] \simeq \frac{\delta \theta^2}{2c}\frac{\dsl\dlo}{\dso}\equiv \frac{\delta \theta^2}{2c}\Deff =\frac{\lambda^2\,\Deff}{4\pi^2 c\,\lpi^2},
\ee 
where the final expression is for $\li<\lpi$ and we have defined the effective distance is $\D=\dsl\dlo/\dso\sim \min(\dsl,\dlo)$ (see \fig{Deff}).

\smallskip
The geometric time delay in an expanding universe can be derived following \citet{schneider92}. Here we assume a flat cosmology. 
The Robertson-Walker metric for a photon along a null geodesics is $ds^2=0=a(\eta)^2(d\eta^2-d\chi^2)$, where $\chi$ is the comoving coordinate, $a(\eta)$ is the expansion factor, and $\eta$ is the conformal time defined by 
\be
    d\eta\equiv \frac{c\, dt}{a(t)}.
    \label{eq:dt_eta}
\ee
Since $ds^2=0$, the geometric conformal time delay is the difference between the paths $\deta=\delta \chi$, giving
\be
    \deta=\chisl+\chilo-\chiso
    \label{eq:deta}
\ee
where $\chisov$, $\chislv$ and $\chilov$ are the comoving vectors which form a triangle.
$\chiso$ is the distance from the source (at redshift $\zs$, point "S" in the top panel of \fig{time_delay}) directly to the observer (at $z_0=0$, point "O" in the figure), $\chisl$ is the distance between the source and a point on the screen (at $z$, point "L") with an angle of $\ths$ with respect to $\chiso$, and $\chilo$ is the distance from point "L" to the observer, after deflection of the ray at point "L" on the screen by an angle $\dth$. The angle between $\chisov$ and $\chilov$ is the observed angle $\tho$.
In a flat universe, the distance between the source and the observer is $\chiso = \chisl \cos{\ths} + \chilo\cos{\tho}$, using this, in the limit of small angles, with \eq{deta} 
\be
    \deta\approx \chisl \frac{\ths^2}{2} + \chilo \frac{\tho^2}{2} 
    \sim \frac{\chisl\chilo}{\chiso} \frac{\dth^2}{2} \frac{\chilo+\chisl}{\chiso}.
\ee
where from the top panel of \fig{time_delay}, $\ths=\dth\chilo/\chiso$ and $\tho=\dth\chisl/\chiso$. Since $(\chilo+\chisl)/\chiso\approx 1$,
\be
    \deta 
    \approx \frac{\chisl\chilo}{\chiso} \frac{\dth^2}{2}.
    \label{eq:deta_fin}
\ee 
In a flat universe, the angular diameter distance is given by
\be
    d_{21} 
    = \frac{c}{1+z_2}\int_{z_1}^{z_2} \frac{dz}{H(z)}
    = \frac{1}{1+z_2}(\chi(z_2)-\chi(z_1))
\ee
where $z_2>z_1$ and $H(z)$ is the Hubble parameter. Namely, we can replace the comoving distances with the angular diameter distances, such that $\chiso=(1+\zs)\dso$, $\chisl=(1+\zs)\dsl$ and $\chilo=\opzo[]\dlo$.
Finally, using \eq{dt_eta} and \eq{deta_fin}, the geometric time delay can be expressed as
\be
    \delta t_\mr{geom} 
    \approx \frac{\dsl\dlo}{\dso}\frac{\dth^2\opzo[]}{2c} =\frac{\lamo^2\,\Deff}{4\pi^2 c\, \opzo[]\,\lpi^2},
\ee
which is equivalent to \eq{time_delay_1} with $\lamo$ replaced by $\lamz=\lamo/\opzo[]$, the wavelength in the rest frame of the screen, the distances $\dsl$, $\dlo$ and $\dso$ replaced by the relevant angular diameter distances (so $\dlo\ne\dso-\dsl$), and $\Deff$ replaced by $\Deff\opzo[]$.

% -----------------------------------------
\subsubsection{Temporal broadening due to multiple screens}
\label{se:tau_multi_screens}
% -----------------------------------------

The geometry involved in multiple scattering screens is sketched in the cartoon in the bottom panel of \fig{time_delay}%
\footnote{See similar illustrations in \citet{schneider92, feldbrugge23} in the context of lensing.}, illustrated for the case of three screens between the source and the observer. The associated time delay for multiple screens is,
\be
    \taut 
    = \sum_{i} \frac{(1+z_{i})}{2c} \frac{d_\mr{\ii,o} d_{i-1,i}}{d_\mr{\ii-1,o}} \dth[\ii]^2 
\ee
where $z_{i-1}>z_{i}$, $d_{j,i}$ is the angular diameter distance between $z_{j}$ and $z_{i}$, and the observed angle of the $i$-th screen is $\tho[,\ii]=\theta_{i}-\theta_{i-1} = \dth[\ii] {d_{i-1,i}}/{d_\mr{\ii-1,o}}$.

% =========================================
\section{Halo profiles}
\label{se:profiles}
% =========================================

% _________________________________________
{
\begin{figure}[ht]%
    \centering%
    \includegraphics[width=0.8\columnwidth]{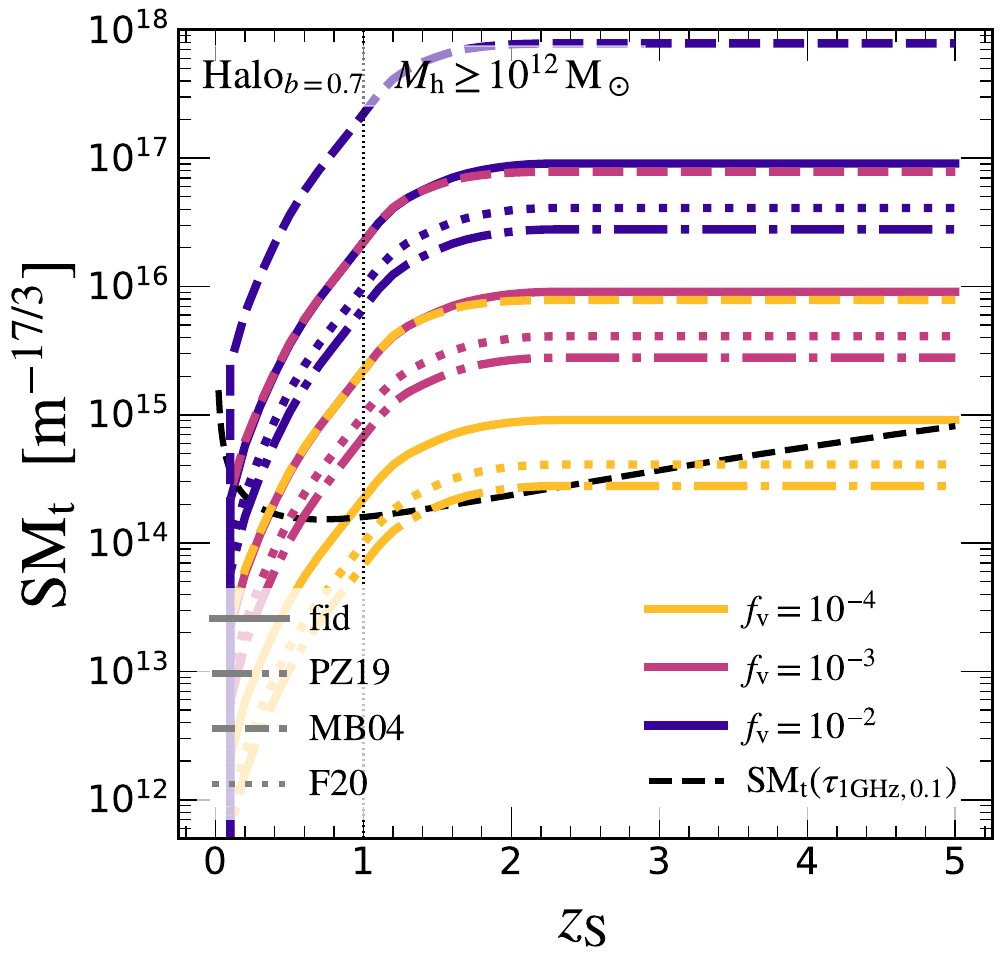}
    \caption{
    Similar to \fig{taut}, here, we assume several halo profiles (left legend) and repeat the estimate of the total $\smt$. The solid lines indicate our fiducial model, the dashed-dot lines assume the fiducial profile of \citet{prochaska19-halo}, the dashed coloured lines represent the profile by \citet{mb04}, and the dotted lines indicate the profile of \citet{faerman20}.
    }
    \label{fig:sm_halo_fv_pro}
\end{figure}
}% fig:sm_halo_fv_pro
% _________________________________________

To estimate the effect of a density and pressure profile on our results, we briefly repeat one of our key results using several profiles and compare it to our fiducial model. 
In \fig{sm_halo_fv_pro}, we repeat the calculation of the cumulative SM for haloes, assuming different profiles.
To account for the SM through a halo with a given profile, we assume our fiducial impact parameter $b=0.7$, and divide the path length $\DL$ through the halo into $8$ bins. We then find the cloudlets' density $\n[,\ii]$ and scale length $\lc[,\ii]$ (as in \fig{clump_prop}) according to the pressure profile in each $i$th bin. We sum up the SM along $\DL$ to find the overall SM of the halo, and finally, calculate the total SM across cosmic time, as we did in \figsiti{smt}{taut}.
Here, the different line styles (see the left legend) indicate different assumptions on the profiles. Our fiducial model is shown in solid (same lines as \fig{taut}), and we compare it to the mNFW profile \citep[marked as PZ19 in the legend][dash-dot lines]{prochaska19-halo}, the profile by \citet{mb04} (MB04, dashed), and the profile by \citet{faerman20} (F20, dotted).%
\footnote{For the mNFW profile, we normalise the temperature profile such that $T(R)=\Tv$, where $R$ obeys $\rho_\mr{dm}(R)=\rhov$. For the other two profiles, we use the normalisation described in the corresponding papers.}
Focusing on the results at $\zs\lsim1$, our fiducial model somewhat overestimates the SM compared to PZ19 and F20 and significantly underestimates the total SM by haloes when compared to MB04. Namely, for PZ19 and F20, our fiducial model roughly corresponds to a somewhat smaller impact parameter $b\sim0.5-0.6$, and a larger one, $b\sim0.9$, for the MB04 profile.

% @@@@@@@@@@@@@@@@@@@@@@@@@@@@@@@@@@@@@@@@@
\end{appendix}
% @@@@@@@@@@@@@@@@@@@@@@@@@@@@@@@@@@@@@@@@@
\label{LastPage}
\end{document}